# Carbon dioxide in silicate melts: A molecular dynamics simulation study.

Bertrand Guillot and Nicolas Sator

*Laboratoire de Physique Théorique de la Matière Condensée, Université Pierre et Marie Curie (Paris 6),*

*UMR CNRS 7600, case courrier 121, 4 place jussieu, 75252 Paris cedex 05, France.*

**ABSTRACT**

We have performed a series of molecular dynamics simulations aimed at the evaluation of the solubility of $CO_2$ in silicate melts of natural composition (from felsic to ultramafic). In making in contact within the simulation cell a supercritical $CO_2$ phase with a silicate melt of a given composition, we have been able to evaluate, (i) the solubility of $CO_2$ in the P-T range 1473-2273K and 20-150 kbar, (ii) the density change experienced by the $CO_2$-bearing melt, (iii) the respective concentrations of $CO_2$ and $CO_3^{2-}$ species in the melt, (iv) the lifetime and the diffusivity of these species and (v) the structure of the melt around the carbonate groups. The main results are the following.

(1) The solubility of $CO_2$ increases markedly with the pressure in the three investigated melts (a rhyolite, a mid-ocean ridge basalt and a kimberlite) from about ~2 wt% $CO_2$ at 20 kbar to ~25 wt% at 100 kbar and 2273K. The solubility is found to be weakly dependent on the melt composition (as far as the present compositions are concerned) and it is only at very high pressure (above ~100 kbar) that a clear hierarchy between solubilities occurs (rhyolite<MORB<kimberlite). Furthermore at a given pressure the calculated solubility is negatively correlated with the temperature.

(2) In $CO_2$-saturated melts, the proportion of carbonate ions ($CO_3^{2-}$) is positively correlated with the pressure at isothermal condition and is negatively correlated with the temperature at isobaric condition (and vice versa for molecular $CO_2$). Furthermore, at fixed (P,T) conditions the proportion of carbonate ions is higher in $CO_2$-undersaturated melts than in the $CO_2$-saturated melt. Although the proportion of molecular $CO_2$ decreases when the degree of depolymerization of the melt increases, it is still significant in $CO_2$-saturated basic and ultrabasic compositions at high temperatures. This finding is at variance with experimental data on $CO_2$-bearing glasses which show no evidence of molecular $CO_2$ as soon as the degree of depolymerization of the melt is high (e.g. basalt). These conflicting results can be reconciled with each other by noticing that a simple low temperature extrapolation of the simulation data predicts that the proportion of molecular $CO_2$ in basaltic melts might be negligible in the glass at room temperature.

 (3) The carbonate ions are found to be transient species in the liquid phase, with a lifetime increasing exponentially with the inverse of the temperature. Contrarily to a usual assumption, the diffusivity of carbonate ions into the liquid silicate is not vanishingly small with respect to that of $CO_2$ molecules: in MORB they differ from each other by a factor of ~6 at 1473K and only a factor of ~2 at 2273K. Although the bulk diffusivity of $CO_2$ is governed primarily by the diffusivity of $CO_2$ molecules, the carbonate ions contribute significantly to the diffusivity of $CO_2$ in depolymerized melts.

(4) Concerning the structure of the $CO_2$-bearing silicate melt, the carbonate ions are found to be preferentially associated with NBO's of the melt, with an affinity for NBOs which exceeds that for BOs by almost one order of magnitude. This result explains why the concentration in carbonate ions is positively correlated with the degree of depolymerization of the melt and diminishes drastically in fully polymerized melts where the number of NBO's is close to zero. Furthermore, the network modifier cations are not randomly distributed in the close vicinity of carbonate groups but exhibit a preferential ordering which depends at once on the nature of the cation and on the melt composition. However at the high temperatures investigated here, there is no evidence of long lived complexes between carbonate groups and metal cations.

# 1. Introduction

The distribution, recycling and storage of carbon in the Earth are of fundamental importance to understand the global carbon cycle (Zhang and Zindler, 1993; Keppler et al., 2003; Hirschmann and Dasgupta, 2009). Based on various methods of estimation (volatiles in mid-ocean ridge basalts and ocean island basalts, mass balance of volcanic and non volcanic gas emissions, fluid inclusions in xenoliths,..) the reported estimates of carbon abundance in the Earth's mantle lie in the range 50-500 ppmw (Javoy et al., 1986; Bottinga and Javoy, 1991; Zhang and Zindler, 1993; Jambon, 1994; McDonough and Sun, 1995; McDonough and Rudnick, 1998; Holloway, 1998; Marty and Tolstikhin, 1998; Deines, 2002; Saal et al., 2002; Cartigny et al. 2008). Degassing of $CO_2$ at mid-ocean ridges (MOR) may give information on the source region. However, owing to the very low solubility of $CO_2$ in tholeiitic basalts (~0.5 ppmw at one bar, Dixon (1997)) nearly all MOR lavas have exsolved their $CO_2$-rich vapor as they approach the ocean floor. Consequently their $CO_2$ contents reflect mostly the pressure of eruption and not that of the source region.

Notable exceptions are the rare $CO_2$-rich tholeiitic glasses known as *popping* rocks collected along the mid-Atlantic ridge (Hekinian et al., 1973). Their vesicularity is very high (more than 10% in volume) with vesicles containing up to ~1 wt% $CO_2$ (Sarda and Graham, 1990; Graham and Sarda 1991; Javoy and Pineau, 1991); these MORB glasses are g potentially those that most faithfully preserve the $CO_2$ contents of the source region. In comparison, MORB melt inclusions, trapped in olivine phenocrysts med in submarines MORBs of from the East Pacific Rise (Siqueiros transform fault), suggest a C-content for the source region as low as 10-70 ppmw (Saal et al., 2002). In contrast the C-content of Kilauea primary magma is estimated about 1900 ppmw (Gerlach et al., 2002) and that of the parental magma associated with the Society hotspot could be as high as ~3800 ppmw (Aubaud et al., 2005). comparison, the solubility The concentration of carbon in upper mantle minerals (e.g. olivine) does not exceed a few tens of ppmw (Keppler et al., 2003; Shcheka et al., 2006) whereas in transition zone minerals (e.g. wadsleyite) and lower mantle silicates (e.g. Mg-Perovskite) the carbon solubility is below the limit of detection (<0.2 ppmw, Shcheka et al. (2006)). Although the reduced forms of carbon (graphite and diamonds) are also present in the mantle, their contribution to the carbon budget is considered of minor importance (however see Stagno and Frost, 2010). So a largely accepted conclusion is that carbonate phases must be the dominant lithology storing the carbon throughout the mantle (Dasgupta and Hirschmann, 2010).$CO_2$ observed at MORs and at hot spot volcanoes is the final step of a suite of mechanisms depending on various parameters (composition of the source region, depth and extent of partial melting, bubble nucleation at depth, style of degassing,..) and beginning deep in the mantle. But how deep is the question.

Petrological and experimental investigations have shown that subducted carbonate phases are stable near the transition zone and at lower mantle conditions (Kraft et al., 1991; Brenker et al., 2007). Moreover the presence of carbonated minerals in upper mantle lithology (e.g. carbonated peridotite) decreases the temperature of the solidus by up to 300°C, the resulting melt evolving continuously with increasing temperature from a carbonatitic composition near the solidus (with ~5wt% $SiO_2$) to a carbon-rich silicate melt ($\geq$ 20 wt% $SiO_2$) at higher temperature (Eggler, 1973; Wyllie and Huang, 1975; Eggler, 1976; Wendlandt and Mysen, 1980; Falloon and Green, 1989; Dalton and Presnall, 1998; Lee and Wyllie, 1998; Wyllie and Lee, 1998; Hammouda, 2003; Gudfinnsson and Presnall, 2005; Dasgupta and Hirschmann, 2006; Brey et al., 2008; Ghosh et al., 2009; Foley et al., 2009). So, there is a growing evidence that the onset of partial melting beneath MORs may well begin at depths approaching 300 km (~90 kbar), producing 0.03-0.3 % carbonatitic liquid (Dasgupta and Hirschmann, 2006).2005), a conductivity anomaly which can be explained by ~0.1% volume fraction of carbonatitic melts percolating in the silicate matrix (Gaillard et al., 2008). Furthermore, the observation of carbonates and carbonatite melts located near the transition zone or the uppermost lower mantle (Brenker et al., 2007; Walter et al., 2008) suggests that a partial melting at these depths could be triggered by the intersection with a geotherm or by a plume going through this lithology. Indeed it has been shown recently (Ghosh et al., 2009) that a magnesiocarbonatite-like melt can be generated by incipient melting of carbonated peridotite at pressure around 200 kbar in the temperature range 1460-1500°C.

Although the existence of carbonatite melts at depth in the mantle is becoming a well supported hypothesis (Bell and Simonetti, 2010) it is a matter of fact that carbonatite lavas are very rare (Mitchell, 2005). An example of a volatile rich magma, dominated by $CO_2$, from a deep origin ($\geq$250 km) is given by kimberlite volcanism (for a review see Sparks et al., 2006). There is much debate about the nature of the parental melt (Ringwood et al., 1992; Edgar and Charbonneau, 1993; Gudfinnsson and Presnall, 2005; Girnis and Ryabchikov, 2005; Keshav et al., 2005; Kopylova et al., 2007; Canil and Bellis, 2008; Mitchell, 2008; Sparks et al., 2009, Brey et al., 2008; Kjarsgaard et al., 2009) because kimberlite samples are contaminated with crustal and mantle xenoliths and are strongly altered by serpentinisation. Based on bulk-rock geochemical analysis and experimental melting of various carbonate-silicate assemblages, reconstructed compositions of kimberlite primitive melts are ultrabasic magmas that are silica-poor (<35 wt%) and MgO+CaO rich (>45 wt%) with $CO_2$ contents of the order of 20 wt% or more.heir trace element characteristics (Keshav et al., 2005; Brey et al., 2008). However a lot of important thermophysical data are still lacking. Neither their thermodynamics properties (equation of state) nor the solubility of $CO_2$ as function of composition,  pressure and temperature are well constrained, parameters which have a direct influence on magma ascent and eruption scenarios (Spera, 1984; Wilson and Head III, 2007; Kavanagh and Sparks, 2009).

Hence , it appears that the melting of various carbonated lithologies at depth in the mantle could be the origin of virtually all C-bearing silicate melts observed at the surface of the Earth. Of course the eruption of these lavas and the associated $CO_2$ emission are the final result of a long suite of events starting at depth with the incipient melting of the primary carbonated rock mantle, the evolution of this magma  fate will depend on the local geotherm, the melting rate of the silicate mantle during upwelling, the occurrence of liquid-liquid immiscibility in $CO_2$-rich melts and the potential recrystallization of the silicate melt, the triggering of bubble nucleation and the degassing path of $CO_2$. In this scenario a key parameter is the solubility of $CO_2$ in the melt and its evolution with pressure (depth) and composition from carbonatitic to basaltic. Moreover, the $CO_2$ -contents may have a strong influence on the physical properties of the melt (e.g. density, viscosity, and wetting properties) and on the dynamics of magma ascent (Spera, 1981; 1984). But these quantities are still poorly constrained for $CO_2$-rich silicate melts (Brearley and Montana, 1989; White and Montana, 1990; Lange, 1994; Minarik, 1998; Bourgue and Richet, 2001; Ghosh et al., 2007).

In the last three decades an important effort has been devoted to the experimental investigation of the solubility of carbon dioxide in silicate melts of various compositions , the starting materials are powdered silicates to which a $CO_2$-giving agent is added in a sufficient amount to ensure $CO_2$ saturation (but not always), either under the form of carbonates ($CaCO_3$, $MgCO_3$,..) or silver oxalate ($Ag_2C_2O_4$). This assemblage is pressurized and held at superliquidus temperature in a piston-cylinder apparatus during a sufficient run time (>10 mns) to make sure that the vapour-melt equilibrium is reached. Quenched glasses (affected by some crystallinity) are recovered by turning off the electrical power to the furnace and are analyzed by [14]C beta-track mapping techniques, gas chromatography or electron microprobe analysis in early works (Mysen et al., 1975, 1976; Mysen, 1976; Brey and Green, 1976; Eggler and Mysen, 1976; Brey, 1976; Mysen and Virgo, 1980a,b; Rai et al., 1983)and mostly by infrared (IR) spectroscopy afterwards (for a review up to the nineties see Blank and Brooker, 1994). IR spectroscopy has been used to estimate determinant in estimating the speciation of $CO_2$ in silicate melts because $CO_2$ molecule and carbonate ion CO32- exhibit well separated vibration bands in the IR absorption spectrum (for a review up to the nineties see Blank and Brooker, 1994). The relative proportion of molecular and carbonate species measured in C-bearing glasses was strongly dependent on the silicate composition, a highly polymerized melt (e.g. albite) favoring $CO_2$ molecules when in a depolymerized melt (e.g. diopside) only carbonate ions are detected. Thus the speciation and $CO_2$ solubility in various binary and ternary oxide mixtures and in natural compositions as well, have been reported within the pressure range 2-20 kbar (Fine and Stolper, 1985a,b; Stolper et al., 1987; Tingle and Aines, 1988; Stolper and Holloway, 1988; Shilobreyeva and Kadik, 1989; Fogel and Rutherford, 1990; Mattey et al., 1990; Brey et al., 1991; Pan et al., 1991; Mattey, 1991; Thibault and Holloway, 1994; Jakobsson, 1997; Brooker et al., 1999, 2001a,b; King and Holloway, 2002; Morizet et al., 2002; Behrens et al., 2004a,b; Botcharnikov et al., 2005, 2006; Behrens et al., 2009; Lesne et al., 2010; Morizet et al., 2010). For natural compositions, molecular $CO_2$ is the dominant species in rhyolitic melts (Fogel and Rutherford, 1990; Blank et al., 1993; Tamic et al., 2001; Behrens et al., 2004a) and is rapidly counterbalanced by carbonate ions with decreasing silica contents along the suite rhyolite → dacite → andesite → phonolite → icelandite (Behrens et al., 2004b; King and Holloway, 2002; Botcharnikov et al., 2006; Morizet et al., 2002; Jakobsson, 1997), with only carbonate ions appearing in IR spectra of basaltic compositions (Stolper and Holloway, 1988; Pan et al., 1991; Mattey, 1991). However this drastic evolution of the speciation with composition is hardly noticeable in the $CO_2$ solubility, and varies little from one composition to the other at given (P,T) conditions, at least for natural compositions and in the limited (P,T) range for which solubility data are available $SCO2 =$ 0.25 wt% in rhyolite, 0.30 wt% in dacite, 0.42 wt% in andesite and 0.43 wt% in basalt, see Fig.8 in at 20 kbar and T=1400°C, $SCO2 = 1.12$ wt% in andesite and 1.53 wt% in phonolite after Brooker et al. (2001a) and 1.49 wt% in basalt after Mattey (1991) and Pan et al. (1991). In fact, according to Brooker et al. (2001a), the solubility is found to be not only a function of the non bridging oxygen (NBO) content of the melt (expressed as NBO/T where T represents tetrahedral network former cations), but also of the nature of the network modifying cations. Thus in polymerized melts the solubility increases with the number of alkalis while the concentration of molecular $CO_2$ decreases rapidly and is compensated by carbonate species increased abundance. In depolymerized melts where only carbonate ions are detected, the solubility varies with the composition in metal cations, the solubility increasing with the following order K>Na>Ca>Mg~Fe (for a thermodynamic treatment of $CO_2$ solubility data, see Spera (1980)). Furthermore, in natural compositions (from felsic to ultrabasic) the $CO_2$ solubility is found to be negatively correlated with the temperature at a given pressure (Fogel and Rutherford, 1990; Pan et al., 1991; Thibault and Holloway, 1994; Brooker et al., 2001a; Morizet et al., 2002).

A prerequisite of the aforementioned experimental investigations is that the speciation of $CO_2$ and its solubility in the silicate liquid equilibrated at high T and P are well preserved in the quenched glasses recovered at room temperature. Although this question has been much debated in the literature it is only recently that specific investigations have been carried out ( Morizet et al., 2001, 2007; Nowak et al., 2003; Porbatzki and Nowak, 2001; Spickenbom et al., 2010). Thus it has been convincingly demonstrated by annealing C-bearing glasses below the glass transition temperature Tg, that the equilibrium between molecular and carbonate species continues to evolve between Tg and room temperature, the chemical equilibrium shifting towards molecular $CO_2$ with increasing temperature. Consequently the speciation observed in C-bearing glasses at room temperature could underestimate significantly the amount of $CO_2$ molecules in melts equilibrated at superliquidus temperature. Another intriguing result is that the diffusion coefficient of $CO_2$ in silicate melts is nearly invariant over a wide range of composition (Nowak et al., 2004). If $CO_2$ molecules predominate in silica rich melts, and only carbonate ions are present in basaltic glasses, either $CO_2$ molecules and $CO_3^{2-}$ ions have similar diffusivities or $CO_2$ molecules diffuse faster than $CO_3^{2-}$ ions but, in such a case, $CO_2$ molecules should be present in a significant concentration in basaltic melts.supports the idea that the speciation observed in quenched glasses may not be representative of the one in silicate melts equilibrated above liquidus temperature.

Our purpose is to evaluate by molecular dynamics (MD) simulation the solubility, speciation, diffusivity and structure of $CO_2$ in silicate melts at thermodynamic conditions related to those existing in the Earth's upper mantle. $CO_2$-bearing silicate melts in the liquid phase at high temperature. To quantify the role played by the melt composition on the $CO_2$ solubility, we have investigated three compositions, a rhyolite, a mid-ocean ridge basalt and a Mg-rich kimberlite over the (P,T) range 20-150 kbar and 1473-2273K.

## 2. Simulation method

The basic idea of our MD simulation is equilibration of a supercritical fluid phase composed of $CO_2$ molecules ($T_c = 304.1$ K) and a silicate melt of a given composition, ystem being equilibrated at fixed temperature and pressure. As the $CO_2$ molecules are free to move across the interface separating the supercritical $CO_2$ phase and the liquid silicate, the $CO_2$ solubility is obtained by evaluating, at equilibrium, the average number of carbon atoms present within the melt in the course of the MD run. The model accounts for thanks to a model potential, the chemical reaction acting between $CO_2$ molecules and the oxygens of the silicate: $CO_2 + (O^{2-})_{melt} <=> (CO_3^{2-})_{melt}$. Hence, with the solubility represented evaluated by summing the average numbers of $CO_2$ and $CO_3^{2-}$ species present in the silicate melt the biphasic system being at equilibrium.

A MD simulation consists of solving iteratively the equations of motion of an assembly of particles interacting via a given force field (see Allen and Tildesley, 1987). The latter one generally is empirical, i.e. adjusted so as to reproduce some properties of the real material (Van Gunsteren and Berendsen, 1990). The success of the MD simulation relies on the accuracy of the interaction parameters. There are three different sets of interaction involved: $CO_2$-$CO_2$ interaction, $CO_2$-silicate interaction and silicate-silicate interaction. The derivation of these interaction potentials is not straightforward and is described in details in Appendix A. The complete set of interaction parameters is listed in Table 1.

sts  The latter one generally is empirical, i.e. adjusted so as to reproduce some properties of the real material (Van Gunsteren and Berendsen, 1990). The interatomic potential for ionic systems (e.g. silicates) is usually modeled by a Born-type analytic term ($\sim e^{-r/\rho}$) for repulsion between atoms combined with a Coulomb term for ion-ion interactions and supplemented by an attractive contribution ($\sim -1/r^6$) accounting for the dispersion forces between atoms. In the case of molecular systems (e.g. $CO_2$), the interaction potential between two molecules is expressed as the sum of a short range contribution describing the repulsion-dispersion interactions between atoms (e.g. the Lennard-Jones potential $\sim 1/r^{12} - 1/r^6$), plus an electrostatic contribution which expresses the multipolar interactions between molecules (e.g. $CO_2$ has a quadrupole moment). When chemical bonds are involved (e.g. the C-O bond in carbonate ion), the association-dissociation reaction can be conveniently described by a Morse potential. An alternative to this empirical description would be to use the *ab initio* MD method which e density functional theory (Parr and Yang, 1995). Although this method is not free of approximations (Perdew et al., 2009), it has the great advantage [3]-10[4] times more expensive in computer time than classical MD) that its use is limited, for the time being, to small system size (a few hundreds of atoms instead of several thousand with classical MD) and short trajectories (~10 ps instead of 10,000 ps). Consequently, an accurate evaluation of $CO_2$ solubility of reach ay computational resources

$CO_2$, whereas we have developed a specific model to describe the interactions between $CO_2$ and the silicate melt. Thus we have made use of a force field recently proposed (Guillot and Sator, 2007a,b). the potential energy $u_{ij}$ between two atoms of the melt (where $i, j$ = Si, Ti, Al, $Fe^{3+}$, $Fe^{2+}$, Mg, Ca, Na, K, and O) is given by,

$$(r_{ij}) = z_i z_j / r_{ij} + B_{ij} e^{-rij/\rho ij} - C_{ij}/r_{ij}^6 \qquad (1)$$

$r_{ij}$ is the distance between atoms $i$ and $j$, $z_i$ is the effective charge associated with the ion $i$, and where $B_{ij}$, $\rho_{ij}$ and $C_{ij}$ are parameters describing repulsive and dispersive forces between the ions $i$ and j (values of parameters are given in Table 1).

between $CO_2$ molecules are described by the model of Zhang and Duan (2005). With this model the $CO_2$ molecules are assumed to be linear and rigid with a C-O bond length, $l_{C-O}$ = 1.162 A. The quadrupole moment of the $CO_2$ molecule is represented by point charges located on C and O atoms. The potential energy $u(1,2)$ between molecules 1 and 2 is then given by,

$$(1,2) = i \in 1 j \in 2 [ \ 4\varepsilon_{ij}[(\sigma_{ij}/r_{ij})^{12} - (\sigma_{ij}/r_{ij})^6] + q_i q_j / r_{ij} \ ] \qquad (2)$$

$i,j$ run over the three atoms of the corresponding molecule (1) or (2), $q_i$ is the effective charge associated with the atom $i$ (C or O), $\varepsilon_{ij}$ and $\sigma_{ij}$ are the Lennard-Jones parameters for the corresponding pair of atoms (C-C, C-O and O-O). When this potential is implemented into a MD code (the potential parameters are given in Table 1), it shows an excellent predictability of thermodynamic, transport and structural properties in a wide (P,T) range encompassing both the liquid state and the supercritical region (Zhang and Duan, 2005). In particular the volumetric properties of supercritical $CO_2$ are very well reproduced all the available experimental data from triple point to 1100 K and 8 kbar. $CO_2$ molecules to react with the oxygens of the silicate melt to form carbonate ions, we have modified the model of Zhang and Duan by allowing the $CO_2$ molecule to be flexible. The following harmonic intramolecular potential for bending has been introduced,

$$_\theta = 1/2 \ k_\theta (\theta - \pi)^2 \qquad\qquad (3)$$

$_\theta$ and $\theta$ are the force constant and bending angle, respectively. To reproduce the $\nu_2$ bending mode of vibration in supercritical $CO_2$ ($\nu_2 \sim 670 \ cm^{-1}$, see Yee et al., 1992), a value of 444 kJ/mol/rd$^2$ was assigned to $k_\theta$. We have checked by MD simulation that the modified potential (with flexibility) leads to the same volumetric properties , Furthermore, several studies have shown that $CO_2$ molecules tend to adopt nonlinear geometries at high temperatures and high pressures (Ishii et al., 1996; Saharay and Balasubramanian, 2004, 2007; Anderson et al., 2009), .

Recently, $CO_2$ in the P-T range 1-80 kbar and 300-700 K. Their data for the isotherm 700 K are reported in Fig.1 and are compared with our MD calculations (computational details are given in the next section). The agreement between the two sets of data is satisfactory Also shown in Fig.1 is the prediction of the EOS of Sterner and Pitzer (1994) for the same isotherm and which is based essentially on the same  EOS of Span and Wagner (1996) but which includes , <P<700 kbar and 1000<T<4000K but with a large uncertainty of several hundred of degrees on the estimated temperature)as also as some data for highly compressed solid $CO_2$ The isotherm predicted by the Sterner and Pitzer EOS is close to al., but as emphasized bylatter onese, in contrast with our simulated fluid which is slightly less compressible (see Fig.1). Keeping these findings in mind we have evaluated the high temperature isotherm 2273K and compare the results with the prediction of the Sterner and Pitzer EOS (see the insert in Fig.1). The two isotherms are close to each other up to ~30 kbar but deviatering the aforementioned tendency of the Sterner and Pitzer EOS to overestimate the density at very high pressure and in absence of reliable experimental data in the P-T range under investigation, the PVT properties of our simulated $CO_2$ fluid are likely close to those of the real fluid at these extreme conditions.

$CO_2$ molecule of the supercritical $CO_2$ phase enters into the silicate melt, its interaction potential energy with the melt is given by,

$$_{CO2...melt} = i=1,..NouCO2...Oi \ + \ j \in N1 \ ,..Nn \ uCO2...Xj \qquad\qquad (4)$$

$O_i$ is the oxygen $i$ of the melt (total number $N_o$) and $X_j$ is the cation of species $j$ ($N_j$). The interactions between the $CO_2$ molecule and the cations of the melt (second term of the right hand side of eqn.(4)) are discussed first. These interactions can be decomposed into pair contributions between C, $O_c$ and $O'_c$ atoms of the $CO_2$ molecule and the cation $X_j$ of species $j$,

$$uCO2...Xj = \ uC...Xj \ + uOC...Xj + \ uOC...Xj' \tag{5}$$

$_C$ = +0.5888 e, see Table 1), the carbon-cation interactions $u_{C-Xj}$ are assumed to be governed only by the electrostatic repulsion between their positive charges. As for the oxygen-cation interactions $u_{Oc-Xj}$ and $u_{O'c-Xj}$, they are modeled by the attractive electrostatic interaction between the oxygen of $CO_2$ ($q_O$ = -0.2944 e, see Table 1) and the cations, supplemented by repulsion-dispersion terms of the same kind than those used to describe the cation-oxygen interactions in the pure silicate melt (see eqn.(1)). In order to reduce the number of adjustable parameters, the repulsion-dispersion paramaters $B_{ij}$, $\rho_{ij}$ and $C_{ij}$ (where i= $O_c$ or $O'_c$ and j is the index of the cation) are assumed to be identical to those describing the oxygen-cation interactions in the oxides of the pure silicate melt. In summary, the potential energy between one $CO_2$ molecule and a cation $X_j$ of the melt is given by,

$$_{CO2...Xj} = i \in [CO2] \ ( \ q_i z_j / r_{ij} + B_{ij} e^{-rij/\rho ij} - C_{ij}/r_{ij}^6 ) \tag{6}$$

$i$ runs over the three atoms of the $CO_2$ molecule (with charges $q_i$), $j$ stands for the cation $X_j$ (with charge $z_j$), and $r_{ij}$ is the atom-cation distance.

$CO_2 + (O^{2-})_{melt} \leftrightarrow (CO32-)_{melt}$, where a $CO_2$ molecule reacts with an oxygen to form a carbonate ion, we introduce a Morse potential between the carbon atom of the $CO_2$ molecule and an oxygen of the melt,

$$uC...OmMorse_e \ [ \ (1-e^{-(r-l)/\lambda})^2 - 1 \ ] \tag{7}$$

$_e$ is the dissociation energy of the C...$O_m$ bond, $l$ is the equilibrium distance of the bond (taken equal to the C-O bond length in $CO_2$ molecule, i.e. $l$ = 1.162A) and $\lambda$ is the effective width of the potential. Values assigned to the potential parameters $D_e$, $l$ and $\lambda$ (see in Table 1) are close to those chosen by Pavese et al. (1996) in their simulation[3]). Thus the Morse potential is effective only at short distances when the C...$O_m$ distance is shorter than ~1.6 A (covalent bonding). At larger distances, the oxygen of the melt interacts with the carbon and oxygen atoms of the $CO_2$ molecule through a non covalent interaction described by a sum of electrostatic, repulsive and dispersive contributions, namely

$$u_{CO2...Om} = uC...OmMorse + i{\in}[CO2] \left( q_i z_{Om}/r_{iOm} + B_{iOm}e^{-riOm/piOm} - C_{iOm}/r_{iOm}^6 \right) \qquad (8)$$

$CO_2$ molecule, $q_i$ is the effective charge associated with the atom $i$ of the $CO_2$ molecule, $z_{Om}$ is the effective charge of the oxide anion (-0.945 e) and $B_{iOm}$, $\rho_{iOm}$ and $C_{iOm}$ are parameters describing repulsive and dispersive forces (values of parameters are given in Table 1 and are discussed hereafter).

It is noteworthy that the chemical reaction, $CO_2 + (O^{2-})_{melt} \leftrightarrow (CO32-)_{melt}$ , is controlled mainly by the energy barrier generated by the repulsion+dispersion terms (~$Be^{-r/\rho}-C/r^6$) between the oxide ion of the melt and the two oxygens of the $CO_2$ molecule. As many other multiply charged anions, CO32- is not electronically stable as isolated species (Boldyrev et al., 1996) because the repulsion between the extra electrons cannot be overcome. This is the confinement in condensed phase which renders this anion to be thermodynamically stable (Stefanovich et al., 1998; Cherginets and Pebrova, 2003). Thus, during the simulation, the condition for a $CO_2$ molecule to capture an oxygen of the melt (forming a carbonate ion) is that the two moving partners find a path to overcome the energy barrier separating each other, a barrier whose the height is modulated by the instantaneous configuration of the atoms of the melt (the opposite situation will take place for the dissociation of CO32-). CO32-/$CO_2$ ratio and the value of the $CO_2$ solubility. Basing one's argument on this finding we have adjusted the potential parameters $B_{OcOm}$ and $C_{OcOm}$ $CO_2$ solubility in a silicate melt of natural composition, namely a tholeiite at 1673 K and 20 kbar ($S_{CO2}$~1.5 wt% after Mattey (1991), Pan et al. (1991) and Dixon (1997)). A systematic investigation of the $CO_2$ solubility in this melt as function of the B and C parameters shows that the range of values leading to the which substantiates the robustness of our simulation results. The final parameters of the potential are listed in Table 1.

the variation of the potential energy of the isolated complex $CO_2$--$O_m$ in equatorial configuration is shown in Fig.2 as function of the $C$--$O_m$ separation and for different angulations $\theta$ of the approaching $CO_2$ molecule. For $\theta$ =180° the potential energy curve exhibits only one minimum at $r_{C--Om}$ ~ 2.6A, a configuration corresponding to a molecule-ion complex stabilized by quadrupole-ion interaction. For $\theta$ =160°, the curve presents two minima at 2.15A and 1.35A, respectively, these minima being separated by an energy barrier culminating near 1.55A. The potential well around 2.15A is characteristic of the molecule-ion complex whereas the narrow potential well around 1.35A corresponds to the emergence of the carbonate ion after barrier crossing. For $\theta$ = 140° only the carbonate ion is observed with an energy minimum located around 1.23A.

When a $CO_2$ molecule is immersed in In practice, the carbonate ion is formed in the silicate melt when the distance between the carbon atom of a $CO_2$ molecule and an oxygen of the melt is less than 1.6A.

The interatomic potentials ve were implemented into the DL_POLY2.0 code for MD simulations (Smith and Forrester, 1996). The equations of motion for ions and molecules were solved with a time step of 1 fs ($10^{-15}$s). The simulation box is parallelepipedic (with $L_x > L_y = L_z$) and is replicated in 3D. The long range coulombic interactions are accounted for by a Ewald sum with $\alpha L_x$=5~7 where $\alpha$ is the width of the charge distribution on each atom. Different system sizes were investigated at once to check the consistency of the results and also because the $CO_2$ solubility in the investigated silicate melts increases drastically at high pressures. The smallest system was constituted by a fluid phase composed of 96 $CO_2$ molecules in contact with a melt formed by ~1,000 ions and the largest system was composed of 1,000 $CO_2$ molecules + 2,000 ions. MD calculations were performed in the microcanonical ensemble (NVE) and were averaged over very long simulation runs (10-50 ns i.e. $10^7$- $5.10^7$ time steps) to reach a sufficient accuracy (~ ±5% for the solubility). Supplementary calculations were performed in the isobaric-isothermal ensemble (NPT) to evaluate the statistical uncertainties on the PVT properties of the system. For the (T,P) range investigated here (1473-2273K and 20-150kbar), the statistical uncertainties of the MD runs are $\Delta P/P$~±5%, $\Delta \rho/\rho$~±1% and $\Delta T/T$~±1%, respectively. Further details will be given in the discussion of the results when necessary.

To cover a representative set of natural compositions, three silicate melts have been investigated (see Table 2): a rhyolite, a MORB and a kimberlite (the Mg-rich composition investigated by Brey et al., 1991). Starting configurations of the simulation runs were prepared as follows. A cubic box of volume $L^3$ (with $L_x=L_y=L_z=L$) containing N molecules of $CO_2$ (e.g. N=300) was equilibrated at given (T,P) conditions (e.g. T=1673K and P=20kbar). For the same thermodynamic conditions, a silicate melt of a chosen composition (e.g. rhyolite) is simulated by performing a MD calculation with N ions (e.g. N~1,000) in a parallelepipedic box of volume $L^2L'$. After a period of equilibration (~1ns), the two boxes are joined together such as the new system is the fluid $CO_2$ in contact with the silicate melt, the all volume being now the sum $L^2(L+L')$ of the initial volumes. The MD calculation is next carried out at fixed (P,T) conditions until a thermochemical equilibrium is reached. The method of simulating, by MD, two phases in contact with each other via an explicit interface is not new and has long been used to investigate liquid-vapor, liquid-liquid and liquid-solid coexistence in molecular, ionic and metallic systems (Belonoshko, 1994; Alejandre et al., 1995; Belonoshko and Dubrovinsky, 1996; Belonoshko et al., 2000; Morris and Song, 2002; Guillot and Guissani, 2002a; Rivera et al., 2003). However we are aware of only one work dealing with chemically reacting species in two-phase simulation (Guillot and Guissani, 2002b).

For each evaluation of the solubility, we have carefully checked to ensure that thermochemical equilibrium was achieved before averaging. At the beginning of a simulation run there are  no $CO_2$ molecules in the silicate phase. Gradually some molecules of the supercritical $CO_2$ phase enter into the melt, diffuse and eventually react with the oxide anions. A stationary state is reached after about 1 ns when $CO_2$ molecules are exchanged continuously between the two phases, the $CO_2 \leftrightarrow_{CO32^-}$ reaction taking place only in the silicate melt or at the fluid-melt interface. The lifetime of $CO32-$ species in the silicate melt is much shorter than the mean residence time of $CO_2$ molecules in this phase (see later on in section 3.3). Every $CO_2$ molecule diffusing into the melt experiences a great number of interconversions before escaping from the melt. More precisely, for a system composed of 300 $CO_2$+1,000 ions, the mean residence time of a $CO_2$ molecule at 2273K is ~100 ps at 20 kbar and ~1ns at 100 kbar and during a 10 ns long simulation run, every $CO_2$ molecule of the fluid phase has passed through the silicate melt several times in experiencing many $CO_{2\leftrightarrow CO32^-}$ transformations.

Concerning the time scale for melt relaxation, in the case of liquid rhyolite, the most viscous liquid investigated here, the viscosity estimated by simulation is ~ 85 Pa.s at 1473K and 20 kbar, and the relaxation time is about ~ 8.5 ns (in using the Maxwell relation, $\tau = \eta/G_\infty$ where the shear modulus at infinite frequency $G_\infty$ is equal to ~ $10^{10}$ Pa.s, after Dingwell and Webb (1989)). Hence our rhyolitic melt is fully relaxed over a run time of 50 ns. In the case of basaltic and kimberlitic melts, their viscosity is much lower (e.g. for MORB, $\eta$ ~ 1.6 Pa.s at 1473K and ~ 0.08 Pa.s at 2273K, whereas for Kimberlite $\eta$ ~ 0.16 Pa.s at 2273K), which implies that these melts are relaxed in less than 1 ns when our run times are 10 ns or more. Notice that the viscosity of rhyolite is underestimated by several orders of magnitude by the simulation at 1473K. This is a defect of the force field used for silicates which tends to overestimate the diffusivity of network former ions in silicic melts for temperatures below ~2500K (for a discussion, see Guillot and Sator, 2007a). But this deviation is immaterial for the evaluation of the $CO_2$ solubility. In the case of basaltic and ultramafic melts, the low viscosity of these fragile liquids (Giordano and Dingwell, 2003; Russell et al., 2003) is well reproduced by the simulations.

The solubility is obtained by evaluating the average number of carbon atoms, which is also equal to the sum of $CO_2$ and $CO3^{2-}$ species present in the silicate phase during the simulation run. This is illustrated in Fig.1 where the running average of the number of carbon atoms into a basaltic melt at T=2273K and P=20 and 80 kbar is shown as function of running time for a 10 ns long simulation run. At P=20 kbar there are on average ~8 carbon atoms in the basaltic melt, the error bar being around ±5% (compare also the instantaneous value of the step and the running average in Fig.1). Notice that 8 molecules in a basaltic melt composed of 1,000 ions correspond to a $CO_2$ solubility of 1.6 wt% (in gCO2/gmelt). As the $CO_2$ solubility increases drastically with the pressure (a factor of ~10 between 20 and 80 kbar, see Fig.1), the statistical uncertainty becomes smaller at high pressures and is estimated about ±2% when P is in the range 50-150 kbar. Finally, we have checked that our simulations are fully reversible with pressure and temperature, the necessary condition that the thermochemical equilibrium is reached.

To evaluate with accuracy the $CO_2$ solubility in the silicate phase, a prerequisite is to have a well identified interface separating the two coexisting phases (melt and supercritical $CO_2$). As a matter of fact, when the solubility becomes high (≥10 wt%), it may happen that all the $CO_2$ molecules of the fluid phase enter into the melt, indicating that the melt is undersaturated in $CO_2$. In that case one has to increase the number of molecules of the fluid phase to reach saturation. Thus when ~100 $CO_2$ molecules are sufficient to reach saturation up to ~50kbar in MORB at 2273K, about ~500 molecules are required to reach saturation at 150kbar. For illustration, the mean density profile along the X-axis of the simulation box is shown in Fig.2 (see curve labeled MORB+$CO_2$) for a basaltic melt composed of 1,000 ions in equilibrium at 2273K and 50kbar with a supercritical $CO_2$ phase formed initially by 300 $CO_2$ molecules. A snapshot of the simulation cell is also shown for better illustration (notice that the melt has two surfaces in contact with the $CO_2$ supercritical fluid). In making an average over a long simulation run (here 10 ns) , the density profile $n(x)$ becomes very smooth and can be accurately fitted by a hyperbolic tangent function as it is commonly used in the literature (Chapela et al., 1977) for describing fluid-liquid interfaces,

$$n(x) = n_f + \tfrac{1}{2}\,(n_f - n_m)[\ \tanh(x-x^+)d - \tanh(x-x^-)d\ ] \tag{1}$$

where $n_f$ and $n_m$ are the bulk densities in the fluid phase and in the $CO_2$-bearing silicate melt, respectively, $x^+$ and $x^-$ are the positions of the Gibbs dividing surfaces associated with each interface and $d$ is the thickness of the interfaces. In addition to the total density profile, re also shown in Fig.2 are the partial density profiles associated either with C species (curve labeled $CO_2$ and expressed in g/cm$^3$ of $CO_2$) or with silicate species only (curve labeled MORB). The $CO_2$ solubility, in g/g, is then given by the ratio of the density of C species in the bulk melt divided by $n_m$ the density of the $CO_2$-bearing melt. As expected there is virtually no silicate moieties in the supercritical $CO_2$ phase. However, from time to time some metal cations (essentially Na and K and a few Si) escape from the melt and go into the fluid phase, a finding which indicates that the corresponding oxides are weakly soluble in supercritical $CO_2$. Besides, the presence of silicate elements in a $CO_2$-rich vapor phase has been inferred by Brey and Green (1976) in studying $CO_2$-bearing olivine melilitite melts.

# 3. Results

### 3.1. Solubility

The solubility of $CO_2$ in the three silicate melts at 2273K are presented in Figs. 3-5 7 as function of pressure (the results are also collected in a supplementary Table B1of the online version, see Appendix B). One important finding is that the solubility increases steadily with the pressure and reaches large values at high pressure ($SCO2 > 20$ wt% for P>90 kbar). It is noteworthy that the $CO_2$ solubility depends weakly on melt composition for pressures up to ~80 kbar. This finding is in a qualitative agreement with $CO_2$ solubility data obtained in the low pressure range 1~20 kbar with natural compositions from rhyolitic to basaltic (Tamic et al., 2001; Brooker et al., 2001a; Botcharnikov et al. 2005; Botcharnikov et al., 2006) and with mafic melts (Dixon, 1997; Brooker et al., 2001a; Behrens et al., 2009; Lesne et al., 2010) as long as the contents in alkalis and CaO are sufficiently low ($Na_2O + K_2O \leq 5$wt% and CaO $\leq 13$wt%) as is the case for MORB and Mg-rich Kimberlite investigated here. At higher pressure the kimberlitic melt becomes more $CO_2$-rich than the basaltic melt, the latter becoming slightly richer in $CO_2$ than the rhyolitic composition (e.g. ~40 wt% $CO_2$ in kimberlite at 120 kbar as compared with ~31 wt% in MORB and ~28 wt% in rhyolite). In contrast, the speciation ($CO_2$ versus $CO32-$) varies strongly with composition and pressure at fixed temperature (see the inserts in Figs. 3-5-7). Thus at fixed (P,T) conditions the molar fraction in carbonate ion, $XCO32- = NCO32-/(NCO32- + NCO2)$, increases gradually from felsic to basic and ultrabasic composition, a general trend which is in a qualitative agreement with IR data of the literature (Fine and Stolper, 1985a,b; Stolper et al., 1987; Stolper and Holloway, 1988; Fogel and Rutherford, 1990; Mattey et al., 1990; Brey et al., 1991; Pan et al., 1991; Mattey, 1991; Thibault and Holloway, 1994; Jakobsson, 1997; Brooker et al., 1999, 2001a,b; King and Holloway, 2002; Morizet et al., 2002; Behrens et al., 2004a,b; Botcharnikov et al., 2005, 2006). Furthermore, for the three simulated melts at isothermal condition (e.g. 2273K) the mole fraction of carbonate ion increases with the pressure. Nevertheless at 2273K, molecular $CO_2$ is found to be the dominant species at 20 kbar ($XCO2$~ 84% in rhyolite, ~71% in MORB and ~57% in kimberlite) and it is only at a higher pressure that carbonate ions become the dominant species (above ~35 kbar in kimberlite, ~50 kbar in MORB and ~85 kbar in rhyolite). At very high pressure (P≥120 kbar) the proportion of carbonate ions reaches 70% or more in the three melts. In order to quantify the influence of $CO_2$-contents on the speciation, we have also investigated $CO_2$-undersaturated melts at 2273K and 100 kbar with a fixed amount of $CO_2$ (2 wt%). As indicated in the inserts of Figs.3-5 , the mole fraction of carbonate ion is greater when the melt is $CO_2$-undersaturated, which implies that for fixed (P,T) conditions the proportion of molecular $CO_2$ in the silicate melt is maximum at saturation.

From a more quantitative point of view, our simulation data seem to be in conflict with IR data on quenched melts. As a matter of fact, in $CO_2$-bearing rhyolitic glasses only $CO_2$ molecules are detected by IR spectroscopy (Fogel and Rutherford, 1990; Blank et al., 1993; Behrens et al, 2004a), while in basaltic glasses only the vibration bands associated with carbonate ions are clearly visible (Pan et al., 1991; Mattey, 1991; Dixon et al., 1995). But in glasses of intermediate composition (e.g. dacite, andesite, jadeite and phonolite) both molecular and carbonate species are identified (Brooker et al., 1999; King and Holloway, 2002; Morizet et al., 2002; Behrens et al, 2004b; Botcharnikov et al., 2006). Furthermore, Morizet et al. (2001, 2007) ,and Porbatzki and Nowak (2001), Nowak et al. (2003) and Spickenbom et al. (2010) in studying various silicate compositions (jadeite, albite, dacite and phonolite) have convincingly demonstrated that the speciation preserved in quenched melts below the glass transition temperature changes with the annealing temperature and is different from the true speciation occurring in the melt in equilibrium at superliquidus temperature. As the concentration of molecular $CO_2$ increases with a higher annealing temperature, the concentration of molecular $CO_2$ in the equilibrated liquid is likely much higher than the one observed in the glass. The conclusion of these studies is that the abundance of molecular species increases and that of carbonate species decreases when the temperature of the melt increases (at constant pressure). To check if the simulation is able to reproduce this behavior we have evaluated the temperature dependence of the speciation in the temperature range 1473-2273K for rhyolitic and basaltic compositions, these two melts being $CO_2$-saturated at 20 kbar. In agreement with the experimental data discussed above the speciation is found to be strongly dependent on temperature, the higher the temperature of the melt the larger the concentration in molecular $CO_2$. For instance at 1473K the molar fraction of $CO_2$ is equal to ~62% in rhyolite and ~37% in MORB whereas at 2273K it amounts to ~84% and ~71%, respectively. Correlatively we have evaluated the temperature dependence of the $CO_2$ solubility in the MORB composition over the pressure range 20-100 kbar. As shown in Fig.6, the calculated solubility is negatively correlated to the temperature (i.e. the solubility tends to decrease when the temperature increases), this trend being more pronounced at high pressure than at low pressure (see the nearly constant solubility curve at P=20 kbar in the insert of Fig.6).

A nearly constant or a negatively correlated temperature dependence has been reported in tholeiitic glasses ( Pan et al., 1991) but also in albite (Stolper et al., 1987), diopside (Rai et al., 1983), olivine-leucitite (Thibault and Holloway, 1994; Holloway and Blank, 1994) , andesite, Mg- and Ca-rich melilitites (Brooker et al., 2001a), and in a haplo-phonolite melt (Morizet et al., 2002). However these studies were run on a rather restricted range of temperature (~1573-1873K) and pressure (~5-25 kbar). Coming back to our simulation data, it is noteworthy that under constant pressure the mole fraction of molecular $CO_2$ in the basaltic melt increases almost linearly with the temperature (see Fig.7). Thus a simple low temperature extrapolation of our high temperature data suggests that the proportion of molecular $CO_2$ could be virtually undetectable in a MORB melt far below the glass transition temperature ($T_g$~950K). This may explain why only carbonate ions are detected in basaltic glasses by IR spectroscopy.

When it is the question to compare our calculations with experimental data, a brief survey of the literature shows that Unfortunately there is a striking lack of $CO_2$ solubility data in natural melts above ~20-25 kbar. In the case of basaltic melts, the Henrian behaviour of the $CO_2$ solubility at very low pressure and in the kbar range is well documented (Stolper and Holloway, 1988; Pawley et al., 1992; Dixon et al., 1995; Jendrzejewski et al., 1997; Botcharnikov et al., 2005), whereas only a few data are available above 10 kbar (a set of experimental data is shown in Fig.6). In particular, for a MORB-like composition at 20 kbar and 1673K  several studies lead to a solubility value around ~1.5 wt% (Pan et al., 1991; Mattey, 1991; Dixon et al. 1995), a value that we have used to calibrate our model potential.

In investigating the melting of carbonated pelites in the pressure range 25-50 kbar, Thomsen and Schmidt (2008) have estimated by electron microprobe analysis and mass balance calculation a $CO_2$ solubility about ~5 wt% for a basaltic composition at 1573K and 35 kbar, a value which compares very well with our calculation for a MORB melt in this pressure range (see Fig.6 for a comparison). In investigating the partial melting of a carbonated eclogite, Hammouda (2003) identified the coexistence of a carbonatitic melt with a silicate melt in the P-T range 50-65 kbar and 1423-1573K. At 1573K and 60 kbar the silicate melt has a basaltic composition (51.5 wt% $SiO_2$, 14.5 wt% $Al_2O_3$ and 31.9 wt% CaO+MgO+FeO) but is richer in Ca and poorer in Mg than a MORB (18.9 wt% CaO and 4.1 wt% MgO instead of 11.9 wt% and 7.8 wt%, respectively). Its $CO_2$-contents is about 12.2 wt%, a value which fits well our simulation result for MORB in the corresponding P-T range, in spite of slight differences between the two melt compositions. However, in the aforementioned partial melting experiments a vapor phase ($CO_2$ bubbles) is rarely detected in the quench. Consequently the $CO_2$ contents measured in the corresponding melts only leads to a lower bound for the $CO_2$ solubility.

In rhyolitic melts, Fogel and Rutherford (1990) measured a $CO_2$ solubility about ~0.11 wt% at 1473K and 2 kbar and ~0.29 wt% at 5.5 kbar, while Tamic et al. (2001) reported very similar values at 1373K namely, ~0.11 wt% at 2 kbar and ~0.28 wt% at 5 kbar. All these values are close to a linear extrapolation based on the Henry's law evaluated at very low pressure (~620 ppmw/kbar, Blank et al. (1993)). So, to compare with these data, extensive calculations have been carried out in the low pressure range 1-10 kbar, where the $CO_2$ solubility in the rhyolitic melt is low and needs very long simulation runs to be accurately evaluated ($t_{run}$~50 ns). At 2273K the solubility amounts to ~0.07 wt% at 1 kbar, ~0.3 wt% at 5 kbar, and 0.6% at 10 kbar, in excellent agreement with the above data (considering that the temperature dependence of the solubility is negligible in this pressure range). Brooker et al. (1999) measured the solubility in a rhyolitic-like melt ($Na_2O-Al_2O_3-12SiO_2$) at 15 kbar in the temperature range 1823-1973K and found ~1.1 wt% $CO_2$, the effect of temperature on the solubility being found to be negligible. For comparison, at 2273K the simulation yields ~0.6 wt% at 10 kbar, and ~1.8 wt% at 20 kbar (see Fig.4), values which bracket very closely the aforementioned data at 15 kbar  (Brooker et al. (1999)). For information, only molecular $CO_2$ was detected by IR spectroscopy in rhyolitic melts up to ~5 kbar (Fogel and Rutherford, 1990; Tamic et al., 2001), but Blank et al. (1993) have pointed out that the vibration bands associate with carbonate ions were detected in a $CO_2$-bearing rhyolitic glass quenched from 1823K and 25 kbar. This is in accordance with our finding that in the rhyolitic melt, the concentration in carbonate ions decreases rapidly at low pressures (e.g. $XCO_{32^-}$ ~ 8% at 10 kbar, 5% at 5 kbar and only 3% at 1 kbar, see Table B1) and should be detectable by IR spectroscopy only at high pressure (may be at ~10 kbar or beyond).

With regard to kimberlitic melts, Brey et al. (1991) have made an estimate of the $CO_2$ solubility in a Mg-rich composition related to Group II kimberlites (Becker and Le Roex, 2006). This is this very composition that we have investigated here by simulation (see Table 2). However the inhomogeneity of the quenched melt, a mixture of glass+silicates+carbonates, yields only a very rough estimate of the true solubility in the equilibrated melt at superliquidus temperature (~5 wt% at 20 kbar and 1923K after Brey et al., 1991).Subsequently, the same composition was investigated by Girnis and Ryabchikov (2005) and a striking increase of the $CO_2$ solubility was pointed out above ~45 kbar with a solubility as high as ~21 wt% at 55 kbar and 1983K. However the composition of the quenched crystals (silicates+carbonates) under IR analysis was not specified, so it is not assured that the reported solubility corresponds to that in a melt with the original kimberlite composition. Alternatively, more accurate solubility data in kimberlitic compositions can be extracted from partial melting experiments of various silicate+carbonate assemblages. For instance the $CO_2$-contents of $CO_2$-bearing liquids in equilibrium with a iron-free garnet lherzolite phase assemblage have been determined by Dalton and Presnall (1998) and Gudfinnsson and Presnall (2005) over a broad range of temperature (1613-2073K) and pressure (30-80 kbar). Although the analyzed melt is an intergrowth of silicate and carbonate phases on quenching, its composition was evaluated with a good accuracy by electron microprobe analysis of the coexisting phases and the amount of $CO_2$ estimated by mass balance calculation but with a larger uncertainty (see the error bars in Fig.5). The composition of the melt evolved rapidly with the (P,T) conditions from carbonatitic near the solidus to more silica-rich at higher temperature. At 2073K and 80 kbar the composition of the melt (39.1 wt% $SiO_2$, 3.0 wt% $Al_2O_3$, 37.7 wt% MgO and 20.5 wt% CaO in a volatile free basis) as given by Gudfinnsson and Presnall (2005) is close to that of the simulated kimberlite, except that the iron deficiency in the experimental melt is compensated by a higher CaO content (20.5 wt% CaO instead of 8.8 wt% CaO + 10.1 wt% FeO in the simulated composition). In this melt the amount of $CO_2$ is nearly the same that in our simulated kimberlite at 80 kbar and 2273K (i.e. about 18.3±9.5 wt% as compared with 19.8 wt%). For a more exhaustive comparison, we have included in Fig.5 other data points extracted from the study of Gudfinnsson and Presnall (2005) and which corresponds to Mg-rich kimberlite compositions. Bearing in mind the various sources of uncertainty (e.g. differences in composition, possible bias induced by the quench and eventual $CO_2$-undersaturation of the investigated melt) the agreement between these experimental data and the simulation results is satisfactory.

### 3.2. Density

The presence of $CO_2$ in silicate melts has a direct influence on their volumetric properties and hence on the buoyancy of magmas at depth in the Earth's mantle. The density variation of $CO_2$-saturated silicate melt with pressure along the saturation curve at 2273K is shown in Fig.8 for the MORB composition and in Fig.9 for the kimberlite composition (see the dots and the dotted curve in the corresponding figure). In both cases the carbonated melt is less dense than the corresponding $CO_2$-free melt, the higher the pressure and the $CO_2$ contents the larger the deviation in density. Notice the peculiar shape of the saturation curve: a S-shaped curve for MORB and a curve exhibiting a reversal in density around 90 kbar for kimberlite. Also shown in Figs.8 and 9 are several compressibility curves associated with undersaturated melts bearing a given amount of $CO_2$ (e.g. 2; 5; 10; 20; 30 and 40 wt %). These curves have been obtained by fitting the simulation data with an analytical theory based upon a statistical model for hard sphere mixtures (a description of this model is beyond the scope of this article and will be given elsewhere). This model calculation reproduces accurately, thanks to fitted parameters, all the simulation data (i.e. solubility, speciation and compressibility of the $CO_2$-bearing silicate melt within the simulation uncertainties), and can be used as an analytical representation of the equation of state of the simulated melt. Notice that Ohtani and Maeda (2001) have measured by sink/float method the density of a dry MORB melt at 2473K and 149 kbar and found 3.52 g/cm$^3$, a value in excellent agreement with our own evaluation by simulation (3.50 g/cm$^3$) at the same (P, T) conditions.

Recently, in an attempt to quantify the stability of carbonated magmas at the top of the 410 km discontinuity n of the low velocity zone in the Earth's upper mantle, Ghosh et al. (2007) have evaluated the density of a carbonated MORB with ~5 wt% $CO_2$ at 195 kbar and 2573K. In using a sink/float method they found a value of 3.57g/cm$^3$ and our the simulation yields a value of 3.59 g/cm$^3$ at the same (P,T) conditions). More generally, our simulation data suggest that carbonated silicate melts could be buoyant near the 410 km discontinuity provided that a MORB-like melt contains at least ~2 wt% $CO_2$ and a kimberlitic melt ~5wt % $CO_2$. These estimations are based on the comparison between the density variation of the $CO_2$-bearing melts along the simulated isotherm 1873K and the mantle density profile *ak135* of Kennett et al. (1995) in the discontinuity region (corresponding roughly to the pressure range 130-140 kbar, see Figs.8-9). Without this $CO_2$-supply, deep seated melts could be trapped around the upper mantle-transition zone (for possible implications on the mantle convection, see Lee et al., 2010).

The partial molar volume **VCO2** is commonly used to estimate the influence of the volume occupied by $CO_2$ in silicate liquids. We have evaluated this quantity along the saturation curve simply by calculating the volume difference between carbonated and dry melt. The results are presented in Fig.10 for MORB and kimberlite. One notices that the error bars are very large at 20 kbar but decrease very rapidly at higher pressures. As a matter of fact, at low pressure the $CO_2$ solubility is low and the density difference between the dry melt and the carbonated one is very small (a few percent). So even a small uncertainty on the absolute value of the density (~ ±1%) has a significant effect on the evaluation of VCO2. At higher pressures the $CO_2$-content of the melt becomes appreciable, which increases the density difference and decreases the uncertainty on VCO2. As a general trend, VCO2 decreases with pressure and is marginally dependent on melt composition. Furthermore, VCO2 in silicate melts is significantly smaller than the corresponding quantity in the pure $CO_2$ fluid. Indeed, because the carbonate ions are tightly bonded to the melt network (see next section), $CO_2$ is much more confined and occupies less room in the melt that molecular $CO_2$ in its own supercritical fluid.

Liu and Lange (2003) have evaluated VCO2 of the CaCO$_3$ component measured in carbonate liquids near atmospheric pressure from the silicate-based values for the oxide component VCaO given in the literature (Lange and Carmichael, 1987). They obtained VCO2 = 25.8 cm$^3$/mol, a value compatible with our data at low pressure (see Fig.10). In using thermodynamic arguments and solubility data, these authors estimated VCO2 to be higher than 19.9 cm$^3$/mol in alkali silicate liquids, in the range 23-28 cm$^3$/mol in basaltic liquid and ~21.5 cm$^3$/mol in Ca-rich leucitite liquid at 1673K and 10-20 kbar (these data are represented by the rectangular box in Fig.10). Here again our results are compatible with these rough estimations. At very high pressure, Ghosh et al. (2007) obtain VCO2 of 21±1 cm$^3$/mol in a carbonated basaltic melt at 195 kbar and 2573K with 5 wt% $CO_2$ while in a MORB at the same thermodynamic conditions we obtain a slightly lower value (15.5±2.0 cm3/mol). Interestingly at 195 kbar and 2573K the molar volume of $CO_2$ in the supercritical fluid is expected to be about ~21 cm$^3$/mol , a value which can be considered as an upper bound for VCO2 in a silicate melt at the same (P,T) conditions.

### 3.3. Lifetime and diffusivity

To better appraise the behavior of the $CO_2$ diffusivity through the silicate network, it is convenient to estimate first the average lifetimes associated with carbonate ion and molecular $CO_2$ in the melt. With the help of a speciation operator for which a given $CO_2$ molecule is counted as a carbonate ion only if the separation between the carbon atom and an oxygen of the melt is less than 1.6A (the results are virtually invariant when the latter value is in the range 1.6-1.8A), we have evaluated for each individual trajectory of $CO_2$ molecules the time sequence produced by the speciation operator. These time sequences are next analyzed to give the lifetime distribution of $CO_2$ and CO32- species in the melt after averaging over all trajectories. The lifetime distributions of $CO_2$ and CO32- in a $CO_2$-saturated MORB at 2273K and 20 kbar are Fig.13. The two distributions are quasi exponential (i.e. P(t)~e$^{-t/\tau}$ , where P(t) is a probability of transition) with τCO2= 14 ps and τCO32- = 6.5 ps, respectively. In the theory of unimolecular reactions, lifetimes with an exponential distribution are said to be random (Forst, 1973), and the average lifetime is nothing but the decay time τ of the distribution. Furthermore it follows that concentrations and average lifetimes are proportional to each other (i.e. $N_{CO_3^{2-}}/N_{CO_2} = \tau_{CO_3^{2-}}/\tau_{CO_2}$ ), a relationship which allows us to check the accuracy of our calculations (e.g. ±10% on the value of $\tau$). All the results are collected in Table 3 for the three melt compositions at a fixed $CO_2$ concentration (2 wt%) and for different temperatures (1473-2273K) and pressures (20 and 100 kbar). One notices that $\tau_{CO_3^{2-}}$ and $\tau_{CO_2}$ increase exponentially with the inverse of temperature (not shown), the corresponding activation energies ($\Delta E_{CO_3^{2-}}$ = 80.6 kJ/mol and $\Delta E_{CO_2}$= 34.5 kJ/mol) being essentially independent of melt composition. The activation energy corresponds roughly to the energy difference between the bottom of the potential well associated with the reference species ($CO_2$ or $CO_3^{2-}$) and the top of the potential energy barrier separating the two species in the melt. On average in the melt $\Delta E_{CO_3^{2-}} > \Delta E_{CO_2}$ illustrates the behavior of the potential energy of the isolated complex $CO_2$---$O$ along the $C$---$O$ reaction coordinate (see Fig.A2). The consequence is that the average lifetime of $CO_3^{2-}$ species increases more rapidly when the temperature decreases than the average lifetime of $CO_2$ species. Correlatively, the proportion of carbonate ion increases drastically when the temperature decreases $\tau_{CO_3^{2-}}$ and $\tau_{CO_2}$ intersect each other the two species are in equal proportion). Concerning the evolution with pressure (at a given temperature and for a constant abundance in $CO_2$), the average lifetime of $CO_3^{2-}$ increases gently with the pressure (a factor of ~1.8 between 20 and 100 kbar, see Table 3) whereas that associated with $CO_2$ species decreases more drastically (a factor of ~5 between 20 and 100 kbar). This contrasted behavior is directly correlated with the strong increase of the concentration in $CO_3^{2-}$ with the pressure (see Figs.3-5). Finally, notice that in the temperature range investigated here (1473-2273K), the average lifetimes of the species never exceed ~100 ps and hence are much shorter than the length of the simulation runs (typically 10,000 ps or more), a finding which ensures, in retrospect, that chemical equilibrium is fully achieved in our simulations.

The self diffusion coefficient associated with the carbon atom has been evaluated from the mean square displacements of the corresponding atoms in the melt,

$$D_C = \lim_{t \to \infty} \frac{1}{N_c} \sum_{i=1}^{N_c} \frac{<(\vec{r_i}(t) - \vec{r_i}(0))^2>}{6t} \qquad (2)$$

where $i$ runs over all the carbon atoms ($N_C$) belonging to $CO_2$ and $CO_3^{2-}$ species of the melt, $r_i$ is the position of the carbon atom $i$ and where the bracket expresses an averaging over many origin times. In eqn.(2) $D_C$ is the self diffusion coefficient of $CO_2$, irrespective of the speciation, the carbon atoms belonging either to carbonate ions or to $CO_2$ molecules. Hence it is convenient to write the relationship,

$$D_C = x_{C2}\, D_{C2} + x_{C3}\, D_{C3} \qquad\qquad (3)$$

where $D_{C2}$ and $D_{C3}$ are the self diffusion coefficients associated with $CO_2$ and $CO_3^{2-}$ species, and $x_{C2} = N_{CO2}/N_C$ and $x_{C3} = N_{CO3^{2-}}/N_C$ their respective concentration. Although eqn.(3) is formally exact, the evaluation of $D_{C2}$ and $D_{C3}$ by simulation is not straightforward because we have to specify for each time step of the MD run the speciation of the carbon atom whose the mean square displacement is under investigation. Thus $D_{C2}$ and $D_{C3}$ are evaluated from the summation of all trajectories during which the speciation of the atoms is not changing. A consequence of this evaluation is that the error bars are rather large for $D_{C2}$ and $D_{C3}$ ($\sim \pm 30\%$) whereas $D_C$ is evaluated with a much greater accuracy ($\sim \pm 5\%$). The results are reported in Table 3 and those obtained at 20kbar are presented in Fig.11 as function of the parameter NBO/T where NBO is the number of non-bridging oxygens and T that of tetrahedral cations. Clearly, the total diffusion coefficient $D_C$ increases smoothly with NBO/T: from 2.2 $10^{-9}$ m$^2$/s in rhyolite at 2273K, to 2.6 $10^{-9}$ m$^2$/s in MORB and 4.8 $10^{-9}$ m$^2$/s in kimberlite. Notice that in the special case of Na aluminosilicate melts, Sierralta et al. (2002), Spickenbom and Nowak (2005) and Spickenbom et al. (2010) have observed that the bulk diffusivity of $CO_2$ does increase with NBO/T. Moreover, albeit the evolution of $D_C$ with NBO/T tends to be steeper when the temperature decreases, at 1473K the value of $D_C$ in the basaltic melt is only a factor of ~2 greater than in the rhyolitic melt. This variation is still small and it is in accord with the experimental data available in the literature for rhyolitic, intermediate and basaltic melt compositions (Watson et al., 1982; Watson, 1991; Fogel and Rutherford, 1990; Zhang and Stolper, 1991; Nowak et al., 2004; Spickenbom and Nowak, 2005) which indicate that carbon dioxide diffusion is only weakly dependent on the volatile-free melt composition at magmatic temperatures (for a review see Baker et al., 2005). As for the activation energy for $CO_2$ diffusion, our simulation data exhibit a slope for

log($D_C$) as function of (1/T) very similar to those measured experimentally (see Fig.12), without a clear dependence on composition. However the simulation seems to overestimate the diffusivity of $CO_2$ by roughly one order of magnitude with respect to the available experimental data (see Fig.12). This deviation could be caused by the overestimation of the ionic diffusivities (especially those of network former ions) in our simulated melts at magmatic temperatures (see Guillot and Sator, 2007a).

An important result obtained from the MD calculations is that the diffusivity of carbonate ions (DC3) is never negligible with respect to that of molecular $CO_2$ (DC2) even if the carbonate diffusivity is always smaller than that of molecular CO₂. This finding is illustrated in Fig.11 where the values of DC3 and DC2 are compared with each other for the three investigated compositions at 20 kbar. For instance at 2273K the ratio DC2/DC3 is about ~2 in MORB and kimberlite and ~3 in rhyolite. However, because of the abundance of molecular $CO_2$ at this temperature (~84% in rhyolite, ~67% in MORB and ~61% in kimberlite), the bulk diffusivity ($D_C$) is mainly governed by DC2. At lower temperatures the proportion of molecular $CO_2$ decreases in the melt and the contribution of the carbonate ions to the bulk diffusivity $D_C$ becomes more significant. Nevertheless $D_C$ is still dominated by the diffusivity of molecular $CO_2$ in a basaltic melt at 1473K (see Table 3). ($O_m$) of the melt gives another piece of information.

As illustrated in Fig.11, the diffusivity of CO32- and $O_m$ are very close to each other in basic and ultrabasic melts whereas in the silicic melt the carbonate ions diffuse more rapidly than the oxygens (a factor of 2~3 according to the temperature). This behavior can be understood by analyzing the structure of the melt around the carbonate ions (a detailed description is given in the next section). In a depolymerized melt like MORB at 1473K and 20 kbar, about 37% of oxygens are found to be non-bridging oxygens (NBO) and ~78% of carbonate ions are linked to these NBO's. In contrast, in rhyolite only ~7% of oxygens are NBO's but ~40% of carbonate ions are part of this population, the remaining ~60% being involved in bridging Si- CO32--Si bonds. Since in a given melt the NBO's are more mobile than BO's, one can infer that on average the diffusivity of carbonate ions and that of the oxygens are similar in NBO-dominated melts (e.g. MORB and kimberlite) as observed in Fig.11, whereas in BO-dominated melts (e.g. rhyolite) the diffusion of carbonate ions follows essentially the quicker diffusion of the small population of NBO's and not the slower diffusion of the majority of oxygens which are BO's. In summary, the diffusivity of bulk $CO_2$ in silicate melts at superliquidus temperatures is mainly dominated by the diffusive motions of $CO_2$ molecules whose the population is notable even in depolymerized melts. However the contribution of carbonate ions is never negligible at superliquidus temperatures and hence is more significant than generally assumed in the literature, because their mobility is closely related to that associated with the NBO's of the melt. These findings explain why the value of the $CO_2$ diffusivity in polymerized and depolymerized melts doesn't differ very much: there is some compensation between concentration ($CO3^2$- versus $CO_2$) and intrinsic diffusivities of these species. It is noteworthy that Nowak et al. (2004) and Spickenbom et al. (2010) in an attempt to interpret the quasi invariance of their $CO_2$ diffusion data with the composition of the melt was lead to the conclusion that molecular $CO_2$ should exist in a significant amount even in depolymerized melts. A conclusion fully supported by our calculations.

### 3.4. Structural parameters in $CO_2$-bearing silicate melts

The structural role of $CO_2$ in silicate melts is a vast topic and our purpose is not to be exhaustive but rather to evaluate some relevant microscopic parameters which characterize the dissolution of carbon dioxide. Thus the average geometry of the carbonate ion is planar with a O-C-O angle (the one associated with the two rigid C-O bonds) about 139° with $\Delta\theta_{1/2} = \pm5°$, and a C--$O_m$ mean distance about 1.23 A between the carbon atom and the oxygen of the melt forming bond. On average, the O-C-O angle is found to be virtually invariant with the melt composition and the thermodynamic conditions investigated here. Although in our model the carbonate ion is slightly asymmetric with two different C-O bonds (1.16 and 1.23 A), we believe that this feature is immaterial in a silicate melt where the ionic environment is chemically and dynamically fluctuating in contrast with crystalline carbonates where the $CO3^2$- groups are barely distorted from $D_{3h}$ symmetry (e.g. calcite, magnesite or sodium carbonate). Moreover even in solid carbonates and in molten alkali carbonates (Chessin et al., 1965; Koura et al., 1996; Santillán et al., 2005) the C-O bond may vary from 1.25 to 1.31A depending on the cation (Na, K, Ca or Mg), the thermodynamic conditions (Kohara et al., 1998 ; Markgraf and Reeder, 1985) and the crystal symmetry as in shortite for instance (Dickens et al, 1971). As for the $CO_2$ molecules in the simulated melt, they are slightly bent ($\theta_{O-C-O}$ ~171° with $\Delta\theta_{1/2} = \pm7°$), the magnitude of the bending being essentially insensitive to the melt composition and to the thermodynamic conditions investigated here.

In analyzing the atomic configurations generated by MD simulation we have quantified the local structure around $CO_2$ and $CO3^{2-}$ species in the melt. For that we have evaluated the pair distribution functions (PDF) between $CO_2$ and $CO3^{2-}$ species and the elements of the melt. As far as the carbonate ions are concerned, the mean distance X-$OCO3^{2-}$ between the oxygen ($OCO3^{2-}$) of the melt bonded to a carbonate ion and the nearest cation of a given species (X = Si, Ti, Al, $Fe^{3+}$, $Fe^{2+}$, Mg, Ca, Na or K) is almost identical to the corresponding cation-oxygen distance X-$Om$ in the $CO_2$-free melt (in the online version see Table B2). 5 where these  distances (evaluated from the position of the first peak exhibited by the corresponding PDF) are listed for the three silicate compositions at 2273K and 20 kbar with 2 wt% $CO_2$. Furthermore, the mean distance between the carbon atom of a carbonate ion labeled $CCO3^{2-}$ in Table 5) and the nearest cation of a given species is generally shorter than the mean distance between the carbon atom of a $CO_2$ molecule led $CCO2$ in Table 5)and the nearest neighbor cation of the same species. The difference is more important when the electrostatic field strength of the ion is high (e.g. in MORB, d(X-$CCO3^{2-}$) = 2.50A for Si and Al, 2.65A for Mg, 3.10A for Ca and 3.25A for Na as compared with d(X-$CCO2$) = 3.65A for Si, 3.60A for Al, 3.05A for Mg, 3.35A for Ca and 3.35A for Na). More details can be seen in Fig.13 where the pair distribution functions $gC-Om(r)$ and $gC-Si(r)$ between the carbon atom of $CO_2$ and $CO3^{2-}$ species and the oxygen and silicon atoms of the melt are shown for the three compositions. With regard to $CO3^{2-}$, the intense first peak centered around 1.25A in $gC-Om(r)$ and 2.5A in $gC-Si(r)$ corresponds to a tilted configuration where the plane of the carbonate ion makes on average, an angle of 120° with the Si-O bond. In the case of molecular $CO_2$ a shoulder is visible on $gC-Om(r)$ at about 2.65A, the integration of which gives the average number of first neighbor oxygens around $CO_2$. This number is equal to ~1 in all composition, a result which means that each $CO_2$ molecule is on average always located near an oxygen of the melt with a $C-O_m$ separation around 2.65A. Although it can be tempting to proceed with the great number of PDF's at our disposal to get more information on the local structure around carbonated moieties, this approach is tedious. So we have preferred to evaluate other structural parameters to get a better understanding on the local structure around the carbonate species.

In particular we have evaluated the proportion of carbonate ions (and $CO_2$ molecules) which are associated with bridging oxygens (BO) and non bridging oxygens (NBO) of the melt. In our calculation a carbonate ion is defined as being associated with a BO if this oxygen, which is an integral part of the $CO_3^{2-}$ moiety, is connected to two network former cations (Si or Al) of the melt. Alternatively, a carbonate ion is associated with a NBO when this oxygen is connected to only one network former cation or to network modifier cations only (it is then a free oxygen). An oxygen of the melt is defined as connected to a network former cation if the cation-oxygen separation is less than the distance corresponding to the position of the first minimum shown by the PDF (i.e. 2.30A for Si-O and 2.55A for Al-O). In the same way, a $CO_2$ molecule is defined as being associated with a BO (or a NBO) when the nearest neighbor oxygen is a BO (i.e. connected to two network former cations) or a NBO (i.e. connected to one network former cation or to network modifier cations only). It is important to keep in mind that, on average, the carbon atom of a $CO_2$ molecule and the nearest neighbor oxygen are separated by ~2.65A (instead of ~1.25A in $CO_3^{2-}$) and hence the $CO_2$ molecules are much more loosely bonded to BO's and NBO's of the melt than the carbonate ions which are forming chemical bonds with them.

In Table 4 are listed for the three compositions the proportion of BO's and NBO's (see the raw data labeled $O_m$), the percentage of $CO_3^{2-}$ and the one of $CO_2$ associated with BO's and NBO's, respectively. For BO's we have distinguished the Si-O-Si, Si-O-Al and Al-O-Al bonds whereas for NBO's we have distinguished Si-O and Al-O bonds from the free oxygens only connected to network modifier cations. In rhyolite at 2273K, about 88.3% of the oxygens of the melt are BO's and ~11.7% are NBO's (with no free oxygens) but only ~43% of the carbonate ions are associated with BO's whereas a majority of them (~57%) are associated with NBO's. The $CO_2$ molecules are more randomly associated with BO's and NBO's of the melt with a proportion of 78.5 and 21.5%, respectively. As a matter of fact if the $CO_2$ molecules were randomly associated with BO's and NBO's, their rate of association should be identical to the proportion of BO's and NBO's in the rhyolitic melt, that is to say 88.3% and 11.7% respectively. To better appraise the deviation with a random association, it is convenient to evaluate from the raw data given in Table 4, the ratio ANBOi/ABOi where ANBOi is the affinity of a given species $i$ ($i =$ CO32- or $CO_2$) for NBO's and ABOi is the affinity for BO's (with ANBOi + ABOi = 1, for a definition of the affinity see the Appendix C). This ratio is equal to ~10 for CO32- and ~2 for $CO_2$ in rhyolite at 2273K (see last column in Table 4) instead of 1 for a random association (in that case ANBOi= ABOi= 1/2). Consequently, in rhyolite the carbonate ions are preferentially associated with NBO's ($A_{NBO} = 0.91$ and $A_{BO} = 0.09$) whereas the association between $CO_2$ molecules and NBO's is moderately favored with regard to the association with BO's ($A_{NBO} = 0.67$ and $A_{BO} = 0.33$).

In basaltic and kimberlitic melts the population of NBO's is much higher than in silicic melts: ~42% in MORB and ~81% in kimberlite at 2273K, with a notable proportion of free oxygens in kimberlite (~16%, see Table 4). So the population of CO32- associated with NBO's increases accordingly in these depolymerized melts and reach values as high as 81.4% in MORB and 97.5% in kimberlite when that of $CO_2$ molecules amounts to 54.5% in MORB and 88.4% in kimberlite. The ratio ANBOi/ABOi is high for CO32- (~6 in MORB and ~9.0 in kimberlite at 2273K) and still greater than 1 for $CO_2$ (~1.7 in MORB and ~1.8 in kimberlite), which indicates that in depolymerized melts as well, the association of carbonate ions with NBO's is greatly favored (a trend more balanced for $CO_2$ molecules). When the temperature is lowered from 2273 to 1473K, the population of CO32- and $CO_2$ associated with NBO's decreases significantly in rhyolite (from ~57% to ~40% for CO32- and from ~21% to ~9.4% for $CO_2$) and more weakly in MORB (from ~81% to ~78% for CO32- and from ~54% to ~43% for $CO_2$, see Table 4). In fact the reduction of these populations originates essentially from the diminution of the number of NBO's when the temperature decreases (from 11.7% to 6.7%, in rhyolite and from 41.7% to 37.3% in MORB, the silicate melt becoming slightly more polymerized upon cooling), and depends very little on the evolution of the ratio ANBOi/ABOi which barely varies with the temperature.

A consequence of a greater affinity of carbonate ions for NBO's is that they are favored in melts that contain melt contains a high number of NBO's. This finding is in agreement with the common view that $CO3^{2-}$ is the dominant oxidized form of carbon in depolymerized melts, a form which tends to disappear (for the benefit of molecular $CO_2$) in highly polymerized melts where NBO's are scarce. However this argument is only qualitative because the respective populations of $CO3^{2-}$ and $CO_2$ species in a given melt is the result of a subtle enthalpy-entropy compensation (temperature and pressure dependent) which is itself a complex function of BO's and NBO's.

In making the distinction between the carbonate ions associated with BO's (hereafter called BCarb for convenience) from those associated with NBO's (hereafter called NBCarb), we have evaluated the occurrence rates $RXBCarb$ and $RXNBCarb$ of BCarb (or NBCarb) associated with a cation of a given species $X$ (e.g. $X$ = Ca, Mg,...) as the first (nearest) neighbor cation. As long as network former cations are concerned, we have excluded Si and Al atoms which are directly connected to the oxygen of the melt forming bond with the carbonate ion. So only the second or the third nearest neighbor Si or Al atom is accounted for in $RSi,AlBCarb$ and $RSi,AlNBCarb$. In Table 5 are presented for the three simulated melts the occurrence rates $RXBCarb$ and $RXNBCarb$ expressed in percentage.

information is obtained in comparing these occurrence rates with those deduced from a simple model for which a uniform distribution of the cations around the carbonate ions is assumed. For this uniform distribution model near a carbonate ion (BCarb or NBCarb) is simply given by the concentration of this cation in the melt ( $NX/XNX$ , where $N_X$ is the number of cation of species $X$).

For a random distribution the occurrence rate of a particular cation X near a carbonate ion (BCarb or NBCarb) is simply given by the concentration of this cation in the melt ( $NX/XNX$ , where $N_X$ is the number of cation of species $X$). The correlated effect of the large affinity of the carbonate ions for NBO's with the non random distribution of the cations around the carbonate ions is clearly visible in the calculated occurrence rates, and display a significant deviation deviate significantly from those predicted by the uniform distribution model (random distribution). Thus in the simulation it is more frequent to observe a BCarb-X or a NBCarb-X association with a network modifier cation (X = Ca, Mg, $Fe^{3+}$, $Fe^{2+}$, Na and K) than predicted by the uniform distribution model (e.g. in rhyolite $RNaNBCarb$ = 29% instead of 7.6% by the model, see Table 5). The deviation between simulation and model is all the more pronounced when the melt is more depolymerized. In contrast, the immediate vicinity of BCarb and NBCarb in the simulated melts is depleted in Si atoms with respect to the prediction of the uniform distribution model, the depletion being larger around NBCarb than about BCarb (this depletion is expected because the model describes a structure less medium and it does not distinguish network former cations from network modifier cations). A similar behavior occurs with Al in MORB and kimberlite but not in rhyolite where this network former cation is in excess around BCarb and NBCarb (an explanation is given later on).

Although useful the above comparison between calculated occurrence rates and those predicted by the uniform distribution model does not allow us to evaluate at first glance the sequence of preferential associations between carbonate groups (BCarb and NBCarb) and cations. Indeed, the magnitude of the occurrence rates is dominated by the concentration of cations in the melt. To clarify this point, we have evaluated the affinities $AXBCarb$ and $AXNBCarb$ between the carbonate ions and the cations of species X using a definition of the affinity based on occurrence rates and concentrations (see Appendix C). In this framework, the affinity measures the propensity of a cation to associate with a carbonate ionvariation of the affinity with the nature of the cation which makes sense. f For the uniform distribution model, there is no preferential association between carbonate ions and cationsany carbonate-cation association is equiprobable) and the affinities are all equal to each other and are given by ($AXBCarb$ = $AXNBCarb$ = $1/N_s$ where $N_s$ is the number of cationic species in the melt)$_s$ = 7 in rhyolite, 9 in MORB and 8 in kimberlite). Values of the affinities are listed in Table 5. A clear hierarchy between cations is now observedhe concentrations. In normalizing the affinities such that the affinity of Si for carbonate ions is equal to 1,  the ordering of the cationic affinities for BCarb's is the following:

Ca (4.1) ~ Na (4.1) > $Fe^{2+}$ (3.0) > K (2.3) > $Fe^{3+}$ (1.7) ~ Al (1.7) > Si (1) in rhyolite,

Na(5.8) > Ca(5.2) > Mg(4.7) > $Fe^{2+}$(4.0) > K(3.2) > $Fe^{3+}$(2.2) > Al(1.7) > Si(1) > Ti(0.8) in MORB,

and  $Fe^{2+}$ (8.1) > Mg (7.9) > K (7.7) > Na (7.5) > Ca (7.3) > Al (2.2) > Si (1) in kimberlite.

In the same way the ordering of the affinities for NBCarb's is,

$Fe^{2+}$ (15.1) > Ca (12.3) > Na (10.5) > $Fe^{3+}$ (10.3) > K (5.7) > Al (4.3) > Si (1) in rhyolite,

Mg(17.9) >Ca(14.4) >$Fe^{2+}$(12.7)>Na(12.4)>Ti(9.4)>$Fe^{3+}$(9.1) >K(4.9)>Al(4.5)>Si(1) in MORB and,

Mg (30.7) > Ca (24.7) > Na (24.2) > $Fe^{2+}$ (20.8) > Ti (13.1) > K (10.0) > Al (5.4) >Si (1) in kimberlite.

In the three melts the affinity of network modifier cations for carbonate ions is generally much larger than the affinity of network formers as expected. This gap increases with the depolymerization of the melt and is enhanced with NBCarb's. For instance in MORB the relative affinity of Mg with respect to Si is ~5 for BCarb and ~18 for NBCarb, values which become, respectively, ~8 and ~31 in kimberlite. Moreover the ordering of the affinities of Na, K, Ca, Mg, $Fe^{2+}$ and $Fe^{3+}$ depends on the melt composition. This variation of the cationic affinities with the degree of depolymerization of the melt has the consequence of increasing the association rates of some cations with respect to others. This is why in rhyolite the association rate of carbonate ions with Na is found to be especially high in comparison with that evaluated from the uniform distribution model (~26% instead of ~8%, in taking into account as a whole the populations associated with BCarb's and NBCarb's, see the raw labeled *Total* in Table 5). In the same way, the association rates observed with K, Ca and $Fe^{2+}$ leads to a significant contribution (~18%) albeit the nominal concentration of these cations in the melt is low (~8%). In MORB, the strong affinity of Mg and Ca for carbonate groups makes dominant their contribution to the global association rate (~54%), follows by the contribution of Carb-Na and Carb-$Fe^{2+}$ (~23%). In kimberlite a great majority of events involves the NBCarb-Mg association (~76%) follows by NBCarb-Ca (~11%) and NBCarb-$Fe^{2+}$ (~8%), the association with Si being negligible (~2% instead of ~31.5% for a random association). Furthermore it is noteworthy that an important proportion of NBCarb's in kimberlite are linked to a free oxygen of the melt (~33.7% in Table 4) and consequently a significant proportion of carbonate ions are connected only to Mg atoms, with magnesium playing a role of structure former as it is observed in *$CO_2$*-free magnesiosilicate melts (Kohara et al., 2004; Wilding et al., 2008, 2010).

In order to have more information about the BCarb-X and NBCarb-X associations, we have evaluated the mean distance <$d_{C-X}$> between the carbon atom of a carbonate ion (in distinguishing

BCarb from NBCarb) and the nearest neighbor cation of species X. In looking at Table 5, it is clear that the mean distance $<d_{C-X}>$ is systematically shorter for NBCarb-X than for BCarb-X configurations. This finding suggests that the cations are more tightly bonded with NBCarb's than with BCarb's. Indeed, in the case of a BCarb-X configuration, the presence of the two connected $TO_4$ tetrahedra prevents a vicinal cation from closely interacting with the BCarb. An interesting situation occurs when an Al atom approaches a NBCarb (the latter one being connected most of the time to a Si atom, see Table 4). In this case the Al atom connects to one of the two pending oxygens ($O_p$) of the NBCarb, the corresponding Al-$O_p$ distance being in the range ~1.9-2.1A and the mean distance $<d_{C-Al}>$ equal to 2.76A in rhyolite, 2.68A in MORB and 2.60A in kimberlite. Further analysis shows that a fourfold coordinated aluminum [4]Al is transformed into a fivefold coordinated [5]Al by connecting with a pending oxygen of a NBCarb. This mechanism leads to the formation of a T-Carb-Al configuration in which the carbonate ion is shared with two network former cations by the mediation of two of its own oxygen atoms. The occurence rate of this configuration is notable in rhyolite (~13.2%) and in MORB (~9.5%) and negligible in kimberlite (~1%). The reason for which Al and not Si forms this peculiar configuration is due to the propensity of Al to be fivefold coordinated in liquid silicates (for a discussion see Guillot and Sator (2007b) and references therein). Thus in our simulations performed at 2273K and 20 kbar the population of [5]Al is as high as ~20% in rhyolite and ~30% in MORB and kimberlite whereas in the case of Si atom, the population of [5]Si is found to be less than a percent in rhyolite and a few percent in MORB and kimberlite.also clearly visible in the PDFs associated with Si-$O_p$ and Al-$O_p$ interactions where $O_p$ is a pending oxygen of a carbonate ion. When a pronounced peak centered about 2.0A is visible on the Al-$O_p$ PDF (not shown), and which corresponds to the formation of a T-Carb-Al configuration, there is no such a peak on the Si-$O_p$ PDF. Besides, the mean distance $<d_{C-Si}>$ separating the carbon atom of a NBCarb and the nearest peripheral $SiO_4$ unit is, on average, larger by 0.2-0.3A than the distance $<d_{C-Al}>$ with the nearest $AlO_4$ tetrahedron (see Table

[3+] merit some comments. Ti is found in fourfold-, fivefold- and sixfold-coordination in MORB and kimberlite with a preponderance for [5]Ti (for a discution see Guillot and Sator (2007b) and references

therein). In analyzing the structure of the melt it is observed that the pending oxygens of NBCarb's tend to form loose bonds with Ti (the corresponding Ti-O$_p$ distance is ~2.25A as compared with ~1.95A for the Ti-O distance in a *CO$_2$*-free melt). Although Ti is sometimes considered as a structure maker cation the T-Carb-Ti configuration cannot be considered equivalent to a T-Carb-Al configuration because the Carb-Ti interaction is too weak. A similar conclusion is reached about Fe$^{3+}$ . This cation is found in fourfold and fivefold coordination in rhyolite and MORB with [4]Fe$^{3+}$ being the most probable coordination, so it is often considered as a potential structure maker cation. However the mean distance between the pending oxygens of a carbonate ion and Fe$^{3+}$ is too large in the simulated melts (~2.30A as compared with ~1.85A for the Fe$^{3+}$-O distance in *CO$_2$*-free melt) to consider Fe$^{3+}$ as a potential structure maker in *CO$_2$*-bearing melts.

The network modifier cations (Fe$^{2+}$, Mg, Ca, Na and K) are characterized by a broad distribution of n-coordinated species with n=4-6 for Mg and Fe$^{2+}$, and n=6-10 for Ca, Na and K (considering the vast literature devoted to this topic, the reader may refer to the discussion in Vuilleumier et al. (2009) and references therein). A broad distribution of n-coordinated species favors the occurrence of the following association reaction between the cations and the carbonate ions,

$$^{[n]}X + 2\,O_p \rightarrow {}^{[n+2]}X \qquad\qquad (4)$$

where a n-coordinated cation of species X interacts with the two pending oxygens of a carbonate ion in a bidented configuration. This is illustrated in Fig.14 where such configurations (and others), issued from the simulations, are shown.

## 4. Discussion

During the last three decades a number of experimental investigations s using Raman, infrared or NMR spectroscopies (Mysen et al., 1976; Eggler and Rosenhauer, 1978; Mysen and Virgo, 1980a,b; Fine and Stolper, 1985a,b; Taylor, 1990; Kohn et al., 1991; Brooker et al., 1999, 2001b; Morizet et al., 2002; Behrens et al., 2004a,b; Botcharnikov et al., 2006) and theoretical calculations as well (De Jong and Brown, 1980; Kubicki and Stolper, 1995; Tossel, 1995), have tried to elucidate the solubility mechanisms of *CO$_2$* in silicate melts. It was initially suggested that in basic melts at least, *CO$_2$* reacts preferentially with NBO's to form CO32- (Eggler and Rosenhauer, 1978). This is precisely what our simulations quantify. Furthermore it was proposed (Holloway et al., 1976; Mysen and Virgo, 1980a) that in liquids on the join CaO-MgO-SiO$_2$-$CO_2$ (e.g. diopside+$CO_2$) for which the solubility of $CO_2$ is positively correlated with Ca concentration, the carbonate ion is associated with Ca to form CaCO$_3$ ion pair. Since then, it has been found for a wide range of melt composition that the $CO_2$ solubility increases with the NBO content of the melt, this increase being all the more important when the Ca concentration is high (Brooker et al., 2001b; Luth, 2006). But the increase of solubility is weaker with natural compositions where the Ca concentration remains low or moderate (<13 wt%) and even if the concentration in (Mg+Fe$^{2+}$) varies noticeably. This behavior is encountered with our simulated melts which exhibit a weak increase of $CO_2$ solubility with the degree of depolymerization he pressure is not too high (see Figs.3-5). Although there is no evidence of a true complexation (i.e. long-lived ion pair) between carbonate groups and alkali or alkali earth cations in the simulations, we do observe preferential associations between carbonate groups and the network modifier cations (see the discussion in section 3.4. about the cationic affinity).

An intriguing aspect of the experimental data is that the vibration spectra of carbonate species in highly polymerized melts (in which molecular $CO_2$ is the dominant oxidized form of carbon) are very distinct from those exhibited by more depolymerized melts. Thus the C-O antisymmetric stretching band ($\upsilon_3$) associated with CO32- appearing in the mid-IR region is split by more than 210 cm$^{-1}$ in Na$_2$O-SiO$_2$ glasses and in jadeite (Fine and Stolper, 1985a,b; Brooker et al., 1999, 2001b) whereas for natural compositions exhibiting a higher degree of depolymerization this splitting is much smaller (<100 cm-1 see Table 2 in Brooker et al., 2001b). The magnitude of the splitting is thought to give an estimation of the degree of distortion from ideal D$_{3h}$ symmetry experienced by CO32- groups within the melt network, the higher the distortion, the larger the splitting. In highly polymerized melts the large splitting of the $\upsilon_3$ stretching band and the chemical shift around 165 ppm in $^{13}$C MAS NMR spectra are assigned to T-Carb-T configurations where two oxygen of a carbonate group are shared with two tetrahedral sites (Fine and Stolper, 1985a,b; Kohn et al., 1991; Kubicki and Stolper, 1995; Tossell, 1995; Brooker et al., 1999). It is significant that this kind of configuration is found in a non negligible abundance in our rhyolitic melt (e.g. RAlNBCarb in Table 5). In depolymerized melts of natural composition the moderate splitting of the C-O vibration band is interpreted by Brooker et al. (2001b) as resulting from the interaction between NBCarb's and $M^{2+}$ cations (with M = Ca, Mg or Fe). This interpretation is fully supported by our simulation results since an overwhelming majority of events in MORB and kimberlite is due to NBCarb-$M^{2+}$ associations.

The local environment around $CO_2$ molecules is poorly known because the $\upsilon_3$ band (asymmetric stretching) of $CO_2$ observed by IR spectroscopy around ~2350 cm$^{-1}$ ( as compared with 2348 cm$^{-1}$ in pure $CO_2$ gas) in weakly depolymerized glasses has no special feature and disappears rapidly as soon as the melt is more depolymerized (Stolper et al., 1987; Tingle and Aines, 1988; Fogel and Rutherford, 1990; Blank et al., 1993; Brooker et al., 1999; Tamic et al., 2001; Brooker et al., 2001b; Morizet et al., 2002; Nowak et al., 2003; Behrens et al., 2004a,b; Botcharnikov et al., 2006). As for the $\upsilon_1$ (symmetric stretching) and $\upsilon_2$ (bending) vibration modes of $CO_2$ they give rise to a Fermi doublet in the Raman spectra of aluminosilicate glasses (Mysen and Virgo, 1980b; Brooker et al., 1999) but in absence of a theoretical guidance it is hard to make a definite conclusion about the incorporation of molecular $CO_2$ into the melt network. The $^{13}$C MAS NMR spectra of albite, jadeite, rhyolite and phonolite glasses (Kohn et al., 1991; Brooker et al., 1999; Morizet et al., 2002) show a resonance peak at 125 ppm (as compared with 124.2 ppm in pure $CO_2$ gas) with spinning side bands which are attributed to dissolved molecular $CO_2$ structurally bonded to the melt network (Kohn et al., 1991). Tossell (1995) and Kubicki and Stolper (1995) have performed ab initio calculations of T-O-T'..$CO_2$ complexes (where T = Si(OH)$_3$Al(OH)$_3$ or SiH$_3$, AlH$_3$) to model the incorporation of $CO_2$ into aluminosilicate glasses. They found that energetically favorable configurations are those where the distance C...O$_b$ between the C atom of a $CO_2$ molecule and the nearest bridging oxygen (Si-O$_b$-Si) lies in the range 2.63-2.90A with a slightly bent $CO_2$ molecule ($\theta_{O-C-O}$ ~168-179°). These configurations are compatible with a $^{13}$C NMR chemical shift of ~125 ppm found in aluminosilicate glasses (Tossell, 1995) and are closely related to those produced by our simulations (e.g. in Fig.14cc). However the chemical shift associated with $CO_2$ molecules is barely sensitive to the melt composition, contrary to the Fermi doublet seen in Raman spectroscopy and whose the intensity varies strongly with the silicate composition (Brooker et al., 1999). The latter observation is in agreement with our own conclusions for which the $CO_2$ molecules are incorporated as interstitial particles and are not randomly distributed into the melt network but are localized in the vicinity of BO's and NBO's with a preference for the latter ones (see Table 4).

At this stage it is important to keep in mind that data from the literature discussed above were obtained from glass samples issued from a $CO_2$-bearing liquid phase. Consequently, the conclusions reached when investigating CO₂-bearing glasses samplesare not necessarily valid at superliquidus temperatures. In fact the structural reorganization induced by the large change in temperature between the liquid and the glass may induce the following modifications. (1) As suggested by several experimental investigations (Porbatzki and Nowak, 2001; Nowak et al., 2003; Morizet et al., 2001, 2007; Spickenbom et al., 2010) and supported by our simulation results, the carbonation reaction favors the carbonate species with regard to molecular $CO_2$ when the temperature decreases, so the carbonate groups are stabilized upon cooling and hence their concentration (versus molecular $CO_2$) are likely much higher in the low temperature glass than in the high temperature liquid. (2) The structure of the glass is not identical to that of the liquid, the glass being slightly more polymerized than the liquid. (3) Correlatively to point (2), the coordination of cations is expected to increase from glass to liquid. For instance, [5]Al and [6]Al coordinated species are expected to be in a higher proportion at liquid temperature (Kiczenski et al., 2005; Stebbins et al., 2008) and therefore it is conceivable that the formation of Si-Carb-Al involving a coordination change for aluminum is favored in the liquid phase. (4) The structural modification of the melt induced by the glass transition may favor the formation of contact ion pairs between carbonate groups and metal cations (e.g. CaCO₃) whereas long lived ion pairs are prevented in the liquid phase owing to the diffusive motions of all the elements. In summary, in absence of experimental data collected in situ at liquid temperature, the comparison between simulation results and data on glass samples has to be approached with caution.

At high pressures our $CO_2$-saturated melts become very $CO_2$-rich (e.g. at 100 kbar and 2273K the $CO_2$ contents are ~24 wt% in rhyolite, ~25 wt% in MORB and ~30 wt% in kimberlite). Yet a visual inspection of the computer-generated atomic configurations shows no evidence of a silicate-carbonate liquid immiscibility in those melts. This qualitative information is confirmed by an analysis of the PDF's which indicates that the distribution of carbon dioxide throughout the melt is homogeneous (not shown). More precisely, carbonate ions and $CO_2$ molecules are distributed between BO's and NBO's of the melt essentially as at lower pressure (e.g. see Table 4), the carbonated species exhibiting a stronger affinity for NBO's. Under pressure one notices also an increase of the cation coordination ([4]Si → [5]Si, [4]Al → [5,6]Al, [4,5,6]Mg → [5,6,7]Mg, ..) and a shortening of the mean distance between the carbon atom of the carbonate groups and the nearest metal cation (by ~ -0.10A between 20 and 100 kbar), a shortening induced by the densification of the melt. As for the geometry of $CO_3^{2-}$ and $CO_2$ into the melt, they are barely modified by an increase of the pressure.

That the simulated $CO_2$-rich melts show no evidence for liquid-liquid immiscibility, merits comment since the existence of a carbonate-silicate immiscible liquid field is well documented in natural lavas (e.g. Peterson, 1989; Kjarsgaard and Peterson, 1991; Panina and Motorina, 2008) and in synthetic melts. In the system $SiO_2$-$Al_2O_3$-CaO-$Na_2O$-$CO_2$ a number of petrological investigations focused on the origin of carbonatites (Kjarsgaard and Hamilton, 1988; Brooker and Hamilton, 1990; Mattey et al., 1990; Lee and Wyllie, 1996; Brooker, 1998) have shown a complex behavior (expansion-reduction) of the two-liquid field with pressure and degree of $CO_2$ undersaturation, the alkali contents controlling the size of the miscibility gap. Although these experiments were restricted to moderate pressures and temperatures (below 30 kbar and 1673K) it appeared that for an alkali poor and silica rich melt as molten rhyolite (with $Na_2O$+$K_2O$ ~ 10 wt%), the $CO_2$ saturated melt is expected to lie in the one liquid field (see Figs. 3 and 4 in Brooker, 1998) even at a pressure as high as 100 kbar, a conclusion that the simulation corroborates.

With regard to silica-undersaturated melts, experimental studies point out that carbonate rich magmas may be generated by melting various carbonate-silicate mineral assemblages (Moore and Wood, 1998; Lee and Wyllie, 1998; Dalton and Presnall, 1998; Hammouda, 2003; Gudfinnsson and Presnall, 2005; Keshav et al., 2005; Luth , 2006;  Dasgupta et al., 2006; Brey et al., 2008; Ghosh et al., 2009; Litasov and Ohtani, 2010). More generally, it has been shown that a low degree of melting produces a carbonatitic liquid whereas a high degree of melting produces kimberlitic like or even basaltic like melts. For instance the melt in equilibrium with a carbonated peridotite exhibits a continuous transition with the degree of melting from a carbonatitic composition with <10 wt% $SiO_2$ to a kimberlitic one with >25 wt% $SiO_2$ (Moore and Wood, 1998; Dalton and Presnall, 1998; Gudfinnssonn and Presnall, 2005; Brey et al., 2008). The composition and $CO_2$ contents of the coexisting silicate melt in the pressure-temperature range 60-100 kbar and 1673-2073K are quite similar to those characterizing the $CO_2$-saturated Mg-rich kimberlitic melt investigated here. This finding supports the view that a kimberlitic like melt under pressure may incorporate a large amount of $CO_2$ without presenting liquid-liquid immiscibility. In contrast, high pressure melting experiments on carbonated eclogite (Hammouda, 2003; Dasgupta et al., 2006) have detected a miscibility gap where a carbonate-rich melt coexists with a silica-undersaturated silicate melt, the two compositions evolving rapidly with pressure and temperature. However the closing of the miscibility gap is expected in increasing the pressure for temperatures higher than ~1673K (see Fig.8 in Dasgupta et al., 2006) and below ~100 kbar (Litasov and Ohtani, 2010). As far as the silicate melt is concerned, at 30 kbar and 1673K its composition is between basic and ultrabasic (~56 wt% of $SiO_2+Al_2O_3+TiO_2$ and ~42 wt% of MgO+CaO+FeO) with ~5 wt% $CO_2$ (Dasgupta et al., 2006) whereas at 60 kbar and 1573K its composition is basaltic (~66 wt% of $SiO_2+Al_2O_3$ and ~32 wt% of MgO+CaO+FeO), and close to a MORB composition with a high $CO_2$-content (~12.3 wt% after Hammouda, 2003). These data are quite compatible with our prediction for the $CO_2$ solubility curve in MORB (see Fig.6). In brief, according to these results a silicate liquid of basaltic to kimberlitic composition is able to incorporate a large amount of $CO_2$ without presenting a liquid-liquid immiscibility provided that pressure and temperature are sufficiently high.

To investigate the absence (or the appearance) of a liquid-liquid miscibility gap with pressure, we have prepared a simulation box filled with a $CO_2$-saturated basaltic liquid at 1873K and 100 kbar and containing ~30 wt% $CO_2$. It was checked that this $CO_2$ bearing liquid was perfectly homogeneous and presented no tendency for phase separation. Subsequently a pressure drop from 100 kbar to 20 kbar was applied at constant temperature. In a few hundreds of ps after the pressure drop, it was observed the formation in the melt of a nanobubble composed exclusively of $CO_2$ molecules (not shown). After an equilibration time, the simulation box exhibited two coexisting phases, a pure $CO_2$ phase and a homogeneous silicate liquid whose the $CO_2$ contents (~2 wt%) coincided with the solubility expected at 1873K and 20 kbar. This finding suggests that in the (P,T) range investigated, the coexistence of a $CO_2$ phase with a $CO_2$-bearing silicate melt prevents a phase separation between a carbonatitic liquid and a $CO_2$-poor silicate liquid. More generally, by placing  contact a silicate liquid of given composition in contact with a pure $CO_2$ phase it is possible, in increasing the pressure, to make a continuous transition between $CO_2$-poor and $CO_2$-rich silicate melt up to $CO_2$ contents matching those of carbonatitic liquids (> 35 wt% $CO_2$).

## 5. Conclusion

By performing a series of molecular dynamics simulations where a supercritical $CO_2$ phase is in contact with a silicate melt of a given composition (rhyolitic, basaltic or kimberlitic) at different temperatures (1473-2273K) and pressures (20-150 kbar), we have been able to evaluate the solubility of $CO_2$, the populations of molecular and carbonate species, their diffusivity through the melt and the local structure. The main results can be summarized at it follows.

(1) The solubility of $CO_2$ increases markedly with the pressure in the three investigated melts, the behavior being super-Henrian above 20 kbar (~2 wt% $CO_2$ at 20 kbar and more than 25 wt% at 100 kbar). Surprisingly, the solubility is found to be weakly dependent on the melt composition (as far as the present compositions are concerned) and it is only at very high pressure (above ~100 kbar) that a clear hierarchy between solubilities occurs, namely rhyolite<MORB<kimberlite. Furthermore at a given pressure the calculated solubility is negatively correlated with the temperature.

(2) In $CO_2$-saturated melts, the proportion of carbonate ions increases under pressure at isothermal condition and decreases when the temperature increases at isobaric condition. Furthermore, at fixed (P,T) conditions the proportion of carbonate ions is higher in $CO_2$-undersaturated melts than in the $CO_2$-saturated melt. Although the proportion of molecular $CO_2$ decreases when the degree of depolymerization of the melt increases, it is still significant in $CO_2$-saturated basic and ultrabasic melts. This finding is at variance with experimental data on $CO_2$-bearing glasses which show no evidence of molecular $CO_2$ as soon as the degree of depolymerization of the melt is high (e.g. basalt).

These conflicting results can be reconciled with each other by noticing that a simple low temperature extrapolation of the simulation data for the solubility of $CO_2$ in MORB predicts that the proportion of molecular $CO_2$ might be negligible in the glass at room temperature. A conclusion also reached by several experimental studies (Porbatzki and Nowak, 2001; Nowak et al., 2003; Morizet et al., 2001, 2007; Spickenbom et al., 2010) which show that an estimation based upon the analysis of quenched glasses tends to underestimate the true contents in molecular $CO_2$ at liquidus temperature.

(3) The carbonate ions are found to be transient species in the liquid phase, with a lifetime increasing exponentially with the inverse of the temperature ( from ~10 ps at 2000K to ~100 ps at 1400K). Therefore carbonate ions are expected to be long-lived species in the glass at room temperature. Contrarily to a usual assumption, the diffusivity of carbonate ions in the liquid silicate is not vanishingly small with respect to that of $CO_2$ molecules: in MORB it is smaller by a factor of ~6 at 1473K and by a factor of ~2 at 2273K. Although the bulk diffusivity of $CO_2$ is governed primarily by the diffusivity of $CO_2$ molecules, the carbonate ions contribute significantly to it in depolymerized melts because they are preferentially associated with the NBO's of the melt which are themselves diffusing faster than the BO's.

(4) Concerning the structure of the melt network around $CO_2$ species, the carbonate ions are preferentially associated with NBO's of the melt, with an affinity for NBOs which exceeds that for BOs by almost one order of magnitude. This result explains why the concentration in carbonate ions is positively correlated with the degree of depolymerization of the melt and is hard to detect in fully polymerized melts where the number of NBO's is close to zero. Furthermore, the network modifier cations are not randomly distributed in the close vicinity of carbonate groups but are ordered in a way which depends at once on the nature of the cation and on the melt composition. But at high temperatures investigated here, there is no evidence of long-lived complexes between carbonate groups and metal cations.

From a geochemical point of view, there is now a body of evidence that incipient melting may occur at depth up to 300 km under oceanic ridges (Dasgupta and Hirschmann, 2006) or in plumes (for a review see Bell and Simonetti, 2010) and carbonate liquids may be at the origin of these melts. In this context it is not immaterial that the present simulation study predicts a very high solubility of $CO_2$ in basaltic and kimberlitic melts at high pressure (as high as 20-25 wt% $CO_2$ at 80 kbar). Although the evaluation of $CO_2$ solubility done here is more like a maximum of what may happen in the source regions, our finding suggests that a great amount of $CO_2$ could be scavenged by interstitial melts (at very low melt fraction) in the upper mantle. So we hope that the present simulation results will promote the development of new degassing scenarios whose the initial step should start in the deep mantle.

**Acknowledgements**

We thank Philippe Sarda for his thoughtful read-through of the manuscript and for his continuing encouragement. The reviewers are acknowledged for their helpful comments.

## APPENDIX A

The derivation of the force field.

The interatomic potential for ionic systems (e.g. silicates) is usually modeled by a Born-type analytic term ($\sim e^{-r/\rho}$, where r is the interatomic distance and $\rho$ a parameter ) for repulsion between atoms combined with a Coulomb term for ion-ion interactions and supplemented by an attractive contribution ($\sim -1/r^6$) accounting for the dispersion forces between atoms. In the case of molecular systems (e.g. $CO_2$), the interaction potential between two molecules is expressed as the sum of a short range contribution describing the repulsion-dispersion interactions between atoms (e.g. the Lennard-Jones potential $\sim 1/r^{12} - 1/r^6$), plus an electrostatic contribution which expresses the multipolar interactions between molecules (e.g. $CO_2$ has a quadrupole moment). When chemical bonds are involved (e.g. the C-O bond in carbonate ion), the association-dissociation reaction can be conveniently described by a Morse potential. An alternative to this empirical description would be to use the *ab initio* MD method which has become popular in the simulation community (e.g.Vuilleumier et al. (2009) and references therein). In this approach, a MD schema is combined with an electronic structure calculation based on the density functional theory (Parr and Yang, 1995). Although this method is not free of approximations (Perdew et al., 2009), it has the great advantage of describing explicitly from first principles the redistribution of electrons in the material as function of temperature and pressure. However its computational cost is so high ($\sim 10^3$-$10^4$ times more expensive in computer time than classical MD) that its use is limited, for the time being, to small system size (a few hundreds of atoms instead of several thousand with classical MD) and short trajectories ($\sim 10$ ps instead of 10,000 ps). Consequently, an accurate evaluation of $CO_2$ solubility requiring large system sizes and long equilibration runs is out of reach with this method considering  present day computational resources.

In the present study we have performed classical MD calculations using empirical potentials of the literature to describe the silicate melt and the fluid $CO_2$, whereas we have developed a specific model to describe the interactions between $CO_2$ and the silicate melt. Thus we have made use of a force field recently proposed for MD simulations of natural silicate melts of felsic to ultrabasic composition (Guillot and Sator, 2007a,b). In practice the potential energy $u_{ij}$ between two atoms of the melt (where $i, j$ = Si, Ti, Al, $Fe^{3+}$, $Fe^{2+}$, Mg, Ca, Na, K, and O) is given by,

$$u(r_{ij}) = z_i z_j / r_{ij} + B_{ij} e^{-r_{ij}/\rho_{ij}} - C_{ij}/r_{ij}^6 \qquad \text{(A1)}$$

where $r_{ij}$ is the distance between atoms $i$ and $j$, $z_i$ is the effective charge associated with the ion $i$, and where $B_{ij}$, $\rho_{ij}$ and $C_{ij}$ are parameters describing repulsive and dispersive forces between the ions $i$ and j (values of parameters are given in Table 1). The thermodynamic and structural properties of the simulated melts reproduce quite satisfactorily the experimental data over a large pressure range (up to ~200 kbar).

The interactions between $CO_2$ molecules in the supercritical fluid phase are described by the model of Zhang and Duan (2005). With this model the $CO_2$ molecules are assumed to be linear and rigid with a C-O bond length, $l_{C-O}$ = 1.162 A. The quadrupole moment of the $CO_2$ molecule is represented by point charges located on C and O atoms. The potential energy $u(1,2)$ between molecules 1 and 2 is then given by,

$$u(1,2) = \sum_{i \in 1} \sum_{j \in 2} [\ 4\varepsilon_{ij}[(\sigma_{ij}/r_{ij})^{12} - (\sigma_{ij}/r_{ij})^6] + q_i q_j/r_{ij}\ ] \qquad (A2)$$

where $i,j$ run over the three atoms of the corresponding molecule (1) or (2), $q_i$ is the effective charge associated with the atom $i$ (C or O), $\varepsilon_{ij}$ and $\sigma_{ij}$ are the Lennard-Jones parameters for the corresponding pair of atoms (C-C, C-O and O-O). When this potential is implemented into a MD code (the potential parameters are given in Table 1), it shows an excellent predictability of thermodynamic, transport and structural properties in a wide (P,T) range encompassing both the liquid state and the supercritical region (Zhang and Duan, 2005). In particular the volumetric properties of supercritical $CO_2$ are very well reproduced with uncertainty on the density generally less than ±2%, compared with the equation of state (EOS) of Span and Wagner (1996) which reviewed all the available experimental data from triple point to 1100 K and 8 kbar. To enable $CO_2$ molecules to react with the oxygens of the silicate melt to form carbonate ions, we have modified the model of Zhang and Duan by allowing the $CO_2$ molecule to be flexible. The following harmonic intramolecular potential for bending has been introduced,

$$u_\theta = 1/2\ k_\theta(\theta\text{-}\pi)^2 \qquad (A3)$$

where $k_\theta$ and $\theta$ are the force constant and bending angle, respectively. To reproduce the $\nu_2$ bending mode of vibration in supercritical $CO_2$ ($\nu_2 \sim$ 670 cm$^{-1}$, see Yee et al., 1992), a value of 444 kJ/mol/rd$^2$ was assigned to $k_\theta$. We have checked by MD simulation that the modified potential (with flexibility) leads to the same volumetric properties as the original model, in the liquid state and the near critical region. Furthermore, several studies have shown that $CO_2$ molecules tend to adopt nonlinear geometries at high temperatures and high pressures (Ishii et al., 1996; Saharay and Balasubramanian, 2004, 2007; Anderson et al., 2009), that can be accomodated by the introduction of the flexibility is then recommended.

We have evaluated the volumetric properties of the new model for thermodynamic conditions up to 2273 K and 150 kbar. Recently, using Brillouin scattering and interferometric measurements, Giordano et al. (2006) have evaluated the EOS of $CO_2$ in the P-T range 1-80 kbar and 300-700 K. Their data for the isotherm 700 K are reported in Fig.A1 and are compared with our MD calculations (computational details are given in the text). The agreement between the two sets of data is satisfactory considering the respective uncertainties (±2%). Also shown in Fig.A1 is the prediction of the EOS of Sterner and Pitzer (1994) for the same isotherm and which is based essentially on the same data as the EOS of Span and Wagner (1996) but which also includes shock-compression data at very high temperature and pressure (100<P<700 kbar and 1000<T<4000K but with a large uncertainty of several hundred of degrees on the estimated temperature). The isotherm predicted by the Sterner and Pitzer EOS is close to that of Giordano et al., but as emphasized by Giordano et al., it describes a fluid slightly more compressible than the real one, in contrast with our simulated fluid which is slightly less compressible (see Fig.A1). Keeping these findings in mind we have evaluated the high temperature isotherm 2273K and compare the results with the prediction of the Sterner and Pitzer EOS (see the insert in Fig.A1). The two isotherms are close to each other up to ~30 kbar but deviate from each other at higher pressure, the simulated fluid being less compressible than the isotherm predicted by the EOS of Sterner and Pitzer. Considering the aforementioned tendency of the Sterner and Pitzer EOS to overestimate the density at very high pressure and in absence of reliable experimental data in the P-T range under investigation, the PVT properties of our simulated $CO_2$ fluid are likely close to those of the real fluid at these extreme conditions.

When a $CO_2$ molecule of the supercritical $CO_2$ phase enters into the silicate melt, its interaction potential energy with the melt is given by,

$$u_{CO2...melt} = \sum_{i=1,..No} u_{CO2...O_i} + \sum_{j \in N_1,..N_n} u_{CO2...X_j} \tag{A4}$$

where $O_i$ is the oxygen $i$ of the melt (total number $N_o$) and $X_j$ is the cation of species $j$ ($N_j$). The interactions between the $CO_2$ molecule and the cations of the melt (second term of the right hand side of eqn.(A4)) are discussed first. These interactions can be decomposed into pair contributions between C, $O_c$ and $O'_c$ atoms of the $CO_2$ molecule and the cation $X_j$ of species $j$,

$$u_{CO2...X_j} = u_{C...X_j} + u_{OC...X_j} + u_{OC...X_j'} \tag{A5}$$

The carbon atom bearing a positive charge ($q_C$ = +0.5888 e, see Table 1), the carbon-cation interactions $u_{C...X_j}$ are assumed to be governed only by the electrostatic repulsion between their positive charges. As for the oxygen-cation interactions $u_{Oc...X_j}$ and $u_{O'c...X_j}$, they are modeled by the attractive electrostatic interaction between the oxygen of $CO_2$ ($q_O$ = -0.2944 e, see Table 1) and the cations, supplemented by repulsion-dispersion terms of the same kind than those used to describe the cation-oxygen interactions in the pure silicate melt (see eqn.(A1)). In order to reduce the number of adjustable parameters, the repulsion-dispersion paramaters $B_{ij}$, $\rho_{ij}$ and $C_{ij}$ (where i= $O_c$ or $O'_c$ and j is the index of the cation) are assumed to be identical to those describing the oxygen-cation interactions in the oxides of the pure silicate melt. In summary, the potential energy between one $CO_2$ molecule and a cation $X_j$ of the melt is given by,

$$u_{CO2...X_j} = \sum_{i \in [CO2]} ( q_i z_j/r_{ij} + B_{ij}e^{-r_{ij}/\rho_{ij}} - C_{ij}/r_{ij}^{6}) \tag{A6}$$

where $i$ runs over the three atoms of the $CO_2$ molecule (with charges $q_i$), $j$ stands for the cation $X_j$ (with charge $z_j$), and $r_{ij}$ is the atom-cation distance.

To model the chemical reaction, $CO_2 + (O^{2-})_{melt} \leftrightarrow (CO3^{2-})_{melt}$, where a $CO_2$ molecule reacts with an oxygen in the melt to form a carbonate ion, we introduce a Morse potential between the carbon atom of the $CO_2$ molecule and an oxygen of the melt,

$$u_{C...O_m}^{Morse} = D_e [ (1-e^{-(r-l)/\lambda})^2 - 1 ] \tag{A7}$$

where $D_e$ is the dissociation energy of the C...$O_m$ bond, $l$ is the equilibrium distance of the bond (taken equal to the C-O bond length in $CO_2$ molecule, i.e. $l = 1.162A$) and $\lambda$ is the effective width of the potential. Values assigned to the potential parameters $D_e$, $l$ and $\lambda$ (see in Table 1) are close to those chosen by Pavese et al. (1996) in their simulation of the structural and elastic properties of calcite ($CaCO_3$). Thus the Morse potential is effective only at short distances when the C...$O_m$ distance is shorter than ~1.6 A (covalent bonding). At larger distances, the oxygen of the melt interacts with the carbon and oxygen atoms of the $CO_2$ molecule through a non covalent interaction described by a sum of electrostatic, repulsive and dispersive contributions, namely

$$u_{CO2...Om} = uC...OmMorse + \sum_{i \in [CO2]} ( q_i z_{Om}/r_{iOm} + B_{iOm}e^{-riOm/piOm} - C_{iOm}/r_{iOm}{}^6 ) \quad (A8)$$

where i runs over the three atoms of the $CO_2$ molecule, $q_i$ is the effective charge associated with the atom $i$ of the $CO_2$ molecule, $z_{Om}$ is the effective charge of the oxide anion (-0.945 e) and $B_{iOm}$, $\rho_{iOm}$ and $C_{iOm}$ are parameters describing repulsive and dispersive forces (values of parameters are given in Table 1 and are discussed hereafter).

It is noteworthy that the chemical reaction, $CO_2 + (O^{2-})_{melt} \leftrightarrow (CO32-)_{melt}$, is controlled mainly by the energy barrier generated by the repulsion+dispersion terms (~$Be^{-r/\rho}$-$C/r^6$) between the oxide ion of the melt and the two oxygens of the $CO_2$ molecule. As is the case for many other multiply charged anions, $CO32-$ is not electronically stable as isolated species (Boldyrev et al., 1996) because the repulsion between the extra electrons cannot be overcome. This is the confinement in condensed phase which renders this anion to be thermodynamically stable (Stefanovich et al., 1998; Cherginets and Pebrova, 2003). Thus, during the simulation, the condition for a $CO_2$ molecule to capture an oxygen of the melt (forming a carbonate ion) is that the two moving partners find a path to overcome the energy barrier separating each other, a barrier whose the height is modulated by the instantaneous configuration of the atoms of the melt (the opposite situation will take place for the dissociation of $CO32-$). The effectiveness of this mechanism will determine the $CO32-$/$CO_2$ ratio and the value of the $CO_2$ solubility. Basing one's argument on this finding we have adjusted the potential parameters $B_{OcOm}$ and $C_{OcOm}$ to reproduce the experimental value of the $CO_2$ solubility in a silicate melt of natural composition, namely a tholeiite at 1673 K and 20 kbar ($S_{CO2}$~1.5 wt% after Mattey (1991), Pan et al. (1991) and Dixon (1997)). A systematic investigation of the $CO_2$ solubility in this melt as function of the B and C parameters shows that the range of values leading to the observed solubility is very narrow. Furthermore, we will see in the following that with this set of parameters one recovers the solubility data for other melt compositions not used for fitting (e.g. rhyolitic and kimberlitic), a finding which substantiates the robustness of our simulation results. The final parameters of the potential are listed in Table 1.

For illustration, the variation of the potential energy of the isolated complex $CO_2$--$O_m$ in equatorial configuration is shown in Fig.A2 as function of the $C$--$O_m$ separation and for different angulations $\theta$ of the approaching $CO_2$ molecule. For $\theta$ =180° the potential energy curve exhibits only one minimum at $r_{C--Om}$ ~ 2.6A, a configuration corresponding to a molecule-ion complex stabilized by quadrupole-ion interaction. For $\theta$ =160°, the curve presents two minima at 2.15A and 1.35A, respectively, these minima being separated by an energy barrier culminating near 1.55A. The potential well around 2.15A is characteristic of the molecule-ion complex whereas the narrow potential well around 1.35A corresponds to the emergence of the carbonate ion after barrier crossing. For $\theta$ = 140° only the carbonate ion is observed with an energy minimum located around 1.23A.

When a $CO_2$ molecule is immersed in the silicate melt, the energy landscape seen by the molecule is much more complex than the simple picture illustrated in Fig.A2, since it depends on the position of all the atoms of the melt. The role of the MD calculation is to generate, in a self consistent way, such microscopic configurations. In practice, the carbonate ion is formed in the silicate melt when the distance between the carbon atom of a $CO_2$ molecule and an oxygen of the melt is less than 1.6A.

## APPENDIX B. SUPPLEMENTARY TABLES

Supplementary Tables associated with this article can be found, in the online version, at doi: xxxxx

**Table B1**

Solubility of $CO_2$ in silicate melts. The solubility $S$ is expressed in $gCO2/\, gmelt$ , $n$ is the density of the $CO_2$-bearing melt and the ratio $NCO32\text{-}/(NCO2+ NCO32\text{-})$ is the molar fraction in carbonate ions. The simulations were performed with 300 $CO_2$ molecules and 1,000 ions for pressures up to 80 kbar, and with 500, 700 or 1,000 $CO_2$ + 2,000 ions at 100 kbar and beyond. The statistical uncertainties are on the last digit for the three quantities tabulated below. For a detailed comparison with data of the literature see text and Figs.4-6.

| Rhyolite | | | |
|---|---|---|---|
| P (kbar) | n (g/cm³) | S (g/g) | NCO32-NCO2+ NCO32- |
| T = 2273K | | | |
| 1 | 2.30 | 0.0007 | 0.03 |
| 5 | 2.37 | 0.003 | 0.05 |
| 10 | 2.42 | 0.006 | 0.08 |
| 20 | 2.52 | 0.018 | 0.16 |
| 50 | 2.86 | 0.074 | 0.38 |
| 80 | 2.91 | 0.18 | 0.48 |
| 100 | 2.96 | 0.24 | 0.55 |
| 120 | 3.00 | 0.28 | 0.62 |
| 150 | 3.10 | 0.34 | 0.70 |
| MORB | | | |
| P (kbar) | n (g/cm³) | S (g/g) | NCO32-NCO2+ NCO32- |
| T = 2273K | | | |
| 20 | 2.76 | 0.016 | 0.29 |
| 50 | 2.96 | 0.067 | 0.50 |
| 80 | 3.05 | 0.167 | 0.63 |
| 100 | 3.07 | 0.25 | 0.68 |
| 120 | 3.10 | 0.31 | 0.71 |
| 150 | 3.16 | 0.37 | 0.76 |

T = 1873K

| 20 | 2.83 | 0.016 | 0.43 |
| 50 | 2.99 | 0.095 | 0.64 |
| 80 | 3.07 | 0.21 | 0.75 |
| 100 | 3.09 | 0.29 | 0.78 |

T = 1673K

| 20 | 2.94 | 0.016 | 0.54 |
| 50 | 3.07 | 0.10 | 0.73 |
| 80 | 3.12 | 0.23 | 0.82 |
| 100 | 3.15 | 0.30 | 0.85 |

T = 1473K

| 20 | 3.01 | 0.016 | 0.63 |

Kimberlite

| P (kbar) | n (g/cm$^3$) | S (g/g) | NCO32-NCO2+ NCO32- |
| --- | --- | --- | --- |

T = 2273K

| 20 | 2.78 | 0.019 | 0.43 |
| 50 | 2.98 | 0.077 | 0.60 |
| 80 | 3.05 | 0.20 | 0.70 |
| 100 | 3.05 | 0.30 | 0.72 |
| 120 | 3.02 | 0.40 | 0.73 |

**Table B2**

Mean distances (in A) between $CO_2$ and $CO_3^{2-}$ species and the elements of the melt as evaluated from the position of the first peak exhibited by the corresponding PDF. In the table, $O_m$ is an oxygen of the melt, $O_{CO_3^{2-}}$ is the oxygen of the melt forming a C-O bond with the carbonate ion, $C_{CO_3^{2-}}$ is the carbon atom of a carbonate ion, and $C_{CO_2}$ the carbon atom of a $CO_2$ molecule. Are also given for comparison the cation-oxygen mean distances ($O_m$-$X$) in the corresponding melt. In the case of $C_{CO_3^{2-}}$-- $O_m$ and $C_{CO_2}$- $O_m$ the two reported values correspond to the mean distance with the first neighbor oxygen and with the first coordination shell, respectively (see also Fig.13). All values were evaluated at 20 kbar and 2273K with $CO_2$ contents of 2 wt%. The uncertainty on the position of the first peak is ±0.025A.

| $X$ | melt | $Si$-$X$ | $Ti$-$X$ | $Al$-$X$ | $Fe^{3+}$-$X$ | $Fe^{2+}$-$X$ | $Mg$-$X$ | $Ca$-$X$ | $Na$-$X$ | $K$-$X$ | $O_m$-$X$ |
|---|---|---|---|---|---|---|---|---|---|---|---|
| $O_m$ | Rhyolite | 1.65 | - | 1.75 | 1.85 | 2.05 | - | 2.40 | 2.45 | 3.0 | 2.65 |
| | MORB | 1.65 | 1.95 | 1.75 | 1.85 | 2.05 | 1.95 | 2.40 | 2.45 | 2.95 | 2.65 |
| | Kimberlite | 1.65 | 1.95 | 1.75 | - | 2.05 | 1.95 | 2.35 | 2.35 | 2.85 | 2.65 |
| $O_{CO_3^{2-}}$ | Rhyolite | 1.65 | - | 1.75 | 1.85 | 2.05 | - | 2.45 | 2.45 | 3.25 | 2.65 |
| | MORB | 1.65 | 1.95 | 1.75 | 1.85 | 2.10 | 1.95 | 2.40 | 2.45 | 3.05 | 2.65 |
| | Kimberlite | 1.65 | 1.95 | 1.75 | - | 2.05 | 1.95 | 2.40 | 2.45 | 3.10 | 2.70 |
| $C_{CO_3^{2-}}$ | Rhyolite | 2.45 | - | 2.45 | 2.50 | 3.00 | - | 3.10 | 3.25 | 3.65 | 1.25; 3.60 |

| | | | | | | | | | | |
|---|---|---|---|---|---|---|---|---|---|---|
| | MORB | 2.50 | 2.95 | 2.50 | 2.50 | 3.00 | 2.65 | 3.10 | 3.25 | 3.50 | 1.25; 3.60 |
| | Kimberlite | 2.55 | 2.95 | 2.50 | - | 3.00 | 2.65 | 3.10 | 3.20 | 3.60 | 1.25; 3.60 |
| CCO2 | Rhyolite | 3.75 | - | 3.55 | 3.50 | 3.25 | - | 3.35 | 3.35 | 3.70 | 2.65; 3.60 |
| | MORB | 3.65 | 4.0 | 3.60 | 3.50 | 3.25 | 3.05 | 3.35 | 3.35 | 3.50 | 2.65; 3.60 |
| | Kimberlite | 3.65 | 4.0 | 3.60 | - | 3.25 | 3.10 | 3.35 | 3.20 | 3.60 | 2.65; 3.60 |

**APPENDIX C**

The affinities $ABOi$ and $ANBOi$ of species i (i = $CO_2$ or $CO3-$) for BO's and NBO's, are defined by the following relationships,

$$ABOi + ANBOi = 1 \qquad (C1)$$

$$ABOi = \text{Norm} \times (PBOi / PBO) \qquad (C2)$$

$$ANBOi = \text{Norm} \times (PNBOi / PNBO) \qquad (C3)$$

$$\text{Norm} = 1/ [\, PBOiPBO \ + \ PNBOiPNBO \,] \qquad (C4)$$

where $PBO$ and $PNBO$ are the populations of BO's and NBO's in the melt (given in columns 5 and 9 of Table 4), and where $PBOi$ and $PNBOi$ are the populations of species *i* which are associated with BO's and NBO's (also given in columns 5 and 9 of Table 4). In this framework, the affinity has the meaning of a probability of association. In the particular case where $PBOi = PBO$ and $PNBOi = PNBO$, the affinities are all equal to 1/2, which means that $CO3-$ (or $CO_2$) has the same probability to be associated with a BO or a NBO. In practice (see Table 4) in the three melts it is observed that $PBOi < PBO$ and $PNBOi > PNBO$ and therefore the affinities are very different from each other.

The affinities $AXBCarb$ and $AXNBCarb$ of a cation of species X (with X = Ca, Na, ..) for the carbonate ions (BCarb or NBCarb) can be defined in a similar way,

$$XAXCarb \ = 1 \qquad (C5)$$

$$AXCarb = \text{Norm} \times (RXCarb / RXUDM) \qquad (C6)$$

$$\text{Norm} = 1/ XRXCarbRXUDM \qquad (C7)$$

where the superscript *Carb* is for BCarb or NBCarb, $RXCarb$ is the occurence rate to observe a cation of species X associated with a BCarb or a NBCarb (see Table 5), and $RXUDM$ is the occurence rate to observe a cation X around a carbonate ion in the Uniform Distribution Model ( $RXUDM = NX/XNX$ , where $N_X$ is the number of cation of species *X* ). In the particular case where $RXCarb =$

RXUDM , the affinities are all equal to $1/N_s$ where $N_s$ is the number of cationic species in the melt (7 in rhyolite, 9 in MORB and 8 in kimberlite). In practice in the three melts it is observed most of the time that for network modifier cations RXCarb > RXUDM whereas for network former cations RXCarb < RXUDM (see Table 5).

**Table 1**

Potential parameters.

Silicate potential (see eqn.(A1))

| X | $z_x(e)$ | $B_{xo}$ (kJ/mol) | $\rho_{xo}$ (A) | $C_{xo}$ (A$^6$ kJ/mol) |
|---|---|---|---|---|
| O | -0.945 | 870570.0 | 0.265 | 8210.17 |
| Si | 1.89 | 4853815.5 | 0.161 | 4467.07 |
| Ti | 1.89 | 4836495.0 | 0.178 | 4467.07 |
| Al | 1.4175 | 2753544.3 | 0.172 | 3336.26 |
| $Fe^{3+}$ | 1.4175 | 773840.0 | 0.190 | 0.0 |
| $Fe^{2+}$ | 0.945 | 1257488.6 | 0.190 | 0.0 |
| Mg | 0.945 | 3150507.4 | 0.178 | 2632.22 |
| Ca | 0.945 | 15019679.1 | 0.178 | 4077.45 |
| Na | 0.4725 | 11607587.5 | 0.170 | 0.0 |
| K | 0.4725 | 220447.4 | 0.290 | 0.0 |

$CO_2$ intramolecular potential (see eqn(A3)): $l_{C-O}$ = 1.162 A, $k_\theta$ = 444 kJ/mol/rd

$CO_2$-$CO_2$ intermolecular potential (see eqn.(A2))

| | q (e) | $\varepsilon$ (kJ/mol) ) | $\sigma$ (A) |
|---|---|---|---|
| C | 0.5888 | | |
| O | -0.2944 | | |
| C-C | | 0.2398 | 2.792 |
| C-O | | 0.4059 | 2.896 |
| O-O | | 0.6872 | 3.000 |

Morse potential between C of $CO_2$ and $O_m$ of the silicate (see eqn(A7)):

$D_e$ = 484 kJ/mol, $l$ = 1.162 A, $\lambda$ = 0.2 A

$CO_2$/silicate potential (see eqns.(A4-A8))

| pair | B (kJ/mol) ) | $\rho$ (A) | C (A$^6$ kJ/mol) |
|---|---|---|---|
| $O_c$-$O_m$ | 435285.0 | 0.265 | 4105.09 |
| $O_c$-Si | 4853815.5 | 0.161 | 4467.07 |
| $O_c$-Ti | 4836495.0 | 0.178 | 4467.07 |
| $O_c$-Al | 2753544.3 | 0.172 | 3336.26 |
| $O_c$-$Fe^{3+}$ | 773840.0 | 0.190 | 0.0 |
| $O_c$-$Fe^{2+}$ | 1257488.6 | 0.190 | 0.0 |
| $O_c$-Mg | 3150507.4 | 0.178 | 2632.22 |
| $O_c$-Ca | 15019679.1 | 0.178 | 4077.45 |
| $O_c$-Na | 11607587.5 | 0.170 | 0.0 |
| $O_c$-K | 220447.4 | 0.290 | 0.0 |

_note_: $O_c$ is an oxygen of a $CO_2$ molecule and $O_m$ an oxygen of the silicate

**Table 2**

Chemical composition (in weight %) on a free-volatile basis of the silicate melts simulated in this study. In parenthesis are indicated the numbers of cations of each species used in the simulations for a total of ~1,000 silicate melt and in the last column is indicated the total number of simulated atoms.

|  | $SiO_2$ | $TiO_2$ | $Al_2O_3$ | $Fe_2O_3$ | FeO | MgO | CaO | $Na_2O$ | $K_2O$ | Total |
|---|---|---|---|---|---|---|---|---|---|---|
| Rhyolite[a] | 74.51 | 0.10 | 13.25 | 0.32 | 1.28 | 0.08 | 0.75 | 4.15 | 5.64 | 100.08 |
|  | (257) | (0) | (54) | (1) | (3) | (0) | (3) | (28) | (25) | (1000) |
| Basalt[b] | 50.59 | 1.52 | 15.11 | 1.15 | 8.39 | 7.77 | 11.87 | 2.94 | 0.13 | 99.47 |
|  | (185) | (4) | (65) | (3) | (26) | (42) | (47) | (21) | (1) | (1000) |
| Kimberlite[c] | 37.17 | 2.02 | 3.53 | 0.0 | 10.32 | 36.16 | 8.99 | 0.81 | 1.01 | 100.01 |
|  | (135) | (6) | (15) | (0) | (31) | (196) | (35) | (6) | (5) | (1001) |

[a] natural obsidian (Gaillard et al., 2003)
[b] basalt from the mid-atlantic ridge (TK21B in Reynolds and Langmuir, 1997)
[c] synthetic average kimberlite (Brey et al., 1991)

**Table 3**

Lifetimes ($\tau$) and diffusivities ($D$) of $CO_2$ and $CO3^2-$ species in silicate melts with 2wt% $CO_2$. Notice that $D_C$ is given by a weighted sum of $DCO2$ and $DCO32-$ (see eqn.(3) in the text) and $DOm$ is the self diffusion coefficient of the oxygens of the silicate. For a comparison with experimental data see text and Fig.12. The uncertainties are ±10% on lifetimes, ±5% on $DC$ and ±30% on $DCO2$ and $DCO32-$.

| melt | T(K) | P(kbar) | $\tau CO2$(ps) | $\tau CO32-$(ps) | $DCO2$($10^{-9}$m²/s) | $DCO32-$($10^{-9}$m²/s) | $DC$($10^{-9}$m²/s) | $DOm$($10^{-9}$m²/s) |
|------|------|---------|---------------|------------------|----------------------|-------------------------|---------------------|----------------------|
| Rhyolite | 2273 | 20 | 27 | 5 | 2.5 | 0.6 | 2.2 | 0.3 |
| | 1873 | 20 | 40 | 12 | 0.64 | 0.08 | 0.53 | 0.04 |
| | 1473 | 20 | 70 | 50 | 0.084 | 0.003 | 0.053 | 0.00085 |
| | 2273 | 100 | 5 | 9 | 1.1 | 0.4 | 0.6 | 0.4 |
| MORB | 2273 | 20 | 14 | 6.5 | 3.14 | 1.5 | 2.6 | 1.37 |
| | 1873 | 20 | 22 | 16 | 1.48 | 0.4 | 1.05 | 0.42 |
| | 1473 | 20 | 35 | 65 | 0.29 | 0.05 | 0.13 | 0.045 |
| | 2273 | 100 | 2.8 | 11 | 2.5 | 0.8 | 1.1 | 1.1 |
| Kimberlite | 2273 | 20 | 12.5 | 9 | 6.1 | 2.7 | 4.8 | 2.7 |
| | 2273 | 100 | 2.5 | 13 | 2.7 | 1.4 | 1.6 | 1.9 |

(in A) between $CO_2$ and $CO_3^{2-}$ species and the elements of the melt as evaluated from the position of the first peak exhibited by the corresponding PDF. In the table, $O_m$ is an oxygen of the melt, $O_{CO_3^{2-}}$ is the oxygen of the melt forming a C-O bond with the carbonate ion, $C_{CO_3^{2-}}$ is the carbon atom of a carbonate ion, and $C_{CO_2}$ the carbon atom of a $CO_2$ molecule. Are also given for comparison the cation-oxygen mean distances ($O_m$-X) in the corresponding melt. In the case of $C_{CO_3^{2-}}$-$O_m$ and $C_{CO_2}$-$O_m$ the two reported values correspond to the mean distance with the first neighbor oxygen and with the first coordination shell, respectively (see also Fig.17). All values were evaluated at 20 kbar and 2273K with $CO_2$ contents of 2 wt%. The uncertainty on the position of the first peak is ±0.025A.

| X | melt | Si-X | Ti-X | Al-X | Fe³⁺-X | Fe²⁺-X | Mg-X | Ca-X | Na-X | K-X | Oₘ-X |
|---|---|---|---|---|---|---|---|---|---|---|---|
| $O_m$ | Rhyolite | 1.65 | - | 1.75 | 1.85 | 2.05 | - | 2.40 | 2.45 | 3.0 | 2.65 |
| | MORB | 1.65 | 1.95 | 1.75 | 1.85 | 2.05 | 1.95 | 2.40 | 2.45 | 2.95 | 2.65 |
| | Kimberlite | 1.65 | 1.95 | 1.75 | - | 2.05 | 1.95 | 2.35 | 2.35 | 2.85 | 2.65 |
| $O_{CO_3^{2-}}$ | Rhyolite | 1.65 | - | 1.75 | 1.85 | 2.05 | - | 2.45 | 2.45 | 3.25 | 2.65 |
| | MORB | 1.65 | 1.95 | 1.75 | 1.85 | 2.10 | 1.95 | 2.40 | 2.45 | 3.05 | 2.65 |
| | Kimberlite | 1.65 | 1.95 | 1.75 | - | 2.05 | 1.95 | 2.40 | 2.45 | 3.10 | 2.70 |
| $C_{CO_3^{2-}}$ | Rhyolite | 2.45 | - | 2.45 | 2.50 | 3.00 | - | 3.10 | 3.25 | 3.65 | 1.25; 3.60 |
| | MORB | 2.50 | 2.95 | 2.50 | 2.50 | 3.00 | 2.65 | 3.10 | 3.25 | 3.50 | 1.25; 3.60 |
| | Kimberlite | 2.55 | 2.95 | 2.50 | - | 3.00 | 2.65 | 3.10 | 3.20 | 3.60 | 1.25; 3.60 |

| CCO2 | Rhyolite | 3.75 | - | 3.55 | 3.50 | 3.25 | - | 3.35 | 3.35 | 3.70 | 2.65; 3.60 |
|------|----------|------|------|------|------|------|------|------|------|------|------------|
| | MORB | 3.65 | 4.0 | 3.60 | 3.50 | 3.25 | 3.05 | 3.35 | 3.35 | 3.50 | 2.65; 3.60 |
| | Kimberlite | 3.65 | 4.0 | 3.60 | - | 3.25 | 3.10 | 3.35 | 3.20 | 3.60 | 2.65; 3.60 |

**Table 4**

Populations (in %) of bridging oxygens (BO) and non-bridging oxygens (NBO) in the simulated melts at 20 kbar with 2 wt% $CO_2$, and percentage of $CO_3^{2-}$ and $CO_2$ associated with BO's and NBO's, respectively. For BO's we have distinguished the Si-O-Si, Si-O-Al and Al-O-Al bonds (the small population of triclusters, if any, is accounted for in BO) whereas for NBO's we have distinguished Si-O and Al-O bonds from the free oxygens only connected to network modifier cations. The ratio ANBOi/ABOi is also indicated, where ANBOi is the affinity of a given species ($i$ = $CO_3^{2-}$ or $CO_2$) for NBO's and ABOi is its affinity for BO's (see text and Appendix C). The BO and NBO populations evaluated from a standard model of glass structure (see Mysen and Richet, 2005) are also indicated for comparison (see the subscript numbers in raw labeled $O_m$).

|  | Si-O-Si | Si-O-Al | Al-O-Al | BO | Si-O | Al-O | free oxygens | NBO | ANBO/ABO |
|---|---|---|---|---|---|---|---|---|---|
| **Rhyolite T = 2273K** | | | | | | | | | |
| $O_m$ | 62.7 | 21.6 | 1.4 | $88.3_{98.4}$ | 10.9 | 0.8 | 0.0 | $11.7_{1.6}$ | |
| $CO_2$ | 54.7 | 21.8 | 1.5 | 78.5 | 20.3 | 1.1 | 0.1 | 21.5 | 2.07 |
| $CO_3^{2-}$ | 21.0 | 19.0 | 2.8 | 42.9 | 52.1 | 4.2 | 0.8 | 57.1 | 10.05 |
| **Rhyolite T = 1873K** | | | | | | | | | |
| $O_m$ | 64.1 | 23.8 | 1.4 | $91.5_{98.4}$ | 7.9 | 0.5 | 0.0 | $8.5_{1.6}$ | |
| $CO_2$ | 62.0 | 22.8 | 1.3 | 86.4 | 12.7 | 0.8 | 0.0 | 13.5 | 1.68 |
| $CO_3^{2-}$ | 25.2 | 22.6 | 2.1 | 49.8 | 46.4 | 3.4 | 0.3 | 50.1 | 10.81 |
| **Rhyolite T = 1473K** | | | | | | | | | |
| $O_m$ | 64.9 | 24.4 | 1.5 | $93.3_{98.4}$ | 6.5 | 0.2 | 0.0 | $6.7_{1.6}$ | |
| $CO_2$ | 64.4 | 22.9 | 2.6 | 90.6 | 9.2 | 0.2 | 0.0 | 9.4 | 1.45 |
| $CO_3^{2-}$ | 16.6 | 32.2 | 10.9 | 59.9 | 38.3 | 1.7 | 0.1 | 40.1 | 9.32 |

---

MORB   T = 2273K

| | | | | | | | | | |
|---|---|---|---|---|---|---|---|---|---|
| $O_m$ | 31.6 | 22.2 | 3.0 | $58.3_{67.0}$ | 33.6 | 5.9 | 2.2 | $41.7_{33.0}$ | |
| $CO_2$ | 23.9 | 18.9 | 2.4 | 45.5 | 44.6 | 6.9 | 3.0 | 54.5 | 1.67 |
| $CO_3^{2-}$ | 6.2 | 9.8 | 2.6 | 18.6 | 60.0 | 13.1 | 8.3 | 81.4 | 6.12 |

MORB   T = 1873K

| | | | | | | | | | |
|---|---|---|---|---|---|---|---|---|---|
| $O_m$ | 31.9 | 24.3 | 3.3 | $60.8_{67.0}$ | 32.2 | 5.4 | 1.6 | $39.2_{33.0}$ | |
| $CO_2$ | 26.3 | 21.5 | 3.2 | 51.3 | 40.7 | 6.2 | 1.9 | 48.8 | 1.48 |
| $CO_3^{2-}$ | 5.3 | 10.3 | 3.7 | 19.3 | 59.1 | 15.0 | 6.6 | 80.7 | 6.49 |

MORB   T = 1473K

| | | | | | | | | | |
|---|---|---|---|---|---|---|---|---|---|
| $O_m$ | 31.6 | 26.5 | 3.5 | $62.7_{67.0}$ | 31.2 | 4.8 | 1.3 | $37.3_{33.0}$ | |
| $CO_2$ | 29.4 | 24.3 | 2.9 | 56.9 | 37.5 | 4.5 | 1.1 | 43.1 | 1.27 |
| $CO_3^{2-}$ | 4.1 | 13.2 | 4.3 | 21.7 | 62.5 | 11.7 | 4.2 | 78.4 | 6.10 |

---

Kimberlite   T = 2273K

| | | | | | | | | | |
|---|---|---|---|---|---|---|---|---|---|
| $O_m$ | 14.4 | 4.1 | 0.2 | $18.8_{5.0}$ | 60.6 | 4.4 | 16.2 | $81.2_{95.0}$ | |
| $CO_2$ | 9.0 | 2.5 | 0.1 | 11.6 | 65.4 | 3.9 | 19.1 | 88.4 | 1.76 |
| $CO_3^{2-}$ | 1.5 | 0.9 | 0.1 | 2.5 | 58.5 | 5.3 | 33.7 | 97.5 | 9.03 |

**Table 5**

Occurence rates ( RXBCarb and RXNBCarb) and affinities ( AXBCarb and AXNBCarb) of silicate cations for the carbonate ions (see text and Appendix C for definitions) in the simulated melts at 2273K and 20 kbar with 2 wt% $CO_2$. The occurrence rates are evaluated on a $CO_2$-free basis and are expressed in percent. Values in parenthesis are obtained from the uniform distribution model which assumes that the distribution of the metal cations around the carbonate ions is uniform and their affinity all equal to each other. The raw labeled *Total* is the occurrence rate for observing a cation of species X in association with a carbonate ion whatever it is (BCarb or NBCarb). The mean distance dC-X is given in A (±0.01A).

| | | $Si$ | $Ti$ | $Al$ | $Fe^{3+}$ | $Fe^{2+}$ | $Mg$ | $Ca$ | $Na$ | $K$ |
|---|---|---|---|---|---|---|---|---|---|---|
| **Rhyolite** | | | | | | | | | | |
| *BCarb-X* | RXBCarb | 47.1 (69.3) | - - | 16.9 (14.6) | 0.3 (0.3) | 1.60 (0.8) | - - | 2.3 (0.8) | 21.3 (7.6) | 10.4 (6.7) |
| | AXBCarb | 0.05 (0.14) | - - | 0.09 (0.14) | 0.09 (0.14) | 0.15 (0.14) | - - | 0.21 (0.14) | 0.21 (0.14) | 0.12 (0.14) |
| | dC-X | 3.28 | | 3.11 | 3.23 | 3.08 | | 3.11 | 3.03 | 3.21 |
| *NBCarb-X* | RXNBCarb | 25.4 (69.3) | - - | 23.0 (14.6) | 1.0 (0.3) | 4.5 (0.8) | - - | 3.7 (0.8) | 29.0 (7.6) | 14.1 (6.7) |
| | AXNBCarb | 0.02 (0.14) | - - | 0.07 (0.14) | 0.17 (0.14) | 0.25 (0.14) | - - | 0.20 (0.14) | 0.17 (0.14) | 0.09 (0.14) |
| | dC-X | 3.04 | | 2.76 | 2.67 | 2.75 | - - | 2.87 | 2.87 | 3.10 |
| *Total* | | 34.7 (69.3) | - - | 20.4 (14.6) | 0.7 (0.3) | 3.3 (0.8) | - - | 3.1 (0.8) | 25.7 (7.6) | 12.5 (6.7) |
| **MORB** | | | | | | | | | | |
| *BCarb-X* | RXBCarb | 19.1 (47.0) | 0.3 (1.0) | 11.7 (16.5) | 0.7 (0.8) | 10.8 (6.6) | 20.4 (10.7) | 25.0 (11.9) | 12.7 (5.3) | 0.3 (0.2) |
| | AXBCarb | 0.03 (0.11) | 0.03 (0.11) | 0.06 (0.11) | 0.07 (0.11) | 0.13 (0.11) | 0.16 (0.11) | 0.17 (0.11) | 0.19 (0.11) | 0.11 (0.11) |
| | dC-X | 3.17 | 3.19 | 3.00 | 2.97 | 2.95 | 2.84 | 2.99 | 2.93 | 3.14 |
| *NBCarb-X* | RXNBCarb | 7.4 (47.0) | 1.5 (1.0) | 11.7 (16.5) | 1.1 (0.8) | 13.1 (6.6) | 29.8 (10.7) | 26.8 (11.9) | 10.3 (5.3) | 0.2 (0.2) |

| | | | | | | | | | |
|---|---|---|---|---|---|---|---|---|---|
| AXBCarb | 0.01 (0.11) | 0.11 (0.11) | 0.05 (0.11) | 0.10 (0.11) | 0.14 (0.11) | 0.20 (0.11) | 0.16 (0.11) | 0.14 (0.11) | 0.06 (0.11) |
| dC-X | 2.90 | 2.66 | 2.68 | 2.73 | 2.69 | 2.58 | 2.75 | 2.71 | 2.88 |
| *Total* | 9.5 (47.0) | 1.3 (1.0) | 11.7 (16.5) | 1.0 (0.8) | 12.7 (6.6) | 28.0 (10.7) | 26.5 (11.9) | 10.7 (5.3) | 0.2 (0.2) |

Kimberlite

| | | | | | | | | | | |
|---|---|---|---|---|---|---|---|---|---|---|
| *BCarb-X* | RXBCarb | 6.0 (31.5) | 0.0 (1.4) | 1.5 (3.5) | - - | 11.1 (7.2) | 68.2 (45.7) | 11.3 (8.2) | 1.7 (1.4) | 1.7 (1.2) |
| | AXBCarb | 0.02 (0.13) | 0.0 (0.13) | 0.05 (0.13) | - - | 0.19 (0.13) | 0.18 (0.13) | 0.17 (0.13) | 0.15 (0.13) | 0.18 (0.13) |
| | dC-X | 3.05 | ... | 2.95 | - - | 3.01 | 2.84 | 2.90 | 2.88 | 3.16 |
| *NBCarb-X* | RXNBCarb | 1.7 (31.5) | 1.0 (1.4) | 1.0 (3.5) | - - | 8.2 (7.2) | 76.4 (45.7) | 11.0 (8.2) | 1.9 (1.4) | 0.6 (1.2) |
| | AXNBCarb | 0.08 (0.13) | 0.10 (0.13) | 0.04 (0.13) | - - | 0.16 (0.13) | 0.23 (0.13) | 0.19 (0.13) | 0.18 (0.13) | 0.07 (0.13) |
| | dC-X | 2.80 | 2.62 | 2.60 | - - | 2.60 | 2.52 | 2.68 | 2.64 | 2.87 |
| *Total* | | 1.8 (31.5) | 1.0 (1.4) | 1.0 (3.5) | - - | 8.3 (7.2) | 76.2 (45.7) | 11.0 (8.2) | 1.8 (1.4) | 0.7 (1. |

## Figure captions

**Fig.A1** Equation of state of supercritical $CO_2$. The *full circles* represent our simulation data, the *dotted curves* are the two isotherms T = 700K and T = 2273K (insert) predicted by the EOS of Sterner and Pitzer (1994) and the two *full curves* is are analytical representations (accounting for error bars) of the Brillouin scattering data of Giordano et al. (2006) for the isotherm T = 700K.

**Fig.A2** Variation of the potential energy of the isolated complex $CO_2$---$O_m$ in equatorial configuration as function of the $C$---$O_m$ separation and for different angulations $\theta$ of the approaching $CO_2$ molecule. The arrows indicate minima of the potential curve. The minima located below $R_{C---O}$ ~1.5A correspond to the formation of CO32- whereas minima located at larger distances correspond to a molecule-ion association.

**Fig.1** Running average of the number of carbon atoms in a simulated MORB melt (with 1,000 ions) in contact with a $CO_2$ phase at T = 2273K and P = 20 and 80 kbar as function of running time (in ns). The step value (in red for the electronic version) is also shown for comparison.

**Fig.2** Mean density profile along the X-axis of the simulation box for a MORB melt composed of 1,000 ions in equilibrium at 2273K and 50kbar with a supercritical $CO_2$ phase formed initially by 300 $CO_2$ molecules (notice that the silicate melt is located in the center part of the box and the $CO_2$ supercritical fluid is on each side). The curve labeled "*MORB + CO_2*" corresponds to the total density evaluated at the abscissa X of the box, the curve labeled "*MORB*" only accounts for the silicate components and the curve labeled "*CO_2*" takes into account only $CO_2$ molecules and carbonate ions. A snapshot of the simulation cell is also shown for comparison (for the electronic version, legend to colors: O (red), Si (yellow), Al (white), Fe (pink), Mg (cyan), Ca (cyan), Na (blue), and C (cyan)). For clarity, the size of the silicate atoms were arbitrarily reduced with respect to that of $CO_2$ molecules.

**Fig.3** Solubility of $CO_2$ in simulated MORB at 2273K as function of pressure. The *full circles* are our simulation data and the lines are just guide for the eyes. Notice the difference in behavior between the contribution coming from molecular $CO_2$ and that associated with CO32- (in color with the electronic

version). The insert shows the evolution of the molar fraction in carbonate ions with pressure. The arrow indicates an increase of the concentration in $CO3^{2-}$ when the melt becomes $CO_2$ undersaturated (the *empty circle* corresponds to a $CO_2$ abundance of 2 wt% at 100 kbar instead of ~25wt% at saturation).

**Fig.4** Solubility of $CO_2$ in simulated rhyolite at 2273K as function of pressure (other details are as in Fig.3). The *dotted line* and the *dashed line* are the low pressure data of Fogel and Rutherford (1990) and Blanck et al. (1993) which follow the Henry's law.

**Fig.5** Solubility of $CO_2$ in simulated kimberlite at 2273K as function of pressure (other details are as in Fig.3). The *squares* (with error bars) represent the $CO_2$ contents in a kimberlitic melt measured by Gudfinnsson and Presnall (2005) over the P-T range 1923-2073K and 47-80 kbar (see text), the *circle* is the solubility measured by Brey et al. (1991) at 20 kbar and 1923K in the presently studied kimberlite composition and the *triangle* is the $CO_2$ abundance measured by Dalton and Presnall (1998) in a related kimberlitic melt at 60 kbar and 1778K.

**Fig.6** Temperature dependence of $CO_2$ solubility in simulated MORB. The *full circles* are our simulation data and the lines (in color with the electronic version) are just guide for the eyes. The experimental data are the following: *squares* (Mattey (1991) at 1673K), *triangle* (Pan et al. (1991) at 1673K), *circle* with error bar (Thomsen and Schmidt (2008) at 1573K), *hexagon* (Hammouda (2003) at 1573K) and the *dotted curve* is the Henry's law observed at low pressure and at 1473K by Dixon et al. (1995) and Jendrzejewsky et al. (1997).

**Fig.7** Molar fraction of molecular $CO_2$ in simulated MORB as function of temperature and pressure. A simple low temperature extrapolation suggests that the concentration in molecular $CO_2$ is likely negligible at room temperaturehe glass transition temperature (~950K).

**Fig.8** Compressibility of $CO_2$-bearing MORB at isothermal condition (2273K). The $CO_2$ content is indicated in wt% on each compressibility curve, the *blue curve* (in the electronic version) starting at zero pressure is the isotherm 2273K for the dry melt, and the $CO_2$ saturation curve is delimited by the *red dots* (the *dotted curve* being just a guide for the eyes). The horizontal double arrow shows the density shift experienced by a MORB melt containing 2 wt% $CO_2$ when the temperature is lowered from 2273K to 1873K. A comparison with the mantle density profile *ak135* of Kennett et al. (1995) (dashed line) shows that this basaltic melt (with 2 wt% $CO_2$) may be buoyant in the 410 km discontinuity region.

**Fig.9** As Fig.8 but for a $CO_2$-bearing kimberlitic melt. In that case it is the compressibility curve for a melt at 1873K with 5 wt% $CO_2$ which intersects the mantle density profile ak135 of Kennett et al. (1995) in the 410 km discontinuity region.

**Fig.10** Partial molar volume of $CO_2$ in MORB and kimberlite along the saturation curve at 2273K. The *squares* and the *full dots* with error bars are our simulation data in MORB and kimberlite, the *dotted line* is the partial molar volume in our simulated pure fluid $CO_2$, the *cross* at P = 0 is the value deduced by Liu and Lange (2003) from carbonate liquids and the *shaded rectangular box* represents rough estimates made by these latter authors for silicate liquids of basaltic composition in the 10-20 kbar range.

Lifetime distributions of $CO_2$ and $CO_3^{2-}$ in a $CO_2$-saturated MORB at 2273K and 20 kbar. The circles are our raw data in a log representation, the two distributions being quasi exponential (i.e. $P(t)\sim e^{-t/\tau}$) with the associated decay times $\tau CO2 = 14$ ps and $\tau CO_3^{2-} = 6.5$ ps, respectively.

Temperature dependence of average lifetimes in MORB and rhyolite. The subscript *2* is for $CO_2$ and *3* is for $CO_3^{2-}$, the *full circles* and the *triangles* are for MORB and the *squares* and the *empty circles* are for rhyolite. The activation energies are expressed in kJ/mol and the lifetimes in ps.

**Fig.11** Diffusion coefficient of $CO_2$ in simulated melts as function of the parameter NBO/T (see text). The *full dots* correspond to the diffusivity of the carbon atoms into the melt, the *crosses* to the diffusivity of $CO_2$ species, the *triangles* to the diffusivity of carbonate ions and the *squares* to the diffusivity of the oxygens of the melt. All data are for melts at 20 kbar with a $CO_2$ abundance of 2 wt %.

**Fig.12** Diffusion coefficient of $CO_2$ in silicate melts of various composition. Our data for the diffusion coefficient of $CO_2$ ($D_C$ in eqn.(3)) in rhyolite (*empty circles*), MORB (*full circles*) and kimberlite (*square*) are compared with data of the literature: Watson (1991) for a rhyolite, Watson et al. (1982) for an aluminosilicate, Zhang and Stolper (1991) for a basalt, Blank et al. (1991) for a rhyolite, Nowak et al. (2004) for a rhyolite, Fogel and Rutherford (1990) for a rhyolite, Sierralta et al. (2002) for albite, Spickenbom and Nowak (2005) for albite.

**Fig.13** Pair distribution functions gC-Om(r) and gC-Si(r) between the carbon atom of $CO_2$ and CO32- species and the oxygen and silicon atoms of the melt (in blue for $CO_2$ and in red for CO32-, the curves associated with MORB and kimberlite being shifted vertically for clarity). Notice the intense peak around 1.25 A in gC-Om(r) associated with CO32- which indicates that an oxygen of the melt has been captured by a $CO_2$ molecule to form a carbonate ion. The shoulder around 2.60 A in gC-Om(r) associated with $CO_2$ characterizes a non bonding ion-molecule association.

**Fig.14** Typical configurations observed in the simulations (snapshots) between cations of the melt and $CO_2$ and CO32- species. The interatomic distances are in A. The captions are the following: (a) a BCarb (Si-Carb-Si) is interacting with a Na atom, (b) a NBCarb (Si-Carb) is interacting with a Ca atom, (c) a $CO_2$ molecule is located near to a BO and interacts with a vicinal Ca atom, and (d) a NBCarb (Si-Carb) is bonding to an Al atom to form a Si-Carb-Al moiety. For clarity, the oxygen atoms not directly involved into these associations are not displayed. Configurations (a), (b) and (c) are taken from a MORB melt and configuration (d) from a rhyolitic melt, both melts being at 2273K and 20 kbar.

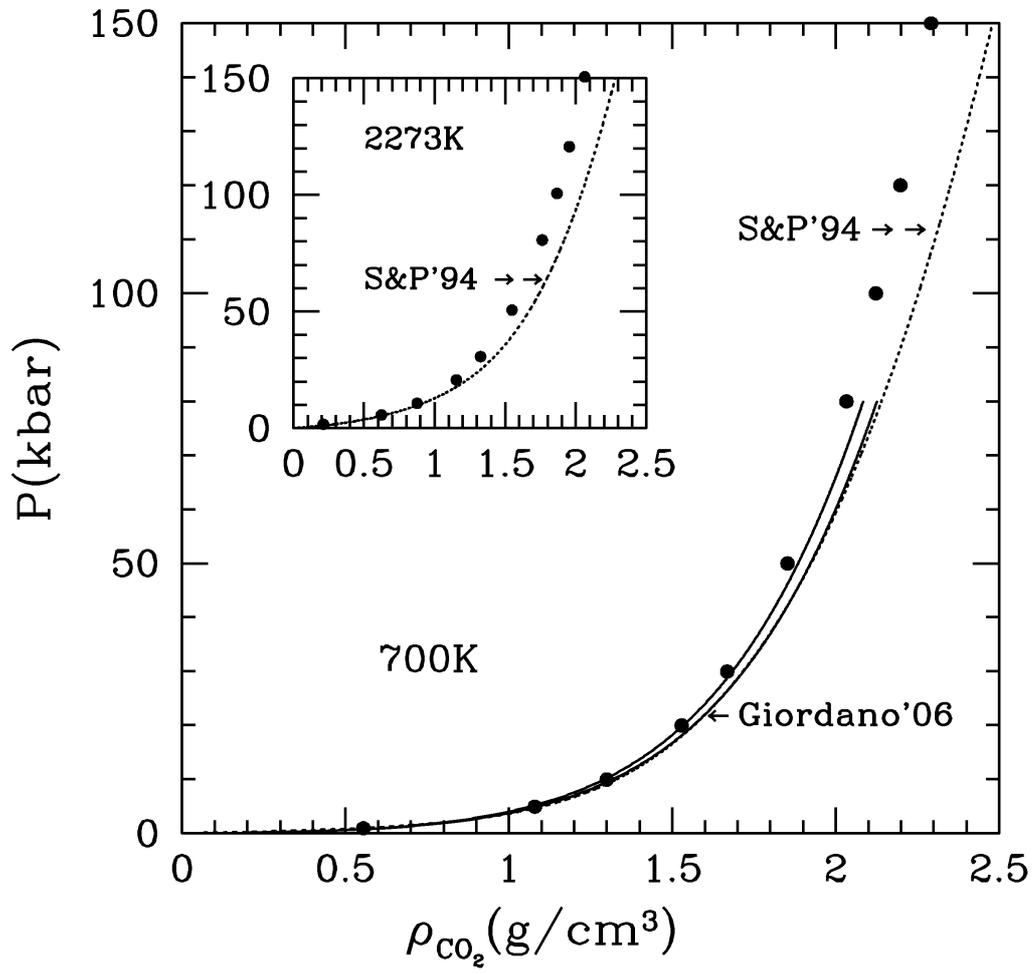

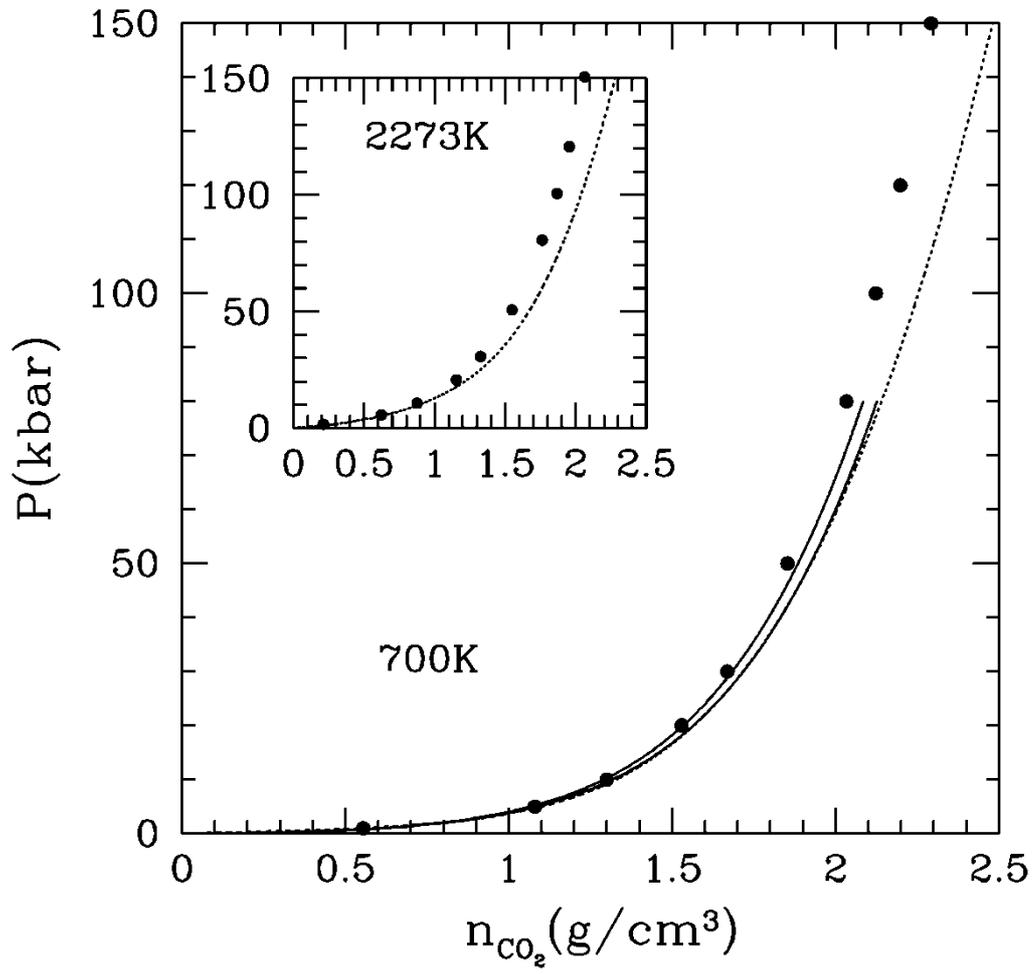

**A1**

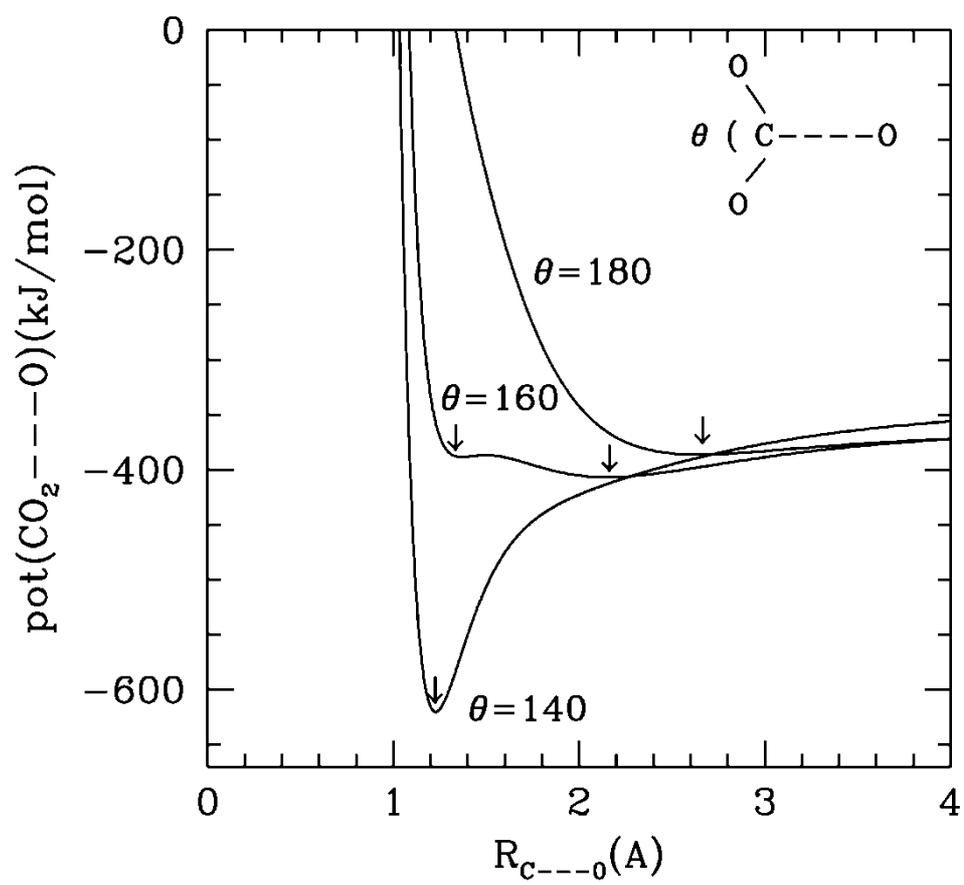

**Fig.A2**

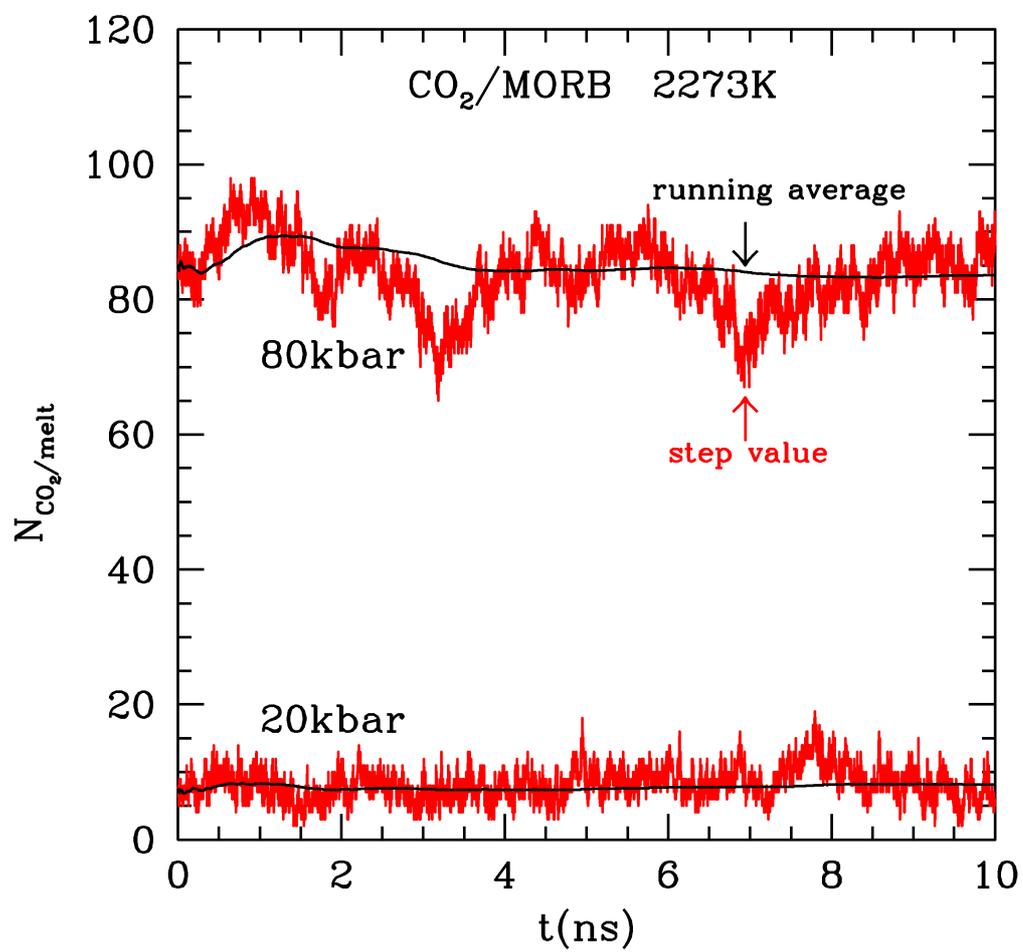

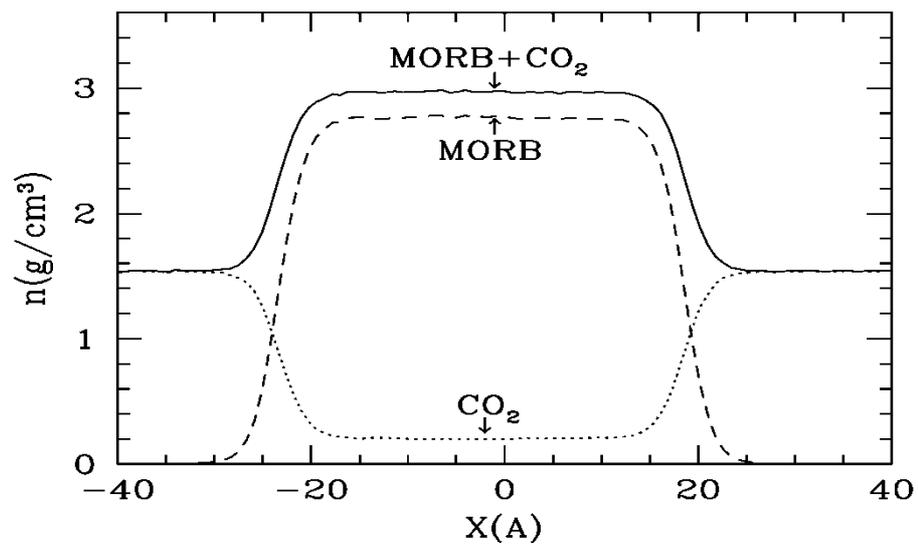

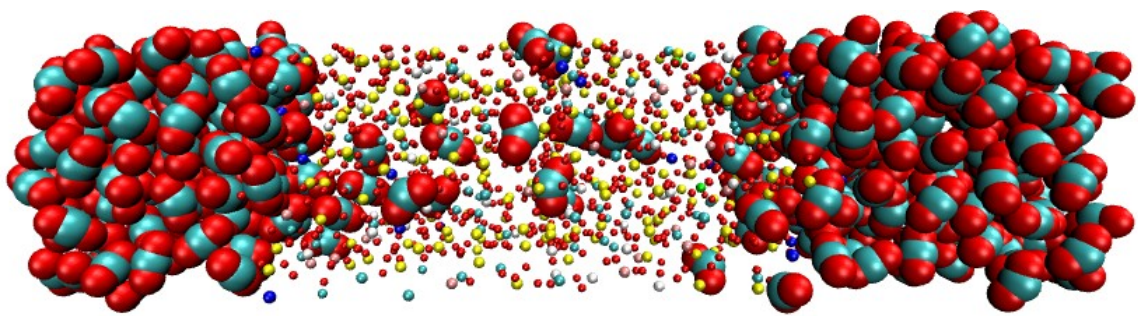

**Fig.2**

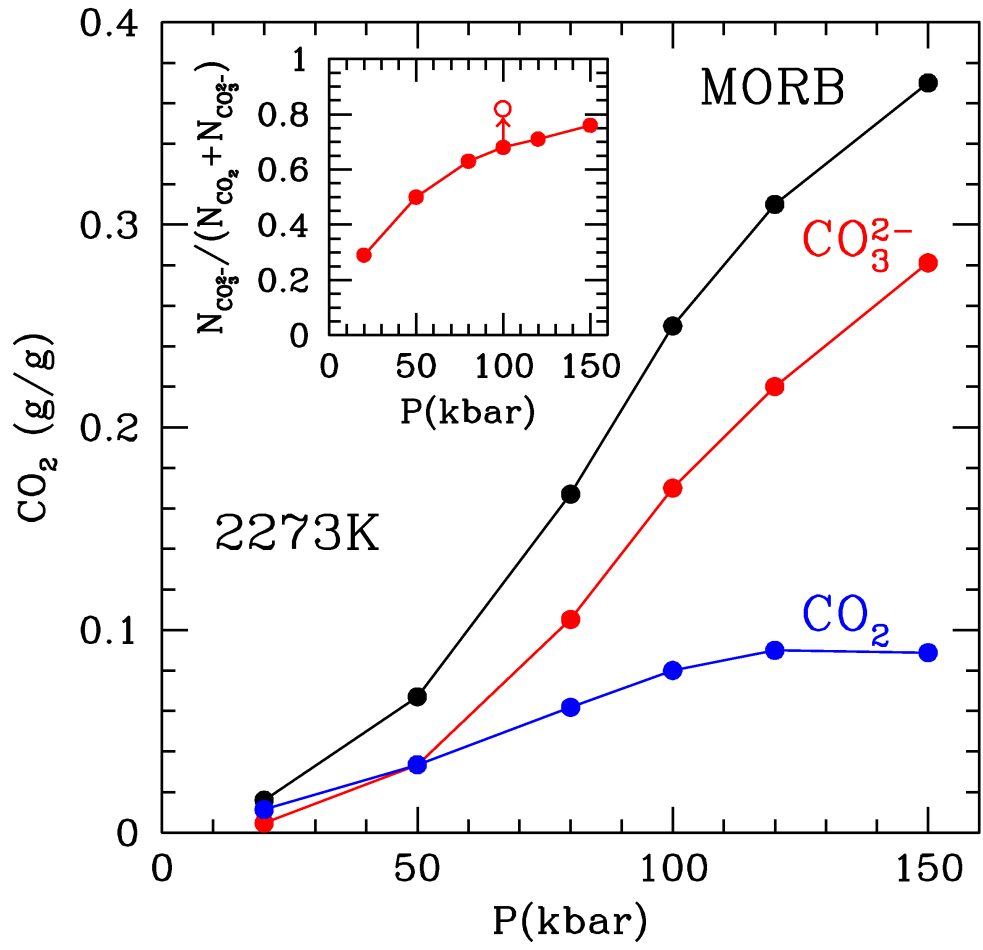

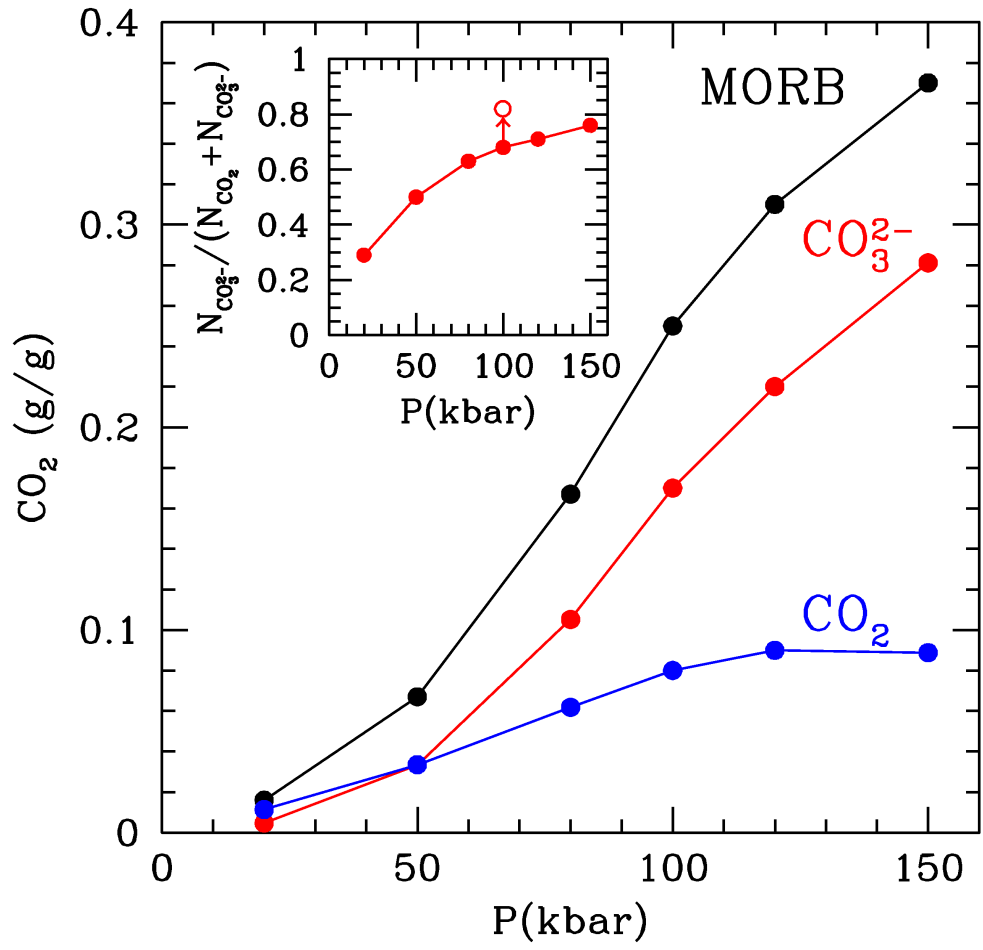

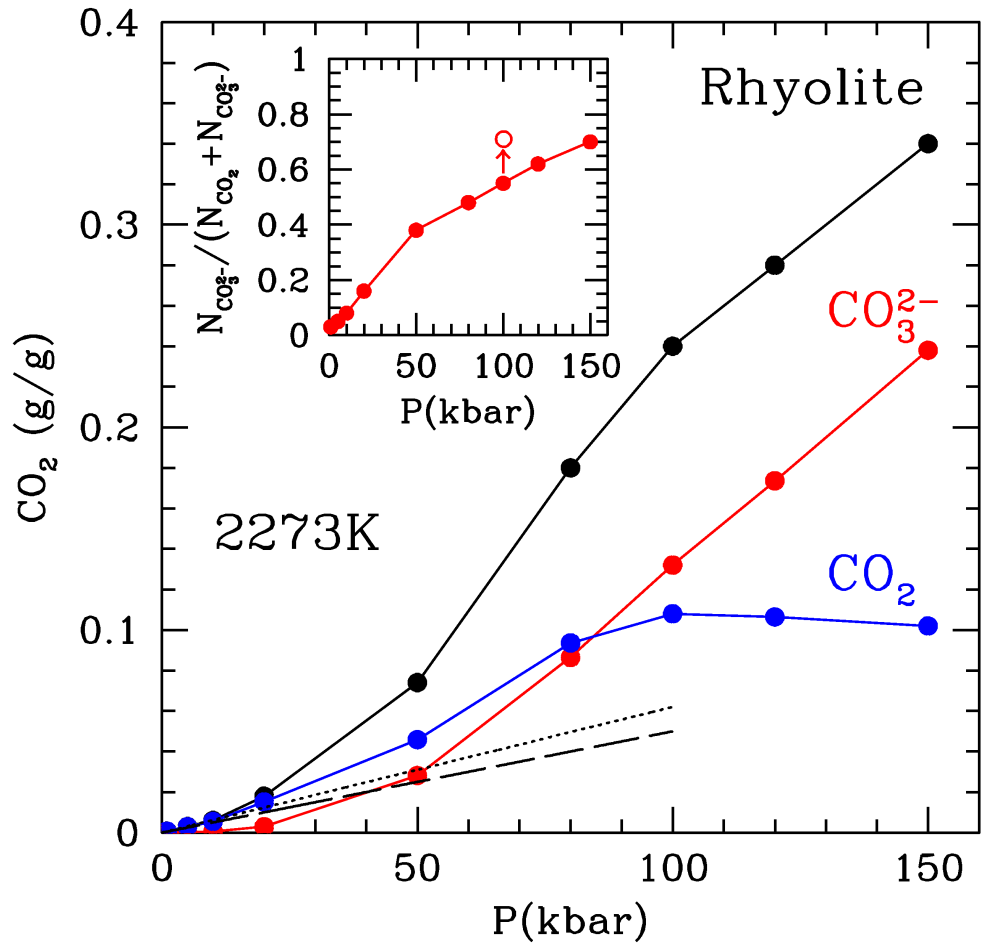

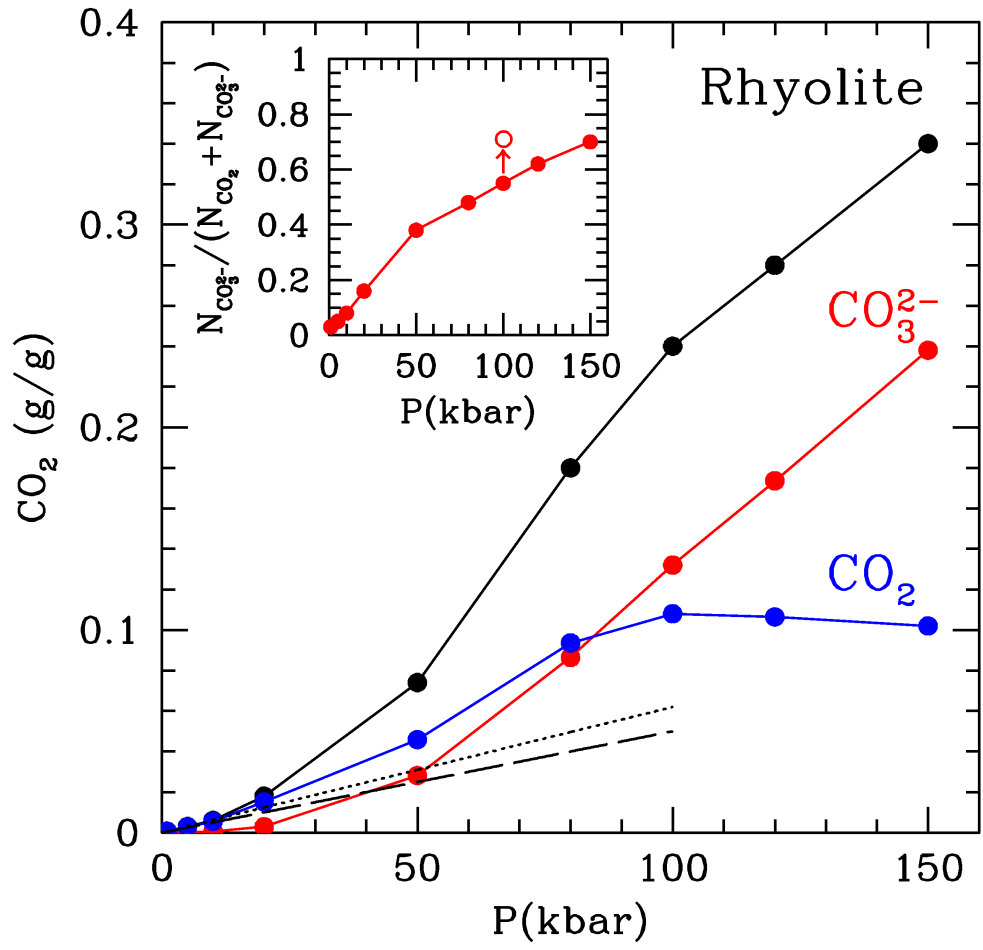

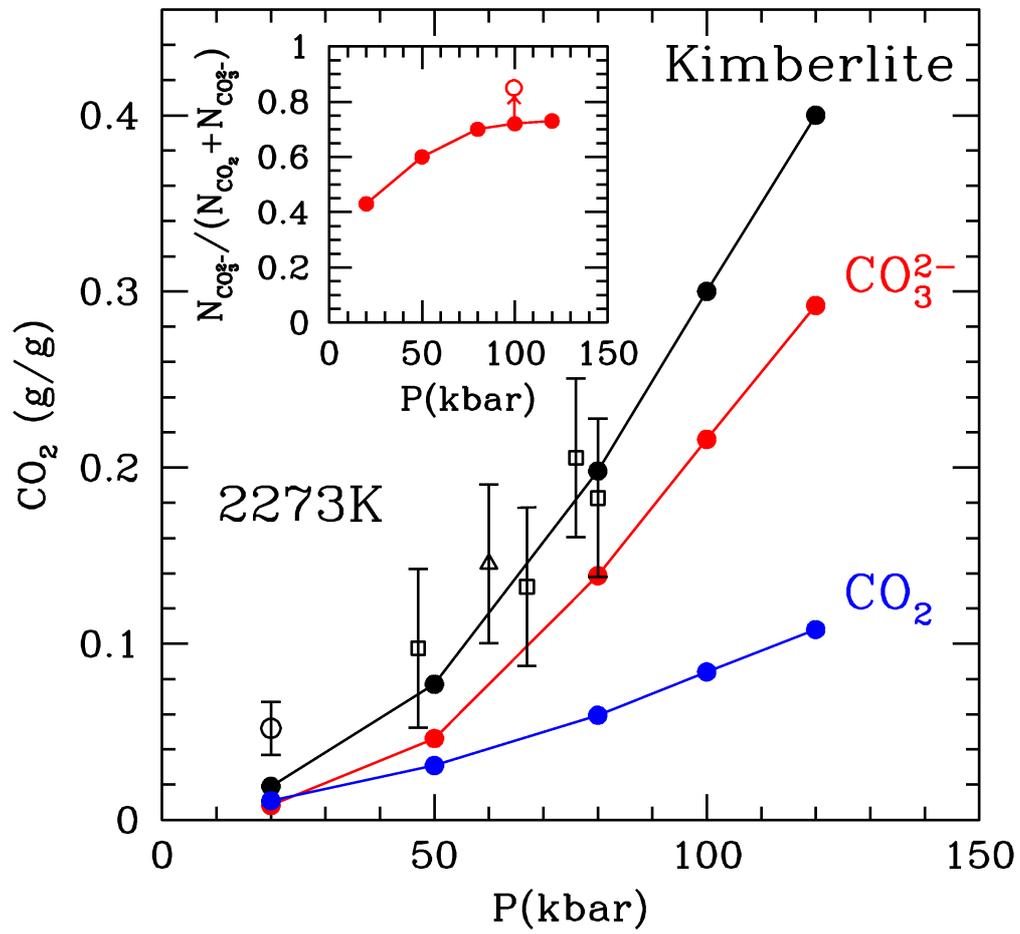

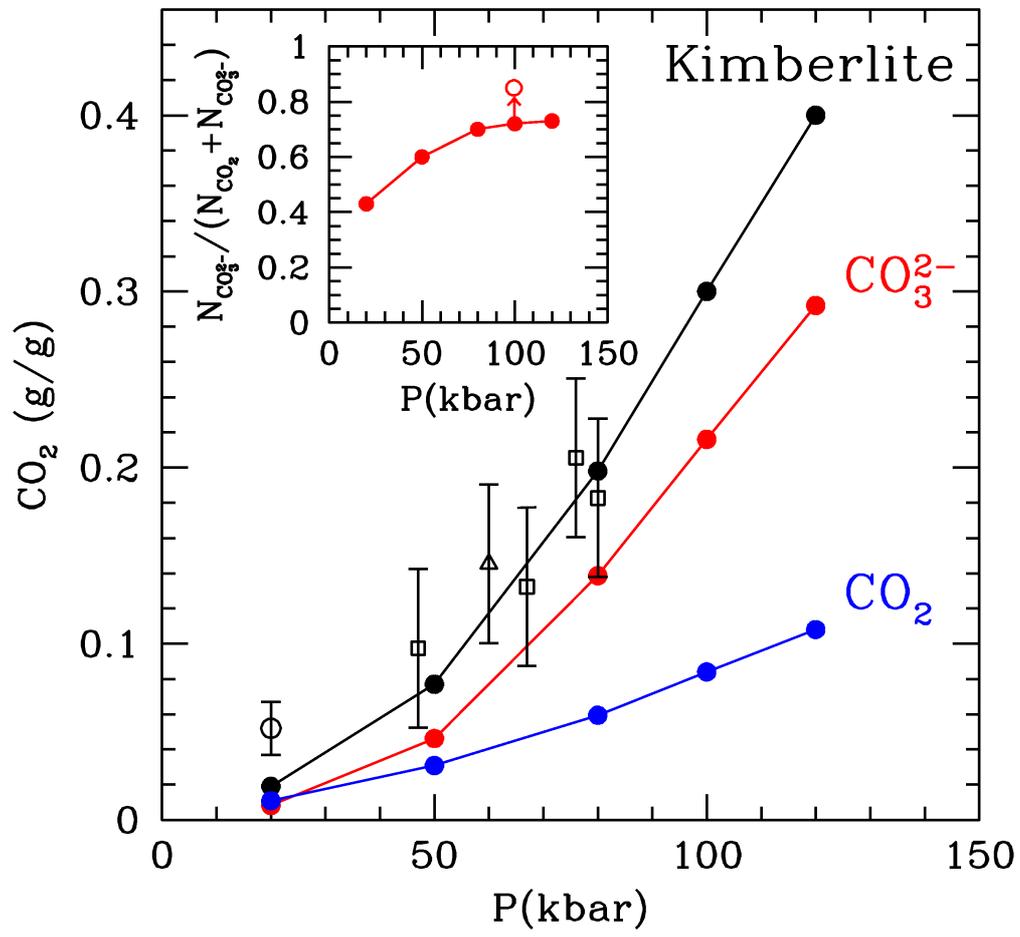

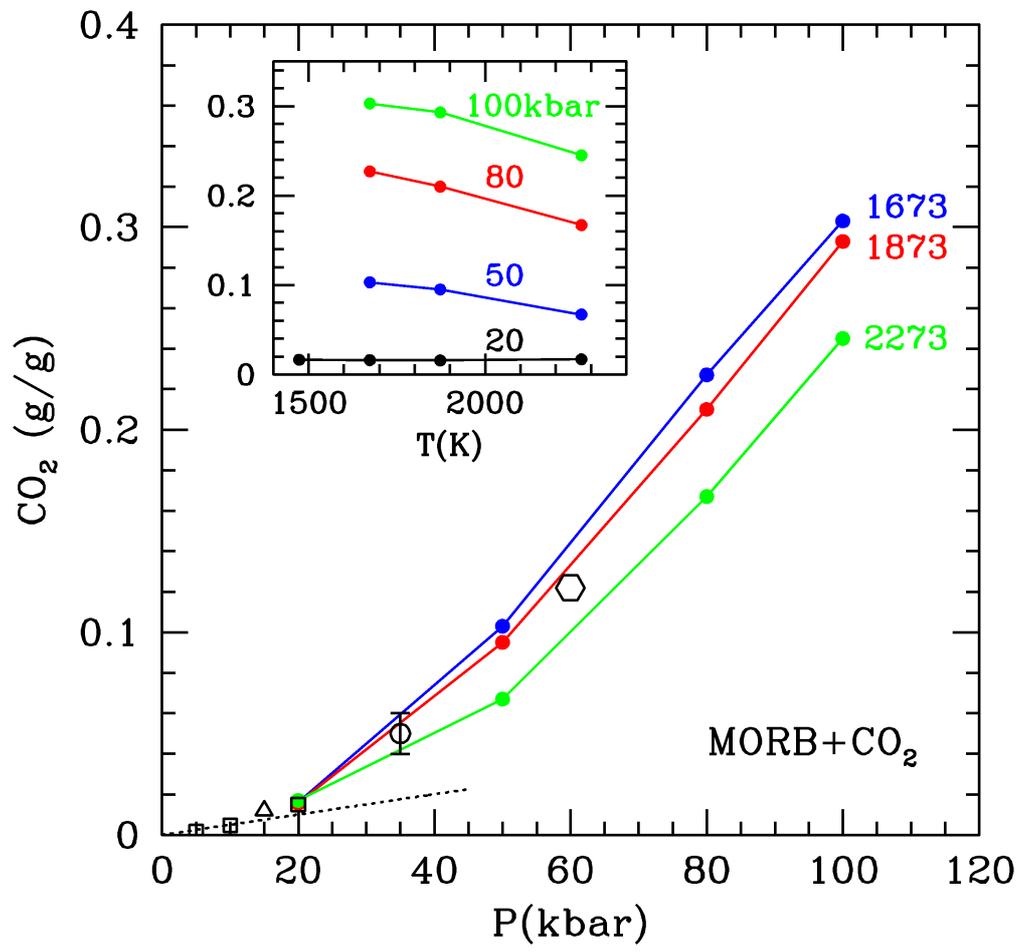

**Fig.6**

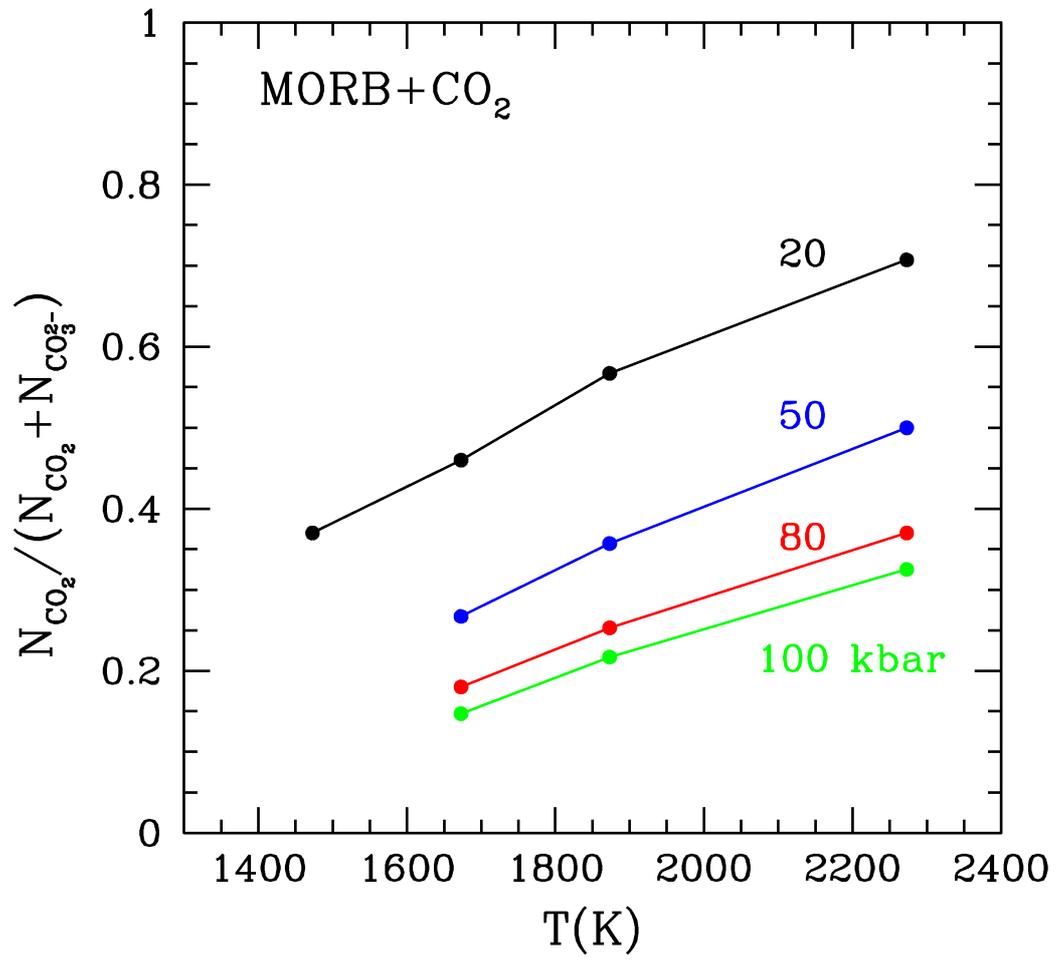

**Fig.7**

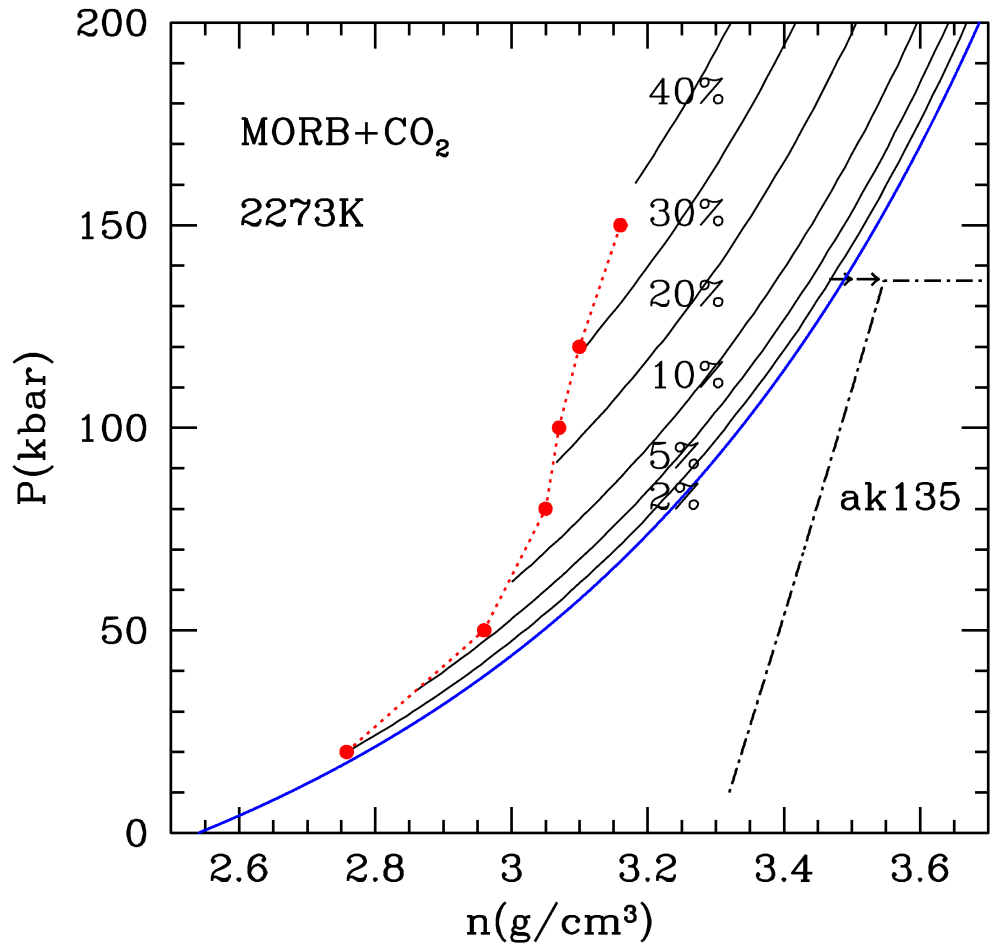

**Fig.80**

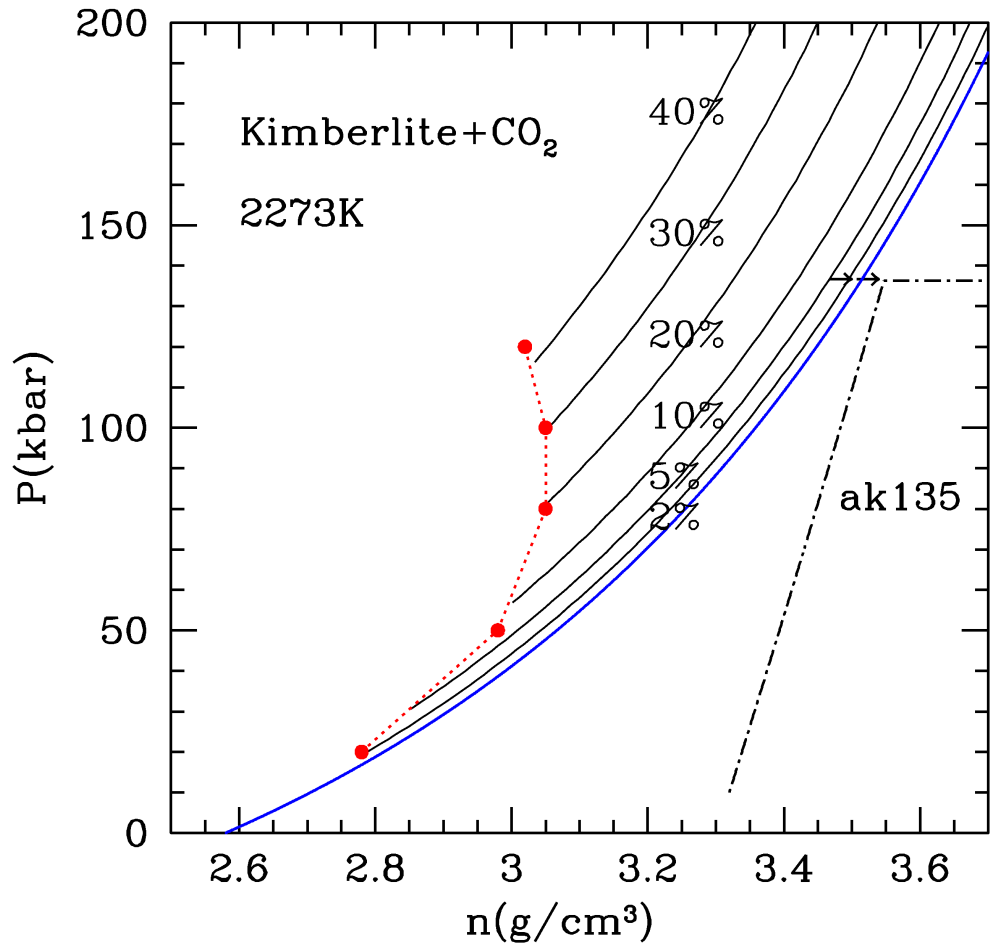

**Fig.91**

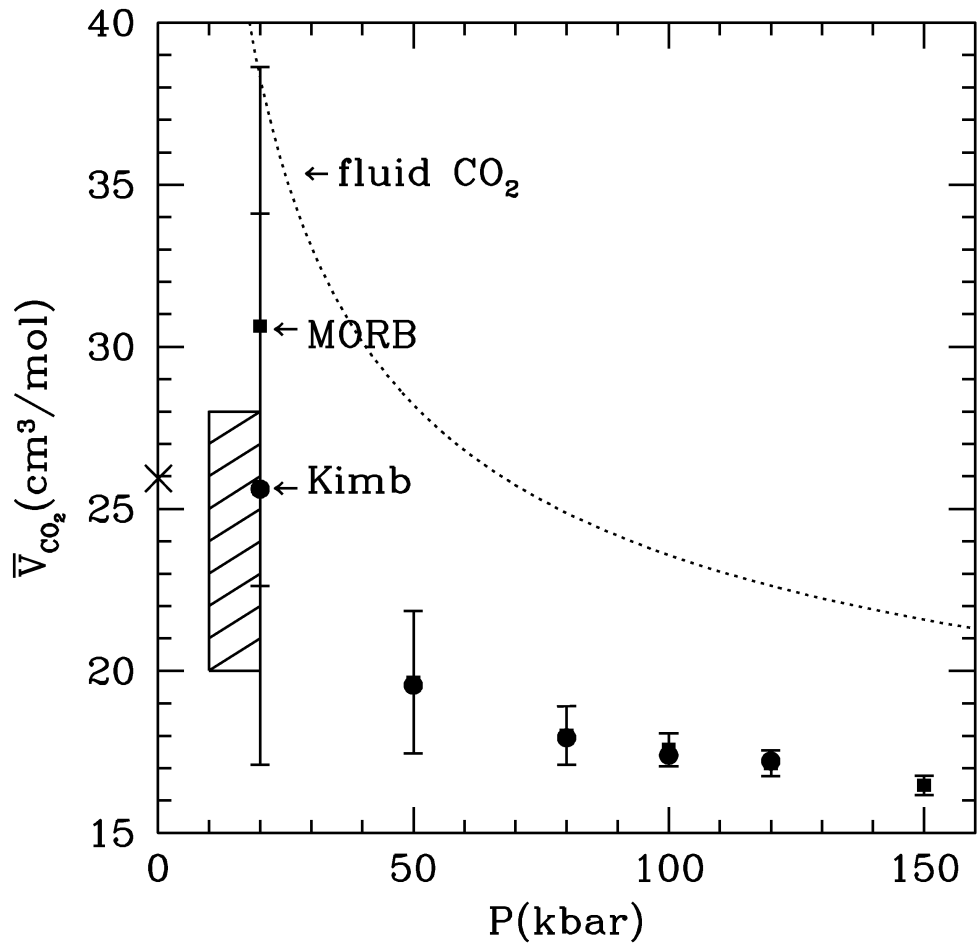

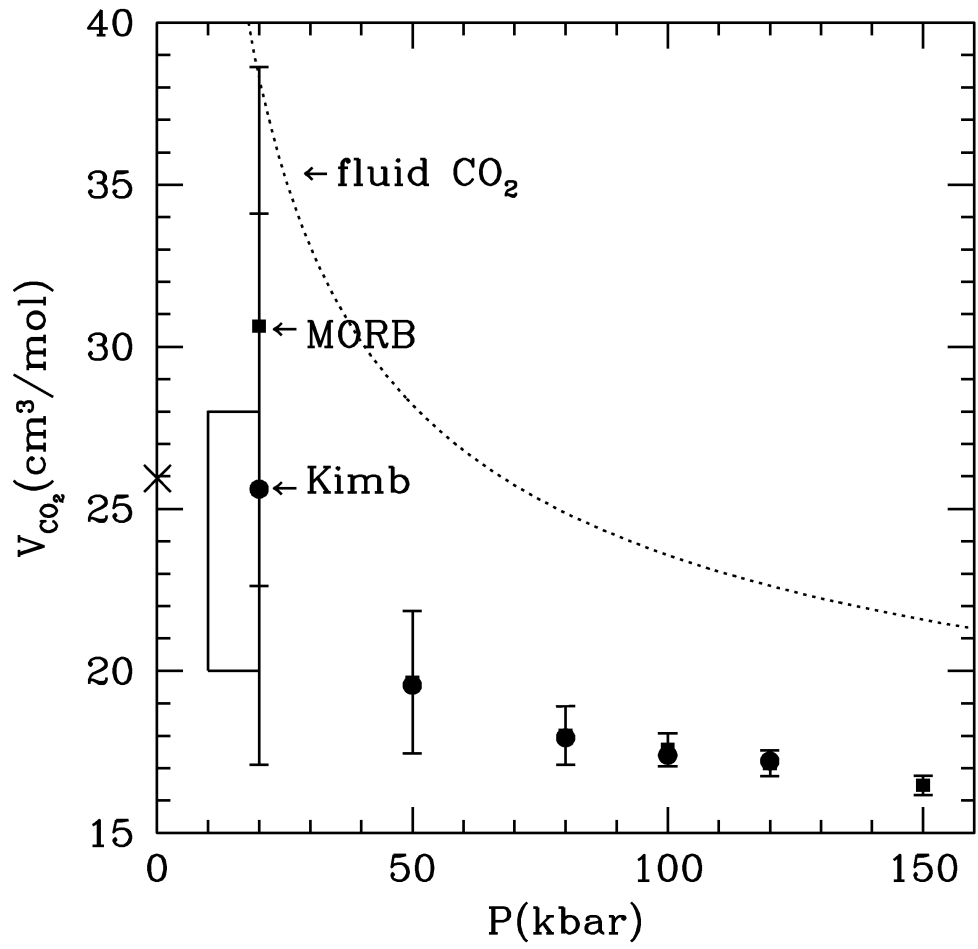

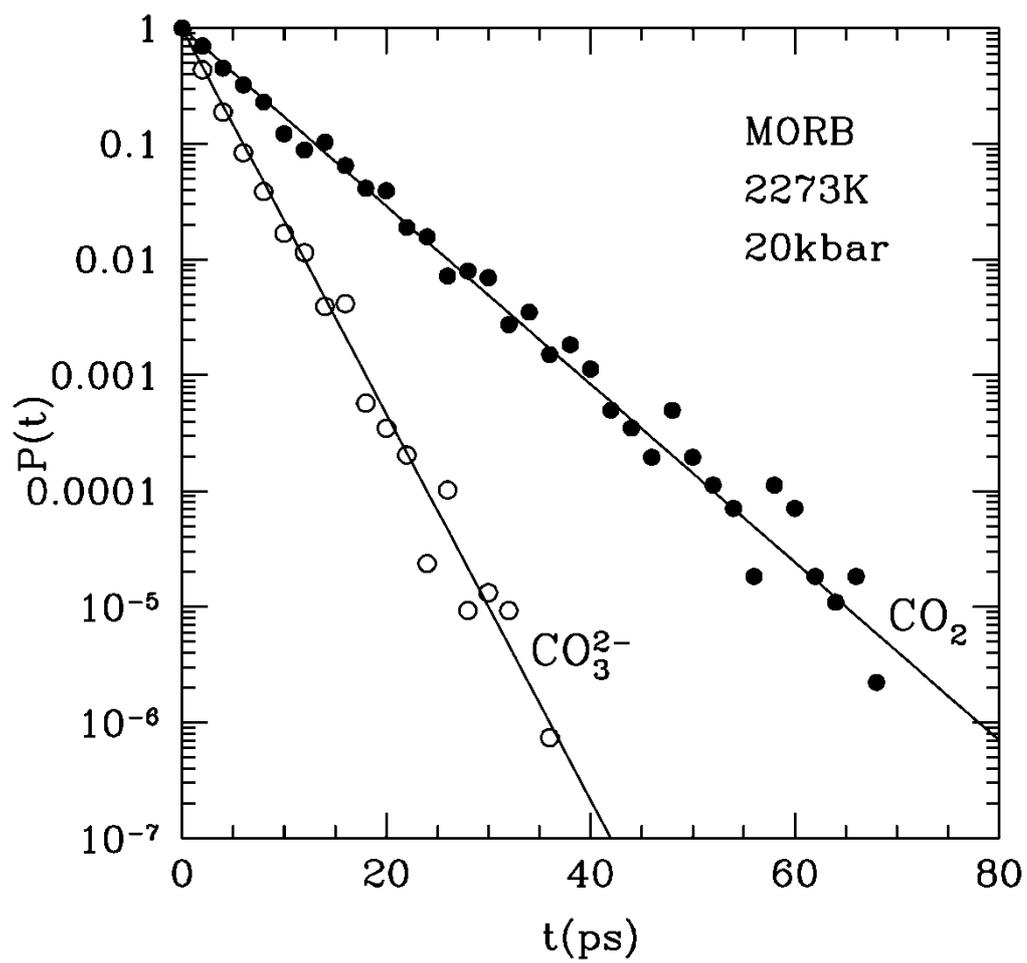

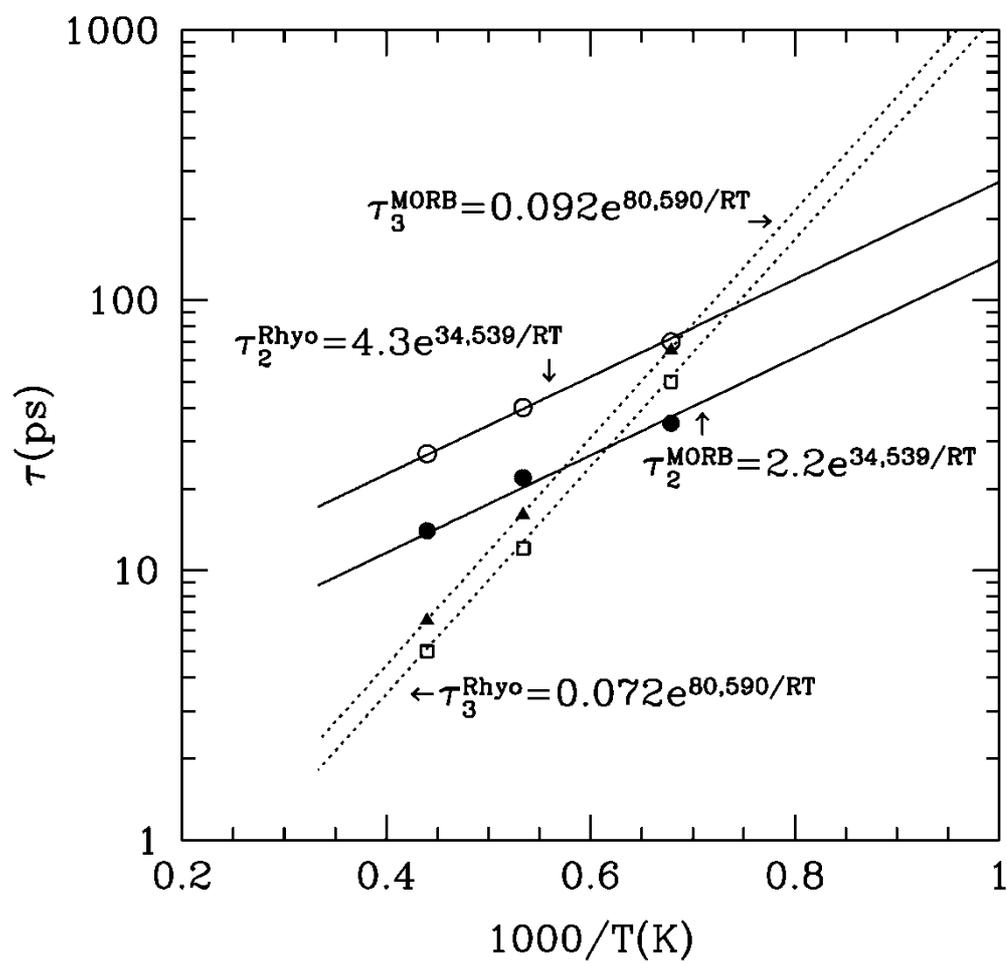

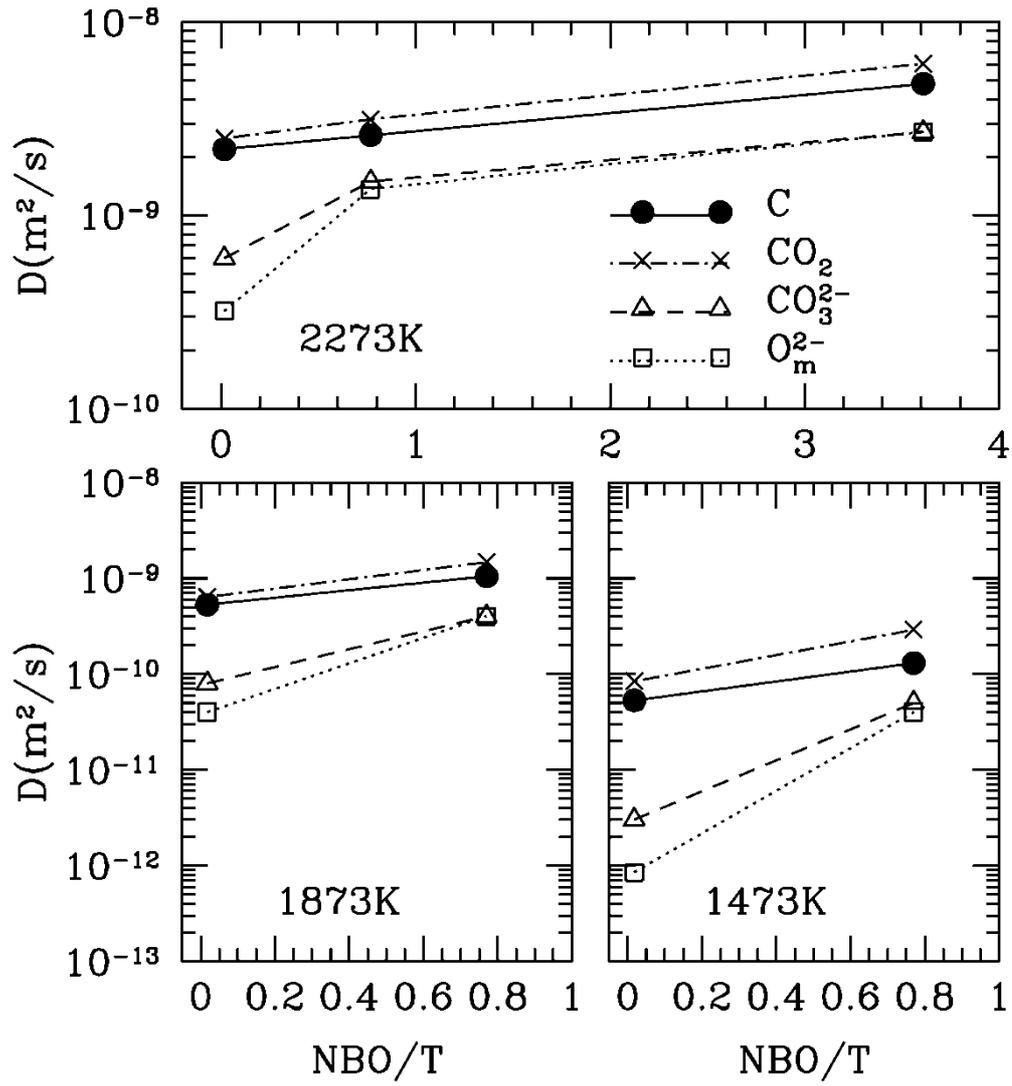

**Fig.115**

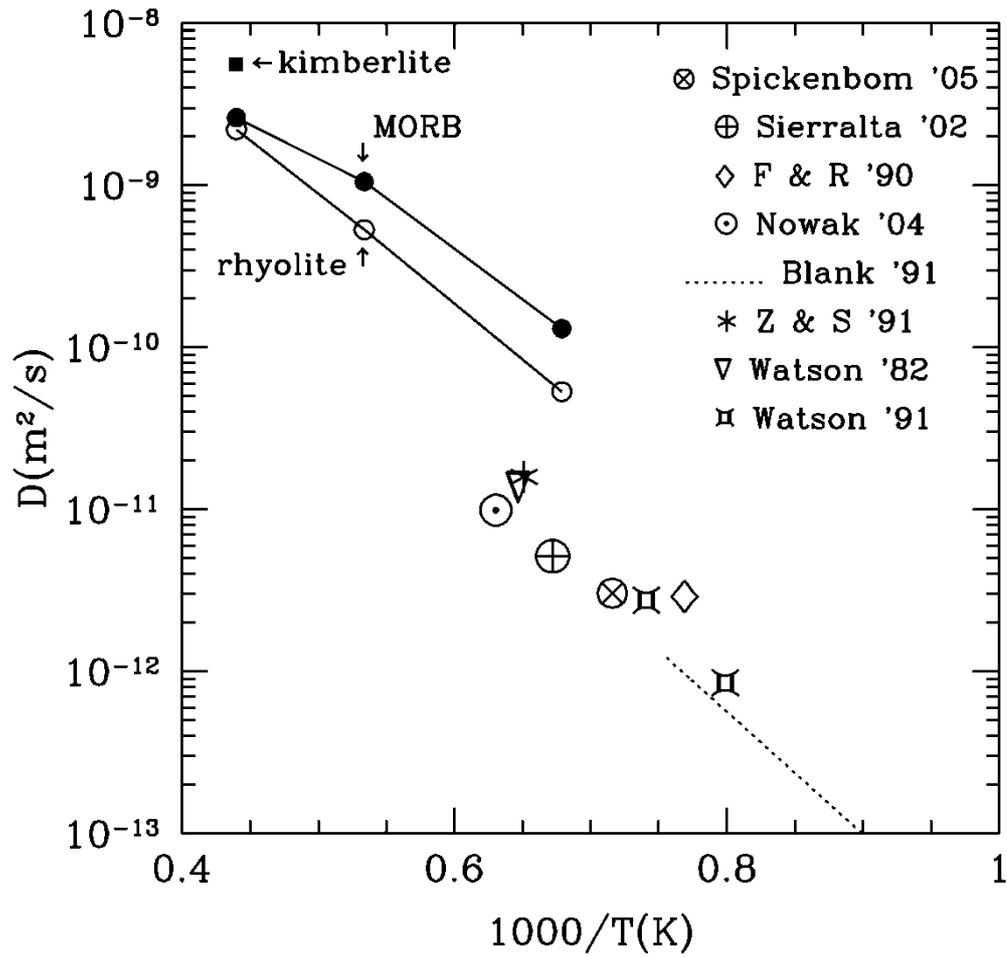

**Fig.126**

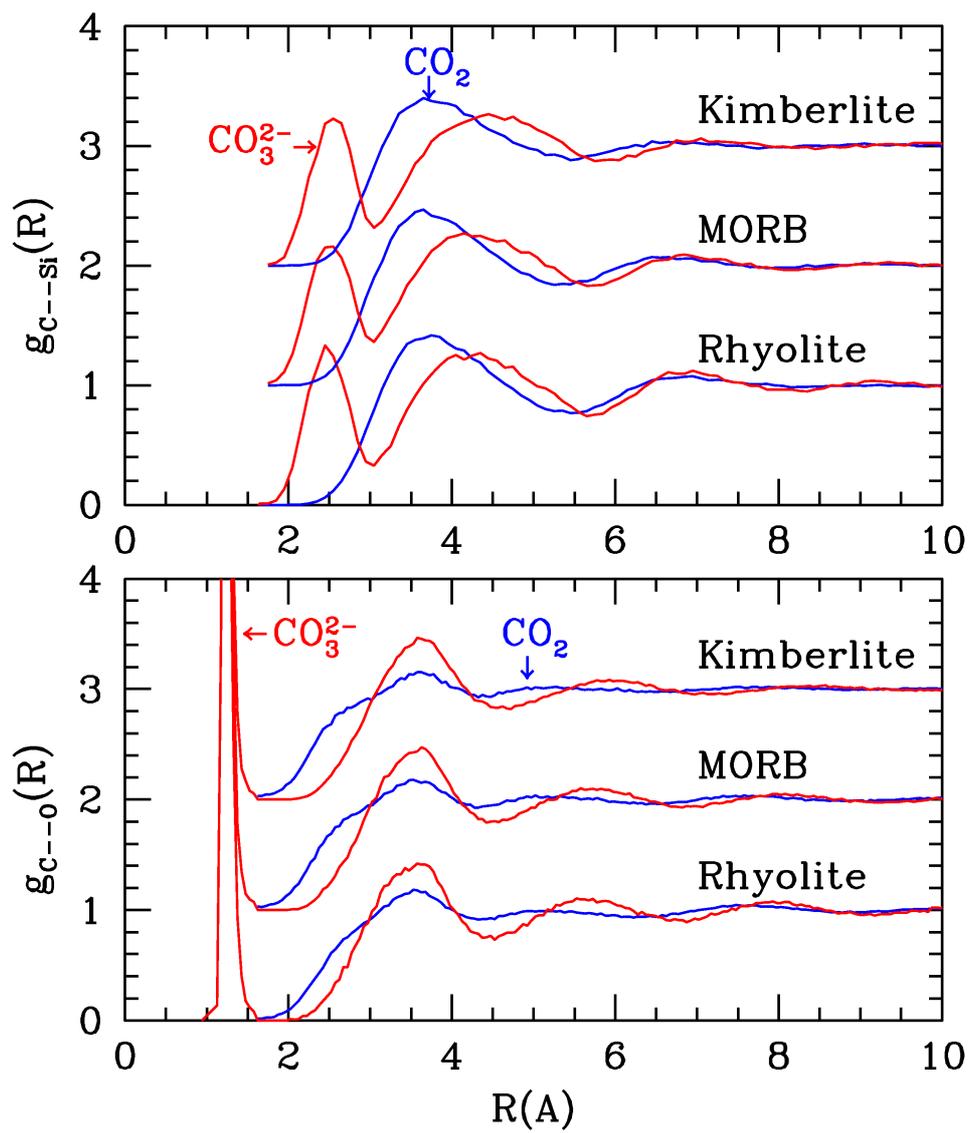

**Fig.137**

**Fig.14**

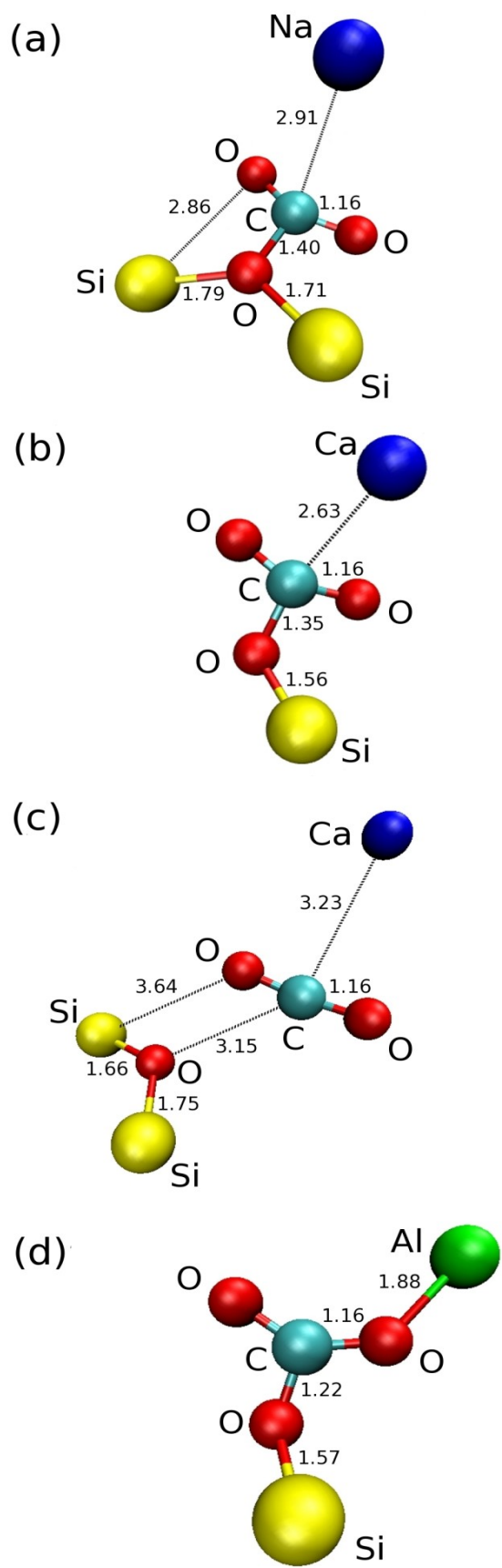

**BIBILIOGRAPHY**

Alejandre J., Tildesley D. and Chapela G.A. (1995). Molecular-dynamics simulation of the orthobaric densities and surface-tension of water. J. Chem. Phys. **102**, 4574-4583.

Allen M.P. and Tildesley D.J. (1987) *Computer simulations of liquids*. Oxford University Press.

Anderson K.E., Mielke S.L., Siepmann J.I. and Truhlar D.G. (2009) Bond angle distributions of carbon dioxide in the gas, supercritical, and solid phases. J. Phys. Chem. A **113**, 2053-2059.

Aubaud C., Pineau F., Hékinian H. and Javoy M. (2005) Degassing of $CO_2$ and $H_2O$ in submarine lavas from the Society hotspot. Earth Planet Sci. Lett. **235**, 511-527.

Baker D.R., Freda C., Brooker R.A. and Scarlato P. (2005) Volatile diffusion in silicate melts and its effects on melt inclusions. Ann. Geophys. **48**, 699-717.

Becker M. and Le Roex A.P. (2006) Geochemistry of South African on- and off-craton, group I and group II kimberlites: Petrogenesis and source region evolution. J. Petrol. **47**, 673-703.

Behrens H., Tamic N. and Holtz F. (2004a) Determination of the molar absorption coefficient for the infrared absorption band of $CO_2$ in rhyolitic glasses. Am. Mineral. **89**, 301-306.

Behrens H., Ohlhorst S., Holtz F. and Champenois M. (2004b) $CO_2$ solubility in dacitic melts equilibrated with $H_2O$-$CO_2$ fluids: Implications for modeling the solubility of $CO_2$ in silicic melts. Geochim. Cosmochim. Acta **68**, 4687-4703.

Behrens H., Misiti V., Freda C, Vetere F., Botcharnikov R.E. and Scarlato P. (2009) Solubility of $H_2O$ and $CO_2$ in ultrapotassic melts at 1200 and 1250°C and pressure from 50 to 500 MPa. Am. Mineral. 94, 105-120

Bell K. and Simonetti A. (2010) Source of parental melts to carbonatites-critical isotopic constraints. Mineral. Petrol. **98**, 77-89.

Belonoshko A.B. (1994) Molecular dynamics of $MgSiO_3$ perovskite at high pressures: Equation of state, structure and melting transition. Geochim. Cosmochim. Acta **58**, 4039-4047.

Belonoshko A.B. and Dubrovinsky L.S. (1996) Molecular dynamics of NaCl (B1 and B2) and MgO (B1) melting: Two-phase simulation. Am. Mineral. **81**, 303-316.

Belonoshko A.B., Ahuja R. and Johansson B. (2000) Quasi-*Ab initio* molecular dynamic study of Fe melting. Phys. Rev. Lett. **84**, 3638-3641.

Blank J.G. and Brooker R.A. (1994) Experimental studies of carbon dioxide in silicate melts: Solubility, speciation, and stable carbon isotope behavior. Rev. Mineral. Geochem. **30**, 157-186.

Blank J.G., Stolper E.M. and Zhang Y. (1991) Diffusion of $CO_2$ in rhyolitic melt. Eos Trans. Am. Geophys. Un. **72**, 312 (abstract).

Blank J.G., Stolper E.M. and Carroll M.R. (1993) Solubilities of carbon dioxide and water in rhyolitic melt at 850°C and 750 bars. Earth Planet. Sci. Lett. **119**, 27-36.

Boldyrev A.I., Gutowski M. and Simons J. (1996) Small multiply charged anions as building blocks in chemistry. Acc. Chem. Res. **29**, 497-502.

Botcharnikov R.E., Behrens H. and Holtz F. (2006) Solubility and speciation of C-O-H fluids in andesitic melt at T = 1100-1300°C and P = 200 and 500 MPa. Chem. Geol. **229**, 125-143.

Botcharnikov R., Freise M., Holtz F. and Behrens H. (2005) Solubility of C-O-H mixtures in natural melts: new experimental data and application range of recent models. Annals of Geophys. **48**, 633-646.

Bottinga Y. and Javoy M. (1991) The degassing of Hawaian tholeiite. Bull. Volc. **53**, 73-85.

Bourgue E. and Richet P. (2001) The effects of dissolved $CO_2$ on the density and viscosity of silicate melts: a preliminary study. Earth Planet. Sci. Lett. **193**, 57-68.

Brearley M. and Montana A. (1989) The effect of $CO_2$ on the viscosity of silicate liquids at high-pressure. Geochim. Cosmochim. Acta **53**, 2609-2616.

Brenker F.E., Vollmer C., Vincze L., Vekemans B., Szymanski A., Janssens K., Szaloki I., Nasdala L., Joswig W. and Kaminsky F. (2007) Carbonates from the lower part of transition zone or even the lower mantle. Earth Planet. Sci. Lett. **260**, 1-9.

Brey G. (1976) $CO_2$ solubility and solubility mechanisms in silicate melts at high pressures. Contrib. Mineral. Petrol. **57**, 215-221.

Brey G.P. and Green D.H. (1976) Solubility of $CO_2$ in Olivine melilitite at high pressures and role of $CO_2$ in the Earth's upper mantle. Contrib. Mineral. Petrol. **55**, 217-230.

Brey G.P. and Ryabchikov I.D. (1994) Carbon dioxide in strongly silica undersaturated melts and origin of kimberlite magmas. N. Jb. Miner. Mh. **10**, 449-463.

Brey G.P., Kogarko L.N. and Ryabchikov I.D. (1991) Carbon dioxide in kimberlitic melts. N. Jb. Miner. Mh. **4**, 159-168.

Brey G.P., Bulatov V.K., Girnis A.V. and Lahaye Y. (2008) Experimental melting of carbonated peridotite at 6-10 GPa. J. Petrol. **49**, 797-821.

Brooker R.A. (1998) The effect of $CO_2$ saturation on immiscibility between silicate and carbonate liquids: an experimental study. J. Petrol. **39**, 1905-1915.

Brooker R.A. and Hamilton D.L. (1990) Three-liquid immiscibility and the origin of carbonatites. Nature **346**, 459-462.

Brooker R.A., Kohn S.C., Holloway, J.R., MCMillan P.F. and Carroll M.R. (1999) Solubility, speciation and dissolution mechanisms for CO2 in melts on the NaAL02-SiO2 join. Geochim. Cosmochim. Acta **63**, 3549-3565.

Brooker R.A., Kohn S.C., Holloway, J.R. and MCMillan P.F. (2001a) Structural controls on the solubility of $CO_2$ in silicate melts Part I: bulk solubility data. Chem. Geol. **174**, 225-239.

Brooker R.A., Kohn S.C., Holloway, J.R. and MCMillan P.F. (2001b) Structural controls on the solubility of $CO_2$ in silicate melts Part II: IR characteristics of carbonate groups in silicate glasses. Chem. Geol. **174**, 241-254.

Canil D. and Bellis A.J. (2008) Phase equilibria in a volatile-free kimberlite at 0.1 MPa and the search for primary kimberlite magma. Lithos **105**, 111-117.

Cartigny P., Pineau F., Aubaud C. and Javoy M. (2008) Towards a consistent mantle carbon flux estimate: Insights from volatile systematics ($H_2O/Ce$, δD, $CO_2/Nb$) in the North Atlantic mantle (14°N and 34°N). Earth Planet. Sci. Lett. **265**, 672-685.

Chapela G.A., Saville G., Thomson S.M. and Rowlinson J.S. (1977) Computer-simulation of a gas-liquid surface.1. J. Chem. Soc. Faraday Trans. II **73**, 1133-1144.

Cherginets V.L. and Rebrova T.P. (2003) On carbonate ion dissociation in molten alkali metal halides at ~ 800°C. J. Chem. Eng. Data **48**, 463-467.

Chessin H., Hamilton W.C. and Post B. (1965) Position and thermal parameters of oxygen atoms in calcite. Acta Cryst. **18**, 689.

Dalton J.A. and Presnall D.C. (1998) The continuum of primary carbonatitic-kimberlitic melt compositions in equilibrium with lherzolite: Data from the system $CaO-MgO-Al_2O_3-SiO_2-CO_2$ at 6 GPa. J. Petrol. **39**, 1953-1964.

Dasgupta R. and Hirschmann M.M. ( 2006) Melting in the Earth's deep upper mantle caused by carbon dioxide. Nature **440**, 659-662.

Dasgupta R. and Hirschmann M.M. (2010) The deep carbon cycle and melting in Earth's interior. Earth Planet. Sci. Lett. **298**, 1-13

Dasgupta R., Hirschmann M.M. and Stalker K. (2006) Immiscible transition from carbonate-rich to silicate-rich melts in the 3 GPa melting interval of eclogite + $CO_2$ and genesis of silica-undersaturated ocean island lavas. J. Petrol. **47**, 647-671.

Deines P. (2002) The carbon isotope geochemistry of mantle xenoliths. Earth-Sci. Rev. **58**, 247-278.

De Jong B.H.W.S. and Brown G.E. (1980) Polymerization of silicate and aluminate tetrahedra in glasses, melts and aqueous solutions - II. The network modifying effects of $Mg^{2+}$, $K^+$, $Na^+$, $Li^+$, $H^+$, $OH^-$, $F^-$, $Cl^-$, $H_2O$, $CO_2$ and $H_3O^+$ on silicate polymers. Geochim. Cosmochim. Acta **44**, 1627-1642.

Dickens B, Hyman A. and Brown W.E. (1971) Crystal structure of $Ca_2Na_2(CO_3)_3$ (shortite). J. Res. Nat. Bur. Stds. A **75**, 129.

Dingwell D.B. and Webb S.L. (1989) Structural relaxation in silicate melts and non-Newtonian melt rheology in geologic processes. Phys. Chem. Minerals **16**, 508-516

Dixon J.E. (1997) Degassing of alkalic basalts. Am. Mineral. **82**, 368-378.

Dixon J.E., Stolper E.M. and Holloway J.R. (1995) An experimental study of water and carbon dioxide solubilities in mid-ocean ridge basaltic liquids Part I: Calibration and solubility models. J. Petrol. **36**, 1607-1631.

Edgar A.D. and Charbonneau H.E. (1993) Melting experiments on a $SiO_2$-poor, CaO-rich aphanitic kimberlite from 5-10 GPa and their bearing on sources of kimberlite magmas. Am. Mineral. **78**, 132-142.

Eggler D.H. (1973) Role of CO2 in melting process in the mantle. Carnegie Inst. Washington Year Book **72**, 457-467.

Eggler D.H. (1976) Does $CO_2$ cause partial melting in the low-velocity layer of the mantle ? Geology **4**, 69-72.

Eggler D.H. and Mysen B.O. (1976) The role of $CO_2$ in the genesis of olivine melilitite: Discussion. Contrib. Mineral. Petrol. **55**, 231-236.

Eggler D.E. and Rosenhauer M. (1978) Carbon dioxide in silicate melts: II. Solubilities of $CO_2$ and $H_2O$ in $CaMgSi_2O_6$ (diopside) liquids and vapors at pressures to 40 kbar. Am. J. Sci. **278**, 64-94.

 Nature **437**, 249-252.

Falloon T.J. and Green D.H. (1989) The solidus of carbonated, fertile peridotite. Earth Planet. Sci. Lett. **94**, 364-370.

Fine G. and Stolper E. (1985a) The speciation of carbon dioxide in sodium aluminosilicate glasses. Contrib. Mineral. Petrol. **91**, 105-121.

Fine G. and Stolper E. (1985b) Dissolved carbon dioxide in basaltic glasses: concentrations and speciation. Earth Planet. Sci. Lett. **76**, 263-278.

Fogel R.A. and Rutherford M.J. (1990) The solubility of carbon dioxide in rhyolitic melts: A quantitative FTIR study. Am. Mineral. **75**, 1311-1326.

Foley S.F., Yaxley G.M., Rosenthal A., Buhre S., Kiseeva E.S., Rapp R.P. and Jacob D.E. (2009) The composition of near-solidus melts of peridotite in the presence of $CO_2$ and $H_2O$ between 40 and 60 kbar. Lithos **112**, 274-283.

Forst W. (1973) Theory of unimolecular reactions (Academic, New York)

Gaillard F., Schmidt B., Mackwell S. and McCammon C. (2003) Rate of hydrogen-iron redox exchange in silicate melts and glasses, Geochim. Cosmochim. Acta **67**, 2427-2441.

Gerlach T.M., McGee K.A., Elias T., Sutton A.J. and Doukas M.P. (2002) Carbon dioxide emission rate of Kilauea volcano: Implications for primary magma and the summit reservoir. J. Geophys Res. **107**, 2189-2203.

Ghosh S., Ohtani E., Litasov K.D. and Terasaki H. (2009) Solidus of carbonated peridotite from 10 to 20 GPa and origin of magnesiocarbonatite melt in the Earth's deep mantle. Chem. Geol. **262**, 17-28.

Ghosh S., Ohtani E., Litasov K., Suzuki A. and Sakamaki T. (2007) Stability of carbonated magmas at the base of the Earth's upper mantle. Geophys. Lett. 34, 22312-1-5.

Giordano D. and Dingwell D.B. (2003) The kinetic fragility of natural silicate melts. J. Phys.: Condens. Matter **15**, S945-S954

Giordano V.M., Datchi F. and Dewaele A. (2006) Melting curve and fluid equation of state of carbon dioxide at high pressure and high temperature. J. Chem. Phys. **125**, 054504-1-8.

Girnis A.V. and Ryabchikov I.D. (2005) Conditions and mechanisms of generation of kimberlite magmas. Geol. Ore Dep. **47**, 524-536.

Graham D. and Sarda P. (1991) Mid-ocean ridge popping rocks: implications for degassing at ridge crests - Comment. Earth Planet. Sci. Lett. **105**, 568-573.

Gudfinnsson G. and Presnall D.C. (2005) Continuous gradations among primary carbonatitic, kimberlitic, melilititic, basaltic, picritic, and komatiitic melts in equilibrium with garnet lherzolite at 3-8 GPa. J. Petrol. **46**, 1645-1659.

Guillot B. and Guissani Y. (2002a) Chemical reactivity and phase behavior of $NH_4Cl$ by molecular dynamics simulations. I. Solid-solid and solid-fluid equilibria. J. Chem. Phys. **116**, 2047-2057.

Guillot B. and Guissani Y. (2002b) Chemical reactivity and phase behavior of $NH_4Cl$ by molecular dynamics simulations. II.The liquid-vapor coexistence curve. J. Chem. Phys. **116**, 2058-2066.

Guillot B. and Sator N. (2007a) A computer simulation study of natural silicate melts. Part I: Low pressure properties. Geochim. Cosmochim. Acta **71**, 1249-1265.

Guillot B. and Sator N. (2007b) A computer simulation study of natural silicate melts. Part II: High pressure properties. Geochim. Cosmochim. Acta **71**, 4538-4556.

Hammouda T. (2003) High-pressure melting of carbonated eclogite and experimental constraints on carbon recycling and storage in the mantle. Earth Planet. Sci. Lett. **214**, 357-368.

Hekinian R., Chaigheau M. and Cheminée J.L. (1973) Popping rocks and lava tubes from the Mid-Atlantic Rift Valley at 36°N, Nature **245**, 371-373.

Hirschmann M.M. and Dasgupta R. (2009) The H/C ratio of Earth's near-surface and deep reservoirs, and consequences for deep Earth volatile cycles. Chem. Geol. **262**, 4-16.

Holloway J.R. (1998) Graphite-melt equilibria during mantle melting: constraints on $CO_2$ in MORB magmas and the carbon content of the mantle. Chem. Geol. **147**, 89-97.

Holloway J.R. and Blank J.G. (1994) Application of experimental results to C-O-H species in natural melts. Rev. Mineral. Geochem. **30**, 187-230.

Ishii R., Okazaki S., Okada I., Furusaka M., Watanabe N., Misawa M. and Fukunaga T. (1996) Density dependence of structure of supercritical carbon dioxide along an isotherm. J. Chem. Phys. **105**, 7011-7021.

Jakobsson S. (1997) Solubility of water and carbon dioxide in an icelandite at 1400°C and 10 kilobars. Contrib. Mineral. Petrol. **127**, 129-135.

Jambon A. (1994) Earth degassing and large-scale geochemical cycling of volatile elements. Rev. Mineral. **30**, 479-517.

Javoy M. and Pineau F. (1991) The volatiles record of a "popping" rock from the Mid-Atlantic Ridge at 14°N: chemical and isotopic composition of gas trapped in the vesicles. Earth Planet. Sci. Lett. **107**, 598-611.

Javoy M., Pineau F. and Delorme H. (1986) Carbon and nitrogen isotopes in the mantle. Chem. Geol. **57**, 41-62.

Jendrzejewski N., Trull T.W., Pineau F. and Javoy M. (1997) Carbon solubility in mid-ocean ridge basaltic melt at low pressures (250-1950 bar). Chem. Geol. **138**, 81-92.

Kavanagh J.L. and Sparks R.S.J. (2009) Temperature changes in ascending kimberlite magma. Earth Planet. Sci. Lett. **286**, 404-413.

Kennett B.L.N., Engdahl E.R. and Buland R. (1995) Constraints on seismic velocities in the Earth from traveltimes. Geophys. J. Int. **122**, 108-124.

Keppler H., Wiedenbeck M. and Shcheka S.S. (2003) Carbon solubility in olivine and the mode of carbon storage in the Earth's mantle. Nature **424**, 414-416.

Keshav S., Corgne A., Gudfinnsson G., Bizimis M., McDonough W.F. and Fei Y. (2005) Kimberlite petrogenesis: Insights from clinopyroxene-melt partitioning experiments at 6 GPa in the CaO-MgO-$Al_2O_3$-$SiO_2$-$CO_2$ system. Geochim. Cosmochim. Acta **69**, 2829-2845.

Kiczenski T.J., DU L.S. and Stebbins J. (2005) The effect of fictive temperature on the structure of E-glass: A high resolution, multinuclear NMR study. J. Non-Cryst. Sol. **351**, 46-48.

King P.L. and Holloway J.R. (2002) $CO_2$ solubility and speciation in intermediate (andesitic) melts: The role of $H_2O$ and composition. Geochim. Cosmochim. Acta **66**, 1627-1640.

Kjarsgaard B. and Hamilton D.L. (1988) Liquid immiscibility and the origin of alkali-poor carbonatites. Mineral. Mag. **52**, 43-55.

Kjarsgaard B. and Peterson T. (1991) Nephelinite-Carbonatite liquid immiscibility at Shombole volcano, East Africa: Petrographic and experimental evidence. Mineral. Petrol. **43**, 293-314.

Kjarsgaard B.A., Pearson D.G., Tappe S., Nowell G.M. and Dowall D.P. (2009) Geochemistry of hypabyssal kimberlites from Lac de Gras, Canada: Comparisons to a global database and applications to the parent magma problem. Lithos **112**, 236-248.

Kohara S., Suzuya K., Takeuchi K., Loong C.-K., Grimsditch M., Weber J.K.R., Tangeman J.A. and Key T.S. (2004) Glass formation at the limit of insufficient network formers. Science **303**, 1649-1652.

Kohara S., Koura N., Idemoto Y., Takahashi S., Saboungi M.L. and Curtiss L.A. (1998) The structure of LiKCO$_3$ studied by ab initio calculations and Raman spectroscopy. J. Phys. Chem. Sol. **59**, 1477-1485.

Kohn S.C., Brooker R.A. and Dupree R. (1991) $^{13}$C MAS NMR: A method for studying CO$_2$ speciation in glasses. Geochim. Cosmochim. Acta **55**, 3879-3884.

Kopylova M.G., Matveev S. and Raudsepp M. (2007) Searching for parental kimberlite melt. Geochim. Cosmochim. Acta **71**, 3616-3629.

Koura N., Kohara S., Takeuchi K., Takahashi S., Curtiss L.A., Grimsditch M. and Saboungi M.L. (1996) Alkali carbonates: Raman spectroscopy, ab initio calculations and structure. J. Mol. Struct. **382**, 163-169.

Kubicki J.D. and Stolper E.M. (1995) Structural roles of CO$_2$ and [CO$_3$]$^{2-}$ in fully polymerized sodium aluminosilicate melts and glasses. Geochim. and Cosmochim. Acta **59**, 683-698.

Lange R.A. (1994) The effect of H$_2$O, CO$_2$ and F on the density and viscosity of silicate melts. Min. Soc. Am. Rev. **30**, 331-369.

Lange R.A. and Carmichael I.S.E. (1987) Densities of Na$_2$O-K$_2$O-CaO-MgO-FeO-Fe$_2$O$_3$-Al$_2$O$_3$-TiO$_2$-SiO$_2$ liquids: new measurements and derived partial molecular properties. Geochem. Cosmochim. Acta **51**, 2931-2946.

Lee W-J. and Wyllie P.J. (1996) Liquid immiscibility in the join NaAlSi$_3$O$_8$-CaCO$_3$ to 2.5 GPa and the origin of calciocarbonatite magmas. J. Petrol. **37**, 1125-1152.

Lee W-J. and Wyllie P.J. (1998) Petrogenesis of carbonatite magmas from mantle to crust, constrained by the system CaO-(MgO + FeO*)-(Na$_2$O + K$_2$O)-(SiO$_2$ + Al$_2$O$_3$ + TiO$_2$)-CO$_2$. J. Petrol. **39**, 495-517.

Lee C.-T. A., Luffi P., Höink T., Li J., Dasgupta R. and Hernlund J. (2010) Upside-down differentiation and generation of a 'primordial' lower mantle. Nature **463**, 930-933.

Lesne P., Scaillet B., Pichavant M. and J.-M. Beny (2010) The carbon dioxide solubility in alkali basalts: an experimental study. Contrib. Mineral. Petrol., doi:10.10007/s00410-010-0585-0

Litasov K. and Ohtani E. (2010) The solidus of carbonated eclogite in the system CaO-Al$_2$O$_3$-MgO-SiO$_2$-Na$_2$O-CO$_2$ to 32 GPa and carbonatite liquid in the deep mantle. Earth Planet. Sci. Lett. 295, 115-126.

Liu Q. and Lange R.A. (2003) New density measurements on carbonate liquids and the partial molar volume of the CaCO$_3$ component. Contrib. Mineral. Petrol. **146**, 370-381.

Luth R.W. (2006) Experimental study of the CaMgSi$_2$O$_6$-CO$_2$ system at 4-8 GPa. Contrib. Mineral. Petrol. **151**, 141-157.

McDonough W.F. and Sun S.S. (1995) The composition of the Earth. Chem. Geol. **120**, 223-253.

McDonough W.F. and Rudnick R.L. (1998) Mineralogy and composition of the upper mantle. Rev. Mineral. **37**, 139-164.

Markgraf S.A. and Reeder R.J. (1985) High-temperature structure refinements of calcite and magnesite. Am. Mineral. **70**, 590-600.

Marty B. and Tolstikhin I.N. (1998) CO$_2$ fluxes from mid-ocean ridges, arcs and plumes. Chem. Geol. **145**, 233-248.

Mattey D.P. (1991) Carbon dioxide solubility and carbon isotope fractionation in basaltic melt. Geochim. Cosmochim. Acta **55**, 3467-3473.

Mattey D.P., Taylor W.R., Green D.H. and Pillinger C.T. (1990) Carbon isotopic fractionation between CO$_2$ vapour, silicate and carbonate melts: an experimental study to 30 kbar. Contrib. Mineral. Petrol. **104**, 492-505.

Minarik W.G. (1998) Complications to carbonate melt mobility due to the presence of an immiscible silicate melt. J. Petrol. **39**, 1965-1973.

Mitchell R.H. (2005) Carbonatites and carbonatites and carbonatites. Can. Mineral. **43**, 2049-2068.

Mitchell R.H. (2008) Petrology of hypabyssal kimberlites: Relevance to primary magma compositions. J. Volcanol. Geoth. Res. **174**, 1-8.

Moore K.R. and Wood B.J. (1998) The transition from carbonate to silicate melts in the CaO-MgO-SiO$_2$-CO$_2$ system. J. Petrol. **39**, 1943-1951.

Morris J.R. and Song X. (2002) The melting lines of model systems calculated from coexistence simulations. J. Chem. Phys. **116**, 9352-9358.

Morizet Y., Kohn S.C. and Brooker R.A. (2001) Annealing experiments on CO$_2$-bearing jadeite glass: an insight into the true temperature dependence of CO$_2$ speciation in silicate melts. Mineral. Mag. **65-66**, 701-707

Morizet Y., Brooker R.A. and Kohn S.C. (2002) $CO_2$ in haplo-phonolite melt: Solubility, speciation and carbonate complexation. Geochim. Cosmochim. Acta **66**, 1809-1820.

Morizet Y., Nichols A.R.L., Kohn S.C., Brooker R.A. and Dingwell D.B. (2007). The influence of $H_2O$ and $CO_2$ on the glass transition temperature: insights into the effects of volatiles on magma viscosity. Eur. J. Mineral. **19**, 657-669.

Morizet Y., Paris M., Gaillard F. and Scaillet B. (2010) C-O-H fluid solubility in haplobasalt under reducing conditions: An experimental study. Chem. Geol., doi:10.1016/j.chemgeo.2010.09.011

Mysen B.O. (1976) The role of volatiles in silicate melts: Solubility of carbon dioxide and water in feldspar, pyroxene, and feldspathoïd melts to 30 kbar and 1625°C. Am. J. Sci. **276**, 969-996.

Mysen B.O. and Virgo D. (1980a) Solubility mechanisms of carbon dioxide in silicate melts: a Raman spectroscopic study. Am. Mineral. **65**, 885-899.

Mysen B.O. and Virgo D. (1980b) The solubility behavior of $CO_2$ in melts on the join $NaAlSi_3O_8$-$CaAl_2Si_2O_8$-$CO_2$ at high pressures and temperatures: a Raman spectroscopic study. Am. Mineral. **65**, 1166-1175.

Mysen, B.O. and Richet P., 2005. Silicate Glasses and Melts: Properties and Structure. Elsevier, Amsterdam.

Mysen B.O., Arculus R.J. and Eggler D.H. (1975) Solubility of carbon dioxide in melts of andesite, tholeiite and olivine nephelinite composition to 30 kbar pressure. Contrib. Mineral. Petrol. **53**, 227-239.

Mysen B.O., Eggler D.H., Seitz M.G. and Holloway J.R. (1976) Carbon dioxide in silicate melts and crystals Part I: Solubility measurements. Am. J. Sci. **276**, 455-479.

Nowak M., Schreen D. and Spickenbom K. (2004) Argon and $CO_2$ on the race track in silicate melts: A tool for the development of a $CO_2$ speciation and diffusion model. Geochim. Cosmochim. Acta **68**, 5127-5138.

Nowak M., Porbatzki D., Spickenbom and Diedrich O. (2003) Carbon dioxide speciation in silicate melts: a restart. Earth Planet. Sci. Lett. **207**, 131-139.

Ohtani E. and Maeda M. (2001) Density of basaltic melt at high pressure and stability of the melt at the base of the lower mantle. Earth Planet. Sci. Lett. 193, 69-75.

Pan V., Holloway J.R. and Hervig R.L. (1991) The pressure and temperature dependence of carbon dioxide solubility in tholeiitic basalt melts. Geochim. Cosmochim. Acta **55**, 1587-1595.

Panina L.I. and Motorina I.V. (2008) Liquid immiscibility in deep-seated magmas and the generation of carbonatite melts. Geochem. Int. **46**, 448-464.

Parr R.G. and Yang W. (1995) Density-functional theory of the electronic structure of molecules. Ann. Rev. Phys. Chem. **46**, 701-728.

Pavese A., Catti M., Parker S.C. and Wall A. (1996) Modelling of the thermal dependence of structural and elastic properties of calcite, $CaCO_3$. Phys. Chem. Minerals **23**, 89-93.

Pawley A.R., Holloway J.R. and McMillan P.F. (1992) The effect of oxygen fugacity on the solubility of carbon-oxygen fluids in basaltic melt. Earth Planet. Sci. Lett. **110**, 213-225.

Perdew J.P., Ruzsinsky A., Constantin L.A., Sun J. and Csonka G.I. (2009) Some fundamental issues in ground-state density functional theory: A guide for the perplexed. J. Chem. Theory Comput. **5**, 902-908.

Peterson T.D. (1989) Peralkaline nephelinites. I. Comparative petrology of Shombole and Oldoinyo L'engai, East Africa. Contrib. Mineral. Petrol. **101**, 458-478.

Porbatzki D. and Nowak M. (2001) Annealing $CO_2$-bearing silicate glasses: A key to quantify $CO_2$ species in silicate melts?. Beih. Eur. J. Mineral. **13**, 143.

Rai C.S., Sharma S.K., Muenow D.W., Matson D.W. and Byers C.D. (1983) Temperature dependence of $CO_2$ solubility in high pressure quenched glasses of diopside composition. Geochim. Cosmochim. Acta **47**, 953-958.

Reynolds J.R. and Langmuir C.H. (1997) Petrological systematics of the mid-Atlantic ridge south of Kane: Implications for ocean crust formation. J. Geophys. Res. B **102**, 14915-14946.

Ringwood A.E., Kesson S.E., Hibberson W. and Ware N. (1992) Origin of kimberlites and related magmas. Earth Planet. Sci. Lett. **113**, 521-538.

Rivera J.L., McCabe C. and Cummings P.T. (2003) Molecular simulations of liquid-liquid interfacial properties: Water-n-alkane and water-methanol-n-alkane systems. Phys. Rev. E **67**, 11603-11612.

Russell J.K., Giordano D. and Dingwell D.B. (2003) High-temperature limits on viscosity of non-Arrhenian silicate melts. Am. Mineral. **88**, 1390-1394

Saal A.E., Hauri E.H., Langmuir C.H. and Perfit M.R. (2002) Vapour undersaturation in primitive mid-ocean-ridge basalt and the volatile content of Earth's upper mantle. Nature **419**, 451-455.

Saharay M. and Balasubramanian S. (2004) *Ab-initio* molecular-dynamics study of supercritical carbon dioxide. J. Chem. Phys. **120**, 9694-9702.

Saharay M. and Balasubramanian S. (2007) Evolution of intermolecular structure and dynamics in supercritical carbon dioxide with pressure: An ab initio molecular dynamics study. J. Phys. Chem. B **111**, 387-392.

Santillán J., Catalli K. and Williams Q. (2005) An infrared study of carbon-oxygen bonding in magnesite to 60 GPa. Am. Mineral. **90**, 1669-1673.

Sarda Ph. and Graham D. (1990) Mid-ocean ridge popping rocks: implications for degassing at ridge crests. Earth Planet. Sci. Lett. **97**, 268-289.

Shcheka S.S., Wiedenbeck, Frost D.J. and Keppler H. (2006) Carbon solubility in mantle minerals. Earth Planet. Sci. Lett. **245**, 730-742.

Shilobreyeva and Kadik A.A. (1989) Solubility of $CO_2$ in magmatic melts at high temperatures and pressures. Geokhimiya **7**, 950-960.

Sierralta M., Nowak M. and Keppler H. (2002) The influence of bulk composition on the diffusivity of carbon dioxide in Na aluminosilicate melts. Am. Mineral. **87**, 1710-1716.

Smith W. and Forester T. (1996) DL_POLY_2.0: a general-purpose parallel molecular dynamics simulation package. J. Mol. Graph. **14**, 136-141.

Span R. and Wagner W. (1996) A new equation of state for carbon dioxide covering the fluid region from the triple-point temperature to 1100 K at pressures up to 800 MPa. J. Phys. Ref. Data **25**, 1509-1594.

Sparks R.S.J., Baker L., Brown R.J., Field M., Schumacher J., Stripp G. and Walters A. (2006) Dynamical constraints on kimberlite volcanism. J. Volcanol. Geoth. Res. **155**, 18-48.

Sparks R.S.J., Brooker R.A., Field M., Kavanagh J., Schumacher J.C., Walter M.J. and White J. (2009) The nature of erupting kimberlite melts. Lithos **112**, 429-438.

Spera F.J. and Bergman S.C. (1980) Carbon dioxide in igneous petrogenesis: I Aspects of the dissolution of $CO_2$ in silicate liquids. Contrib. Mineral. Petrol. **74**, 55-66

Spera F.J. (1981) Carbon dioxide in igneous petrogenesis: II Fluid dynamics of mantle metasomatism. Contrib. Mineral. Petrol. **77**, 56-65

Spera F.J. (1984) Carbon dioxide in petrogenesis III: role of volatiles in the ascent of alkaline magma with special reference to xenolith-bearing mafic lavas. Contrib. Mineral. Petrol. **88**, 217-232

Spickenbom K. and Nowak M. (2005) Diffusion of Ar and $CO_2$ in Na aluminosilicate melts. Geophys. Res. Abs. **7**, 3436-3437.

Spickenbom K., Sierralta M. and Nowak M. (2010) Carbon dioxide and argon diffusion in silicate melts: Insights into the $CO_2$ speciation in magmas. Geochim. Cosmochim. Acta **74**, 6541-6564

Stagno V. and Frost D.J. (2010) Carbon speciation in the asthenosphere: Experimental measurements of the redox conditions at which carbonate-bearing melts coexist with graphite or diamond in peridotite assemblages. Earth Planet. Sci. Lett. **300**, 72-84.

Stebbins J.F., Dubinsky E.V., Kanehashi K., Kelsey K.E. (2008) Temperature effects on non-bridging oxygen and aluminum coordination number in calcium aluminosilicate glasses and melts. Geochim. Cosmochim. Acta **72**, 910-925.

Stefanovich E.V., Boldyrev A.I., Truong T.N. and Simons J. (1998) Ab initio study of the stabilization of multiply charged anions in water. J. Phys. Chem. B **21**, 4205-4208.

Sterner S.M. and Pitzer K.S. (1994) An equation of state for carbon dioxide valid from zero to extreme pressures. Contrib. Mineral. Petrol. **117**, 362-374.

Stolper E. and Holloway J.R. (1988) Experimental determination of the solubility of carbon dioxide in molten basalt at low pressure. Earth Planet. Sci. Lett. **87**, 397-408.

Stolper E., Fine G., Johnson T. and Newman S. (1987) Solubility of carbon dioxide in albitic melt. Am. Mineral. **72**, 1071-1085.

Tamic N., Behrens H. and Holtz F. (2001) The solubility of $H_2O$ and $CO_2$ in rhyolitic melts in equilibrium with a mixed $CO_2$-$H_2O$ fluid phase. Chem. Geol. **174**, 333-347.

Taylor W.R. (1990) The dissolution mechanism of $CO_2$ in aluminosilicate melts - infrared spectroscopic constraints on the cationic environment of dissolved $[CO_3]^{2-}$. Eur. J. Mineral. **2**, 547-563.

Thibault Y. and Holloway J.R. (1994) Solubility of $CO_2$ in a Ca-rich leucite: effects of pressure, temperature, and oxygen fugacity. Contrib. Mineral. Petrol. **116**, 216-224.

Thomsen T.B. and Schmidt M.W. (2008) Melting of carbonated pelites at 2.5-5.0 GPa, silicate-carbonatite liquid immiscibility, and potassium-carbon metasomatism of the mantle. Earth Planet Sci. Lett. **267**, 17-31.

Tingle T.N. and Aines R.D.(1988) Beta track auto radiography and infrared spectroscopy bearing on the solubility of $CO_2$ in albite melt at 2 GPa and 1450°C. Contrib. Mineral. Petrol. **100**, 222-225.

Tossell J.A. (1995) Calculation of the $^{13}C$ NMR shieldings of the $CO_2$ complexes of aluminosilicates. Geochim. Cosmochim. Acta **59**, 1299-1305.

Van Gunsteren W. and Berendsen H.J.C. (1990) Computer simulation of molecular dynamics: Methodology, applications, and perspectives in chemistry. Angew. Chem. Int. **29**, 992-1023.

Vuilleumier R., Sator N. and Guillot B. (2009) Computer modeling of natural silicate melts: What can we learn from ab initio simulations. Geochim. Cosmochim. Acta **73**, 6313-6339.

onatite melt from deeply subducted oceanic crust. Nature **454**, 622-626.

Watson E.B. (1991) Diffusion of dissolved $CO_2$ and Cl in hydrous silicic to intermediate magmas. Geochim. Cosmochim. Acta **55**, 1897-1902.

Watson E.B., Sneeringer M.A. and Ross A. (1982) Diffusion of dissolved carbonate in magmas: experimental results and applications. Earth Planet. Sci. Lett. **61**, 346-358.

Wendlandt R.F. and Mysen B.O. (1980) Melting phase relations of natural peridotite + $CO_2$ as a function of degree of partial melting at 15 and 30 kbar. Am. Mineral. **65**, 37-44.

White B.S. and Montana A. (1990) The effect of $H_2O$ and $CO_2$ on the viscosity of sanidine liquid at high-pressures. J. Geophys. Res. B **95**, 15683-15693.

Wilding M.C., Benmore C.J. and Weber J.K.R. (2008) In situ diffraction studies of magnesium silicate liquids. J. Mat. Sci. **43**, 4707-4713.

Wilding M.C., Benmore C.J. and Weber J.K.R. (2010) Changes in the local environment surrounding magnesium ions in fragile $MgO-SiO_2$ liquids. Euro. Phys. Lett. **89**, 26005-1-5.

Wilson L. and Head III J.W. (2007) An integrated model of kimberlite ascent and eruption. Nature **447**, 53-57.

Wyllie P.J. and Huang W.L. (1975) Peridotite, kimberlite and carbonatite explained in the system $CaO-MgO-SiO_2-CO_2$. Geology **3**, 621-624.

Wyllie P.J. and Lee W.J. (1998) Model system controls on conditions for formation of magnesiocarbonatite and calciocarbonatite magmas from the mantle. J. Petrol. **39**, 1885-1893.

Yee G.G., Fulton J.L. and Smith R.D. (1992) Fourier transform infrared spectroscopy of molecular interactions of heptafluoro-1-butanol or 1-butanol in supercritical carbon dioxide and supercritical haxane. J. Phys. Chem. **96**, 6172-6181.

Zhang Y. and Stolper E.M. (1991) Water diffusion in a basaltic melt. Nature **351**, 306-309.

Zhang Y. and Zindler A. (1993) Distribution of carbon and nitrogen in Earth. Earth Planet. Sci. Lett. **117**, 331-345.

Zhang Z. and Duan Z. (2005) An optimized molecular potential for carbon dioxide. J. Chem. Phys. **122**, 214507-1-15.